%% file: ICDE.tex
\documentclass[conference]{IEEEtran}
\IEEEoverridecommandlockouts
\usepackage{balance}
\usepackage{cite}
\usepackage{amsmath,amssymb,amsfonts}
\usepackage{graphicx}
\usepackage{textcomp}
\usepackage{xcolor}
\usepackage[hidelinks]{hyperref}

\def\BibTeX{{\rm B\kern-.05em{\sc i\kern-.025em b}\kern-.08em
    T\kern-.1667em\lower.7ex\hbox{E}\kern-.125emX}}

\usepackage{graphicx}
\usepackage{subcaption}
\usepackage[noend]{algpseudocode}
\usepackage{algorithm}
\usepackage{amsthm}
\usepackage{amssymb}
\usepackage{amsmath}
\usepackage{array}     
\usepackage{booktabs} 
\usepackage{makecell}
\usepackage{float}  % 在导言部分添加此行
\usepackage{xspace}

\definecolor{darkred}{RGB}{128,0,0}
\newcommand{\rref}[2]{\hyperref[#1]{\textcolor{darkred}{#2}}}

\newcommand{\kw}[1]{{\ensuremath {\mathsf{#1}}}\xspace}

\newcommand{\ACQPP}{\kw{ACS}-\kw{PP}}

\newcommand{\stitle}[1]{\vspace{1ex} \noindent{\bf #1}}

\newcommand{\squishlisttight}{
 \begin{list}{$\bullet$}
  { \setlength{\itemsep}{0pt}
    \setlength{\parsep}{0pt}
    \setlength{\topsep}{0pt}
    \setlength{\partopsep}{0pt}
    \setlength{\leftmargin}{2em}
    \setlength{\labelwidth}{1.5em}
    \setlength{\labelsep}{0.5em}
 } }

\newcommand{\squishnumlist} {
\newcounter{qcounter}
\begin{list}{\arabic{qcounter}.~}{\usecounter{qcounter}} 
{  \setlength{\itemsep}{0pt}
    \setlength{\parsep}{0pt}
    \setlength{\topsep}{0pt}
    \setlength{\partopsep}{0pt}
    \setlength{\leftmargin}{2em}
    \setlength{\labelwidth}{1.5em}
    \setlength{\labelsep}{1.5em}
}}

\newcommand{\squishend}{
  \end{list}
}

\newcommand{\blue}[1]{\textcolor{black}{#1}}
\newtheorem{problem}{Problem}
\newtheorem{definition}{Definition}
\newtheorem{example}{Example}

\newtheorem{observation}{Observation}

\begin{document}

\twocolumn

\title{Efficient Community Search on Attributed Public-Private Graphs
% {\footnotesize \textsuperscript{*}Note: Sub-titles are not captured for https://ieeexplore.ieee.org  and
}

\author{\IEEEauthorblockN{Yuqi Chen}
\IEEEauthorblockA{\textit{Hong Kong Baptist University} \\
Hong Kong, China \\
csyqchen@comp.hkbu.edu.hk}
\and
\IEEEauthorblockN{Weihan Zhang}
\IEEEauthorblockA{\textit{Sun Yat-sen University} \\
Guangzhou, China \\
zhangwh79@mail2.sysu.edu.cn}
\and
\IEEEauthorblockN{Xin Huang}
\IEEEauthorblockA{\textit{Hong Kong Baptist University} \\
Hong Kong, China \\
xinhuang@comp.hkbu.edu.hk}
}

\maketitle

\begin{abstract}
Public-private graph, where a public network is visible to everyone and every user is also associated with its own small private graph accessed by itself only, widely exists in real-world applications of social networks %, collaboration networks, 
and financial networks.    
%combines public and private information related to a query node while preserving privacy. 
Most existing work on community search, finding a query-dependent community containing a given query, only studies on a public graph, neglecting the privacy issues in public-private networks. However, considering both the public and private attributes of users enables community search to be more accurate, comprehensive, and personalized to discover hidden patterns. 
%By integrating private connections and attributes with public data, this approach reveals hidden patterns, enhances the understanding of node relationships, and provides deeper insights for applications such as social interaction and personalized recommendations. 

In this paper, we study a novel problem of \textit{attributed community search in public-private graphs} (\ACQPP), aiming to find a connected $k$-core community that shares the most keywords with the query node. This problem uncovers structurally cohesive communities, such as interest-based user groups or core teams in collaborative networks. To optimize search efficiency, we propose an integrated scheme of constructing a public global graph index and a private personalized graph index. For the private index, we developed a compact structure of the PP-FP-tree index. %and the corresponding conditional PP-FP-tree. 
The PP-FP-tree is constructed based on the public and private neighbors of the query node in the public-private graph, serving as an efficient index to mine frequent node sets that share the most common attributes with the query node. %The conditional PP-FP tree acts as a supplementary structure, enabling a comprehensive exploration of the shared attributes between the node set and the query node, thereby ensuring attribute maximality for the identified communities. 
Extensive experiments on real public-private graph datasets validate both the efficiency and quality of our proposed PP-FP search algorithm against existing competitors. The case study on public-private collaboration networks %demonstrates
provides insights into the discovery of public-private communities. 
\end{abstract}

\maketitle

\section{Introduction}
\input{intro}

\section{Related Work}
\input{related}

\input{problem}

\input{basic}

\input{index}

\input{experiment}

\section{Conclusion} 
%In this paper, we study a novel \ACQPP problem, which aims to identify cohesive public-private core communities containing a query node. We propose several fast querying approaches that first mine frequent attributes and node sets in the local public-private neighborhood of the query node and then search the entire graph for a $k$-core community that maximizes shared attributes. %To achieve this, 
%To further improve efficiency, we develop a new private index based on the PP-FP-tree and the conditional PP-FP-tree to support efficient and high-quality community queries. As an effective approximate algorithm, our PP-FP algorithm has been validated through extensive experiments, demonstrating its efficiency and effectiveness. A case study further demonstrates the benefits of community search on public-private graphs, particularly in enabling early detection of communities and revealing more comprehensive and up-to-date information.

This paper formulates and studies a novel \ACQPP problem to find cohesive communities containing query nodes over public-private graphs. We develop an index-based framework by designing new private index of public-private frequent pattern trees, which can support the fast search of common attributes. Extensive experiments are conducted on real-world public-private graphs with ground-truth communities to show the superiority of our approaches against state-of-the-art competitors. A case study on public-private collaboration graphs reveals that our model can help early identify the underlying communities working on an ongoing project.  

\section{AI-Generated Content Acknowledgement} 
The authors confirm that no content in this paper was generated by AI tools.

\bibliographystyle{IEEEtran} % 指定使用 IEEEtran.bst 样式文件
\bibliography{reference}

\end{document}

%% file: intro.tex
%Community search in traditional social networks primarily focuses on the connections between nodes, but it receives challenges in more complex scenarios. On the one hand, the scale of social network graphs has expanded dramatically, leading to high complexity; on the other hand, node features have also begun to be considered. 
%Consequently, recent works have increasingly focused on attributed graphs \cite{huang2016attribute, fang2017effective, chen2019contextual, zhu2020structure, jiang2022effective}. These graphs not only contain the topological structures but also combine the attributes of the nodes to depict the social network more comprehensively. However, these works assume that all node attributes and their connections are public, overlooking the possibility that certain relationships or attributes may be hidden. This limitation makes them insufficient in uncovering hidden communities and meeting privacy protection needs influenced by private information.

%public-private graph model
%Graph is a widely used model 
Graphs are widely used to depict entities and their complex relationships, such as social networks, collaboration networks, and so on \cite{luo2021detecting, ma2022convex, chen2021efficient, dong2021butterfly, jiang2021efficient, fang2020survey}. However, not all edge information may be publicly accessible in various graph data. For example, users may prefer to hide their friend relationships, e.g., in the Facebook social network, 52.6\% of 1.4 million New York City users chose to hide friend lists~\cite{dey2012facebook}. Such a privacy-preserving network is modeled as a public-private network~\cite{chierichetti2015efficient, yu2021public}, in which a public network is visible to all users and each user is associated with their own small private graph. 
Public-private networks widely exist in various applications of the Weibo social network and academic collaboration networks. 
%For example, on the Weibo network, users have the following relationships to track their interested posts. However, the network platform provides a function of ``secretly following" to identify those users, which are only visible to its users but cannot be accessed by other users.
%In collaboration networks, two scholars have co-authored a research paper regarding having a collaboration edge. 
In collaboration networks, an edge between two researchers indicates that they have co-authored a research paper.
All public edges are regarded as the collaboration of published papers in the existing literature.  However, one ongoing research work that has not been published yet is treated as the private collaborations for these authors. 
In terms of the view of each user, the whole network is the union of the public network and its private graph.

\begin{figure}[!t]
\includegraphics[width=\linewidth]{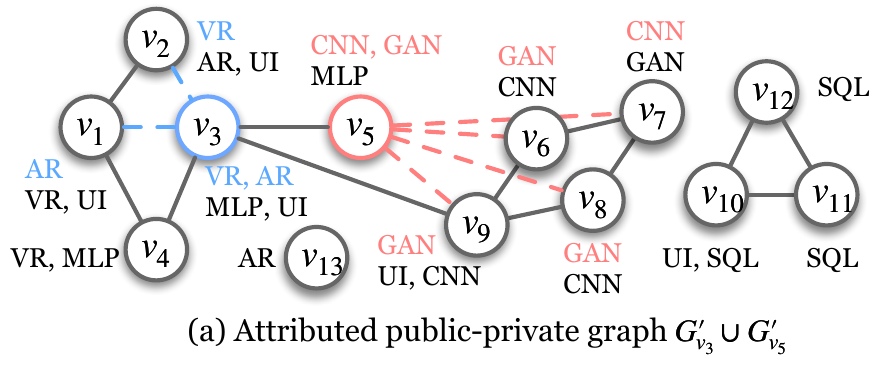}
\label{fig:1-1}
\vspace{-0.9cm}
\end{figure}

\begin{figure}[t]
\includegraphics[width=\linewidth]{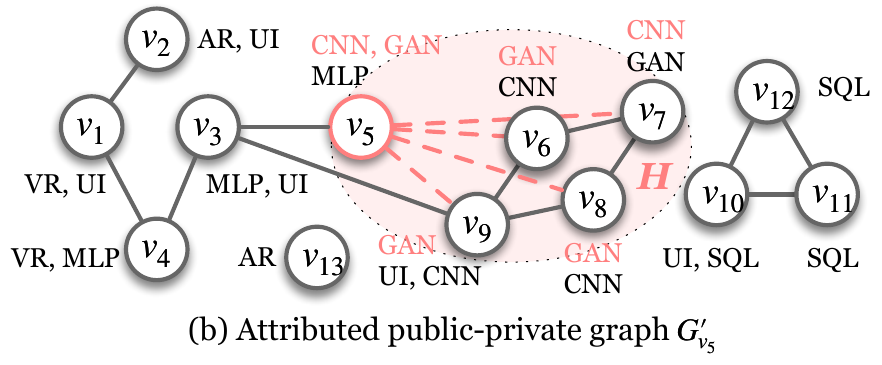}
\vspace{-0.3cm}
\caption{An attributed public-private graph has two private star-graphs $G_{v_3}$ and $G_{v_5}$. For a query $q=v_5$ and $k=3$, the 3-core attributed community is the pink area $H$ in Figure~\ref{AC-PP}(b).}
\label{AC-PP}
\vspace{-0.3cm}
\end{figure}   

%pp-graph analytics 
The public-private graph affects not only the view of the network structure but also the task results of different algorithms. 
% As a result, public-private graph analytics is essentially important to attract significant studies on various tasks, 
As a result, public-private graph analytics is essential for attracting various research across various tasks, including correlation \cite{bansal2004correlation}, clustering coefficient \cite{chierichetti2015efficient}, keyword search \cite{jiang2020ppkws}, truss mining \cite{ebadian2019fast},  all-pair shortest paths \cite{das2010sketch}, etc.

% \begin{figure}[!t]
%     \centering
%     % 第一个子图，占比 70%
%     \begin{minipage}{0.7\linewidth}
%         \centering
%         \includegraphics[width=\linewidth]{figures/example-1-3.pdf}
%         \label{fig:1-3}
%     \end{minipage}
%     \hfill
%     % 第二个子图，占比 28%
%     \begin{minipage}{0.28\linewidth}
%         \centering
%         \includegraphics[width=\linewidth]{figures/example-1-4.pdf}
%         \label{fig:1-4}
%     \end{minipage}
    
%     \vspace{-0.3cm}  % 控制图片与标题的间距    
% \end{figure}

%community search 
Community search aims to find a query-dependent community containing a given set of query nodes in graphs, which has wide applications in social relations discovery and personalized recommendations. 
In the literature, many dense subgraph-based community models have been studied in different types of networks, including simple graphs~\cite{sozio2010cocktail}, directed graphs, attributed graphs, multilayer graphs, and heterogeneous graphs~\cite{fang2020effective, liu2023significant}.
Unfortunately, community search over public-private networks has not been investigated, due to the technical challenges of indexing private information for efficiency. As each user has its own private network, it is an expensive cost to build precomputed information for all users' public-private network in a real large-scale graph with millions of users. 
%keywords
Moreover, except for the topological structure, nodes are usually associated with attributes to represent the node properties, e.g., gender, interests, check-in information, and so on. 
%Imporantce of communities with more shared attributes.
%For an attributed community, it is good to have not only dense connections between community members, but also all community members have more common attributes, indicating they share more homogeneous properties and close relationships. 
Thus, it is important to discover query-dependent attributed communities over such public-private attributed graphs.

In this paper, we study a new problem of \underline{a}ttributed \underline{c}ommunity \underline{s}earch in \underline{p}ublic-\underline{p}rivate graphs (\ACQPP) based on $k$-core, where each node has at least $k$ neighbors within it~\cite{fang2017effective, li2020efficient, lu2022time}.
For an attributed public-private graph, the public graph is represented by the black edges in Figure~\ref{AC-PP}(a), while the blue and red edges correspond to the private graphs of $v_3$ and $v_5$, respectively. 
Given a query node $q$, our novel model seeks to find a connected $k$-core containing $q$ that maximizes the number of common attributes in the union of the public graph and $q$'s private graph. For example, when the query $q=v_5$, the public-private community $H$ is the 3-core containing $v_5$ with the largest number of two common attributes  \{``CNN", ``GAN"\} shown in Figure~\ref{AC-PP}(b). 
To tackle \ACQPP, one straightforward method is to first enumerate the possible common attributes 
%and check whether there exists a feasible community containing the searched attributes, 
and then 
return one  feasible community with the largest attributes. 
%We can enumerate the attributes starting from small to large or adopt the binary search strategy. 
The enumeration can process in ascending order of attribute size or adopt a binary search strategy.  
However, both methods are time-consuming due to an exponential number of validations of feasible communities.  

To reduce the search space and accelerate $k$-core validation, we explore indexing-based methods for accelerating \ACQPP. Due to the complex attributed network and the sensitive information of private graphs only owned by its users, 
designing an effective indexing structure to support queries on any node $q$ is challenging.
%it brings challenges to designing an effective indexing structure to support querying on any node $q$.
The reason is threefold. First, the community structure and shared keywords may be independent of each other in the community model, which brings two-dimensional quality requirements to consider simultaneously.
Second, for different users, it may contain quite different private structures in the graph and node attributes, in terms of size and distribution. 
Third, designing indexes for each public-private graph is impractical for a large network with millions of nodes.
%, as it needs to build indexes millions of times. %Moreover, most of these computations are redundant, as they repeatedly process the large part of the same public graph.

To tackle these, we design a new index of public-private attribute-based frequent pattern tree (PP-FP-tree).
For each node $v$, we build the PP-FP-tree on a small neighborhood graph by integrating its neighbors' public edges and its associated private graph. The idea of PP-FP-tree is to preserve the common attributes that frequently appear in its neighborhood for fast retrieval to quickly identify the possible number of maximum common attributes. 
% In addition, we equip the PP-FP tree with an auxiliary structure of node coreness, which keeps records of the largest $k$-core that the node can involve in public graphs. 
In addition, we equip the PP-FP-tree with an auxiliary structure, the graph coreness index, which keeps records of the largest $k$-core that the node can be involved in public graphs. 
Based on the PP-FP tree, we propose an efficient algorithm for \ACQPP, which first selects a small number of candidates %based on our compact PP-FP tree index
and then performs $k$-core validation through local exploration.  
% local exploration ?
To summarize, this paper makes the following contributions:
\squishlisttight
\item 
%We formulate and study a novel problem of attributed community search over public-private graphs. The designed community model satisfies the cohesive community structure of $k$-core and admits the largest common attributes to achieve the same property in public and private graphs \textbf{(Section 3)}. 
We formulate and study a novel problem of attributed community search over public–private graphs
%. The proposed community model preserves the $k$-core structural cohesiveness and maximizes the overlap of common attributes in public and private graphs 
\textbf{(Section 3)}. 
We also present two basic solutions that combine fundamental $k$-core decomposition along with the handling of candidate attribute sets \textbf{(Section 4)}.

%The problem we study is to identify a highly relevant and cohesive core community related to a query node. We define the ACQ-PP problem to find a community in public-private graphs that maximizes the shared attributes with the query node while satisfying user-given $k$-core constraints (Section 3). 

\item We propose an integrated scheme of constructing \emph{public global graph index} and \emph{private personalized graph index}. For the private index, we design a compact PP-FP-tree index for each node with a private graph. These indexes allow us to efficiently search for communities in the public-private graph while protecting the privacy of each private graph user \textbf{(Section 5)}. 
%introduce our public-private index construction, which involves building the coreness tree index and the graph attribute map for the large public graph, as well as constructing a dedicated PP-FP tree index for each node with a private graph. These indexes allow us to efficiently search for communities in the public-private graph while protecting the privacy of each private graph user (Section 5). 
Based on the PP-FP-tree index, we propose our efficient indexing-based search algorithm, PP-FP. %First, the PP-FP tree index identifies the largest attribute set and the matching nodes. Second, the graph attribute map retrieves nodes from the public graph that match this attribute set, forming $AttrSet$. Third, the coreness tree index identifies $CoreSet$, the nodes that could potentially form a $k$-core with the query node. Finally, we obtain the intersection of $AttrSet$ and $CoreSet$ to 
In this way, we fast extract key candidates and perform community validation for searching communities \textbf{(Section 6)}.

\item Experiments demonstrate the  superiority of our PP-FP algorithm in addressing the \ACQPP problem, showcasing significant advantages in query efficiency and community quality for large-scale public-private graphs
%. Our case study, exploring attribute-maximized communities in both public-private and public graphs, reflects the advantages of our proposed \ACQPP community model and the applications of discovering up-to-date private communities 
\textbf{(Section 7)}.
%highlights the advantage of community searches in the public-private context and supports the novelty of our ACQ-PP problem definition, highlighting 
\squishend

%% file: related.tex
%Our work closely relates to community discovery, attributed community search, and public-private graph analytics. 

\stitle{Community discovery.} The studies of community discovery are usually categorized into two types of community detection and community search. 
Community detection \cite{bothorel2015clustering, luo2021detecting, ma2022convex, li2018community} identifies cohesive subgroups throughout the graph, while community search \cite{chen2021efficient, dong2021butterfly, huang2014querying, li2018persistent, cui2014local, yuan2017index, jiang2021efficient, fang2020survey} is a query-based process that aims to find the most relevant community for a given node or set of nodes. 
Fang et al. \cite{fang2020survey} review community search algorithms based on structural cohesiveness metrics such as $k$-core \cite{cui2014local, fang2018effective}, $k$-truss \cite{huang2014querying, akbas2017truss, huang2015approximate}, $k$-clique \cite{yuan2017index, wang2017query} and $k$-edge-connected components \cite{chang2015index, hu2017minimal}, and compare them on different graph types, including keyword-based, temporal \cite{li2018persistent} and spatial graphs \cite{fang2017effective1, zhu2017geo}. 

%\stitle{Attributed community search.} Attribute community search focuses on finding communities in a graph that have a cohesive structure and attribute similarity. Related studies include \cite{huang2016attribute, fang2017effective, chen2019contextual, zhu2020structure, jiang2022effective}. 
%Huang et al. \cite{huang2016attribute} propose a method to identify attributed truss communities, which are densely connected subgraphs that maximize attribute relevance. Fang et al. \cite{fang2017effective} propose the CL-tree index, which organizes $k$-core and keyword into a hierarchical tree structure, making attribute-based community search more efficient on large-scale graphs. \cite{chen2019contextual} introduces a contextual community search model that is parameter-free, allowing users to search for communities within large social networks based on a set of query keywords. Unlike traditional methods that specify community structural parameters, this model focuses on both structural and contextual cohesiveness.Different from \cite{huang2016attribute}, which primarily focuses on maximizing attribute similarity, cohesive attributed community search  \cite{zhu2020structure} emphasizes the structural connectivity of the community, ensuring tight connections between internal nodes by avoiding cut-edges and cut-vertices. Jiang et al. \cite{jiang2022effective} develop an efficient community search framework for large star-schema heterogeneous information networks.
%, using meta-path-based core models to find communities without user-specified constraints. \par

\stitle{Attributed community search.} Attributed community search seeks subgraphs that are both structurally cohesive and attribute similar \cite{huang2016attribute, fang2017effective, chen2019contextual, zhu2020structure, jiang2022effective}. Huang et al. \cite{huang2016attribute} propose attributed truss communities that maximize attribute relevance within dense subgraphs. Fang et al. \cite{fang2017effective} design the CL-tree, a hierarchical tree index that combines $k$-core and keyword information for efficient attributed community search. Chen \cite{chen2019contextual} et al. introduce a parameter-free contextual model that retrieves communities by query keywords without preset structural parameters. Zhu et al. \cite{zhu2020structure} focus on cohesive attributed communities by maintaining structural connectivity.
Jiang et al. \cite{jiang2022effective} develop an efficient community search framework for large star-schema heterogeneous information networks.

%\stitle{Public-Private graph analytics.}
%Chierichetti et al. \cite{chierichetti2015efficient} develop scalable and efficient algorithms to compute key metrics such as reachability \cite{archer2017indexing}, centrality \cite{boldi2013hyperball}, and shortest paths \cite{das2010sketch} in public-private social networks. Archer et al. \cite{archer2017indexing} focus on developing efficient algorithms to solve the reachability problem in directed graphs with public, private, and protected nodes. Huang et al. \cite{huang2018pp} generate real-world datasets of public-private networks from DBLP records, called PP-DBLP, to support the analysis and evaluation of public-private network algorithms. Ebadian et al. \cite{ebadian2019fast} develop an efficient algorithm for discovering $k$-truss in public-private graphs. Jiang et al. \cite{jiang2020ppkws} propose an efficient framework, called public-private keyword search, which consists of three major steps: partial evaluation, answer refinement, and answer completion. Yu et al. \cite{yu2021public} propose a public-private graph model for simulating social networks with hidden relationships and introduce the concept of the pp-core number to assess user importance. 
%, where each user has a personalized view of the network, combining both public and private edges. 

\stitle{Public-private graph analytics.}
Chierichetti et al. \cite{chierichetti2015efficient} develop scalable and efficient algorithms to compute key metrics such as reachability \cite{archer2017indexing}, centrality \cite{boldi2013hyperball}, and shortest paths \cite{das2010sketch} in public-private social networks. Archer et al. \cite{archer2017indexing} focus on reachability for directed graphs with public, private, and protected nodes. Huang et al. \cite{huang2018pp} publish the PP-DBLP dataset derived from DBLP to support evaluation. Ebadian et al. \cite{ebadian2019fast} develop an efficient algorithm for discovering $k$-truss in public-private graphs. Jiang et al. \cite{jiang2020ppkws} propose a three-step public-private keyword search framework. %(partial evaluation, answer refinement, and answer completion). 
Yu et al. \cite{yu2021public} introduce a PP-graph model and the pp-core number for measuring user importance. To the best of our knowledge, this is the first work to formulate and study community search on attributed public-private networks. %Our algorithm provides more personalized and meaningful community insights by considering user attributes and connections in public and private networks, and is more efficient and lightweight than prior methods for the public-private graph scenario. 

%To the best of our knowledge, this study is the first work to formulate and study community search on attributed public-private networks. Our algorithm provides more personalized and meaningful community insights by considering user attributes and connections in public and private networks, and is more efficient and lightweight than prior methods for the public-private graph scenario. 
%Compared to previous works, our approach provides a more efficient and lightweight community search method for the public-private graph scenario. 

%% file: problem.tex
\section{Preliminaries}
\label{secpreliminaries}
In this section, we present useful preliminaries and identify the desiderata for communities. Then, we present our problem of public-private attributed community search.

\subsection{Public-Private Community}

% \begingroup
% \setlength{\parskip}{0.5em} % 在此设置段落间距
% \setlength{\parindent}{0pt} % 取消段落缩进

% We consider an undirected, unweighted attributed graph $G = (V, E, A)$, where $V$ is the set of vertices, $E$ is the set of edges, and $A$ is the universal set of attributes. Each vertex $v \in V$ is associated with a set of attributes $attr(v) \subseteq A$.

\textbf{Public attributed graph.} 
We consider a simple model of attributed graph, i.e., an  undirected, unweighted, public attributed graph $G = (V, E, A)$, where $V$ represents the set of vertices, $E$ denotes the set of edges, and $A$ is the universal set of attributes that can be assigned to the vertices. The public neighbors of $v$ is denoted as $N(v)=\{u\in V: (v, u)\in E\}$. 
For each vertex $v \in V$, it is associated with a specific set of attributes, denoted as $attr(v) \subseteq A$. %This means that each vertex can possess multiple attributes, which serve to characterize its properties and relationships within the graph.
This means that each vertex can possess multiple attributes, with $attr(v)$ serving to characterize the vertex's properties and relationships within the graph, such as the users' interests, location check-ins, 
and educational backgrounds in social networks.
Here, the whole structure of $G$ and the attributes $attr(v)$ for any vertex $v$ are publicly accessible to everyone.
However, users can hide their personal contact and properties in their own graphs, leading to a private graph as follows.  

\begin{definition} [Private Star-Graph]
Given a vertex $u\in V$ in a public attributed graph $G$, the private star-graph of $u$ is denoted as $G_u = (V_u, E_u, A_u)$ where $u$ is the center node. Here, the private edge $E_u\subset \{u\}\times V$ is a star structure with  $E_u \cap E = \emptyset$.
Thus, $V_u$ is the set of $u$'s private neighbors, i.e., $V_u= N_p(u)=\{v\in V: (v, u) \in E_u\}$. 
%$E_u \subseteq \{(u,v) \mid v\in V_u\}$ is the set of private edges, which denote the private relationships between $u$ and its private neighbors,
For each vertex $v\in V_u$, $v$ is also associated with private attributes denoted by 
$attr_p(v)$, where $attr_p(v) \subseteq A$ and no common attributes in public and private simultaneously, i.e., $attr_p(v) \cap attr(v)=\emptyset$. 
%is the set of private attributes associated with $u$ and its private neighbors.
\end{definition}

% \xin{Add an example here.}
\begin{table}[t]
    \centering
    \begin{tabular}{>{\arraybackslash}p{2.4cm} >{\arraybackslash}p{5.5cm}} 
    \toprule
    Notation & Meaning \\ \midrule
    $q$ & The query vertex and also the root node of the PP-FP-tree index \\
    $G_q/G'_q$ & The private graph/The public-private graph of the query node $q$ \\
    %$\Theta_{\alpha}$ & The vertex set of attribute $\alpha$ in 1-hop neighborhood\\
    %$\theta_{\alpha}$ & The vertex set of attribute $\alpha$ in public graph $G$\\
    $\Theta_{\alpha}/\theta_{\alpha}$ & The vertex set of attribute $\alpha$ in 1-hop neighborhood/in public graph $G$ \\
    $attr(v)/attr'(v)$ & The public attributes/The public-private attributes of the node $v$ \\
    $privateS/publicS$ & The nodes selected from the private graph index/the public graph index\\
    $S$ & The combination of the sets $publicS$ and $privateS$ for candidate community validation\\
    $attr'(S)$ & The maximum shared attributes of the set $S$
    \\
    %$deg_q$ & The degree of the query node $q$\\
    $N(q)/N'(q)$ & The public neighbors/The public-private neighbors of the query node $q$\\
    $P[v]_x$ & The tree node in the PP-FP-tree $T$\\
    $\kw{Prefix}$-$\kw{Path}(u)$ & All prefix paths ends with node $u$ in tree $T$\\
    $\kw{Prefix}$-$\kw{Path}(P[u]_x)$ & The prefix path ends with node $P[v]_x$ in tree $T$\\
    $coreT$ & The coreness tree index of the public graph $G$\\
    % $gam$ & The graph attribute map of the public graph $G$ \\
    \bottomrule
    \end{tabular}
    \caption{Frequently used notations in this paper}
    \label{symbol}
    \vspace{-0.5cm}
\end{table}

Combining the public graph $G(V, E, A)$ and all private graphs $G_u$ for $u\in V$,  a real-world public-private graph can be
represented as $\mathcal{G}=G\cup\bigcup_{v\in V}G_u$. %For example, 
%\xin{Add an example here G v3 v5.}.
%\begin{example}
In Figure~\ref{AC-PP}(a), a real-world public-private graph is the union of public graph $G$ and all private graphs including $G_{v_3}$ and $G_{v_5}$, represented as $\mathcal{G}=G \cup G_{v_3}\cup G_{v_5}$.
%\end{example}

In the view of different center node, the whole graph structure and attributes are quite different as follows.

\begin{definition} [Public-private Graph]
Given a vertex $u\in V$ in a public attributed graph $G$, the public-private graph in the view of $u$ is the union of public graph $G$ and its private graph $G_u$, i.e., $G\cup G_u$.
\end{definition}
%\begin{example}
As in Figure~\ref{AC-PP}(b), the public-private graph in the view of $v_5$ is formed by the public graph $G$ with its private graph $G_{v_5}$, which is expressed as $G'_{v_5}$ = $G \cup G_{v_5}$.
%\end{example}

%\xin{Add an example here G v3 v5.}.

\noindent \textbf{Dense community structure of k-core\cite{fang2017effective}.} %A $k$-core of a graph $G$ is defined as a maximal subgraph $H$ in which each vertex $v\in H$ has at least degree of $k$.
%As a result, the $k$-core captures densely connected regions of the graph, 
%providing valuable insights into the structural properties and community formations, 
%the context of public-private graph research.
A $k$-core of a graph $G$ is a maximal subgraph $H$ of $G$ such that every vertex $v\in V(H)$ has degree at least $k$ within $H$. In other words, $\forall v\in H$,  $deg_{H}(v) \geq k$. This structural property makes k-cores particularly useful for capturing dense subgraphs, which has been widely used in many core-based community models~\cite{li2020efficient, fang2016effective, lu2022time}. However, the hidden relationships bring significant challenges for identifying cohesive $k$-core in  public-private graphs.  Table~\ref{symbol} lists the notations usually used in the paper.

\noindent \textbf{Desiderata of good  public-private queried community.}
%\textbf{Criteria of a good attributed community:} Given a query node $q$, construct its attributed public-private graph $G'_q = G \cup G'_q = (V\cup V_q, E\cup E_q, A \cup A_q)$, an attributed community is a connected subgraph $H = (V_H, E_H) \subseteq G'_q$ that satisfies:
We analyze the criteria of a good attributed community, w.r.t., a given query vertex $q$ in the public-private graph $G\cup G_q$. Specifically, we regard a public-private connected community $H$ should satisfy the following properties: 

\squishlisttight
\item \textbf{(I) Query node participation.} Since the queried community is highly relevant to the query node, it needs query node participation, i.e., $q\in V(H)$. 
%\item{(II) Small-scale.} The smaller the number of nodes associated with query node in the community, the more closely related the community is, indicating that a community is truly highly related to query node. 
\item \textbf{(II) Private relevant.} The private graph is only accessed to query vertex $q$, which needs to be involved in the online process of community search by sharing homogeneous attributes with $q$ in public and private graphs. 
%The smaller the number of nodes associated with query node in the community, the more closely related the community is, indicating that a community is truly highly related to query node. 
\item \textbf{(III) High attribute correlation.} %The smaller the number of nodes associated with query node in the community, the more closely related the community is, indicating that a community is truly highly related to query node.
The more the number of common attributes, the higher the correlation of the vertices in the community.
\item \textbf{(IV) Structure cohesiveness.} A tightly-knit structure indicates that there are more connections between the nodes within the community, reflecting a higher level of activity among its members.
\squishend

% \textbf{(I) Query node participation.} Since the queried community is highly relevant to the query node, it needs query node participation, ideally positioning it as a central node.
% \\
% \textbf{(II) Small-scale.} The smaller the number of nodes associated with query node in the community, the more closely related the community is, indicating that a community is truly highly related to query node.
% \\
% \textbf{(III) High attribute correlation.} The higher the correlation of the attributes in the community, indicating that the interests or characteristics of the users in the community are more similar. This contributes to a closer community.
% \\
% \textbf{(IV) Structure cohesiveness.} A tightly-knit structure indicates that there are more connections between the nodes within the community, reflecting a higher level of activity among its members.
% \\
% \textbf{\textit{(\romannumeral 5)}
% Low communication cost.} Low communication cost is considered a criterion for a good community because it reflects efficient interactions among the members within the community.

For the criteria of a good queried community, we can guarantee the (I), while for the subsequent (II) to (IV), we aim to find a balance that maximally satisfies these standards simultaneously. To quantify the common attributes in public-private graphs, we begin with a new definition as follows.  
% We then provide a precise definition of the attributed community query in public-private graphs based on these criteria, which also states the main problem studies in this paper.
%We then give a precise definition of the attributed community query in public-private graphs based on these criteria in next section. 
%in Section 3.3, which further introduces the main problem studied in this paper.   

% \subsection{Problem Statement}

% \begin{definition}[Graph Attribute]
%?
\begin{definition}[Common PP-Attributes]
    Given an attributed public-private graph \( H = (V, E, A) \). For each node \( v \in V \), let \( attr'(v) \) represent the attribute set of the node \( v \). The attribute of the graph \( H \) is defined as the intersection of the attribute sets of all nodes in the graph, as
    \[
    attr'(H) = \bigcap_{v \in H} attr'(v).
    \]
\end{definition}
%\begin{example}
In Figure~\ref{AC-PP}(b), consider the graph $H$ consisting of nodes $\{v_5, v_6, v_7, v_8, v_9\}$ with corresponding edges between them, the common pp-attributes of $H$ is the intersection of the attributes of these nodes, that is, $attr'(H)$ = \{CNN, GAN\}.
%\end{example}

\subsection{Problem Statement}
Based on the common pp-attributes, we can give a new community model to capture the homogeneous attributes in a public-private graph, which is different from the existing core-based attribute community \cite{li2020efficient, fang2016effective, lu2022time}. Because the private information of edge connections and node attributes is only known to the query itself, the existing algorithms \cite{fang2016effective, huang2016attribute} cannot precompute the information required for index construction. Hence, they are not directly applicable, which motivates our formulation. Therefore, we define the problem of $k$-core community search on attributed public-private graphs %studied in this paper is formulated 
as follows.
%Given an undirected attributed graph $G = (V, E, A)$, where $V$ is the vertex set, $E \subseteq V\times V$ is the edge set, and $A$ is the attribute function such that $\forall v \in V, A(v) \subseteq I$, while $I$ is the set of all keywords.

\begin{problem}[\ACQPP] Given a public attributed graph $G = (V, E, A)$, a query vertex $q$, construct a queried pp-graph $G'_q = G \cup G_q = (V\cup V_q, E \cup E_q, A \cup A_q)$, a parameter $k \in \mathcal{Z}^+$, the problem is to find a subgraph $H \subseteq G'_q$ satisfying the following properties:
\squishlisttight
    \item \textbf{Query participation.} $q \subseteq V(H)$;
    \item \textbf{Public-private structure cohesiveness.} $H$ is a connected $k$-core within $G'_q$;
    \item \textbf{Common attribute maximality.} 
    The size of $attr'(H)$ is maximal. 
    %\cap attr'(q)
\squishend
%\end{itemize}

\end{problem}

\stitle{\blue{Problem hardness.}}
\blue{The \ACQPP problem can be shown to be NP-hard problem as follows. First, one special instance of \ACQPP is equivalent to the ACQ problem \cite{fang2017effective}, when the private information is empty, i.e., $G_q=\emptyset$. Thus, both \ACQPP  and ACQ problems find the maximum common attributes in a $k$-core community containing the query vertex $q$. 
However, the problem ACQ is NP-hard by reducing from the NP-complete problem of maximum clique discovery in a graph $G$, which is turned out to find a complete subgraph with the largest common attributes. 
%However, the problem ACQ is NP-hard, which can be reduced from the NP-complete problem of maximum clique discovery in graph $\hat{G}(\hat{V}, \hat{E})$ by constructing a graph instance for ACQ with the parameter $k=1$, i.e., $\bar{G}(\bar{V}, \bar{E}, \bar{A})$ where $\bar{G}$ is a complete graph and $\bar{A}(v)= \hat{N}(v) \cup \{v\}$. 
}

%\begin{example}
%Consider the attributed public-private graph $G'_{v_3}\cup G'_{v_5}$ for two query nodes, $q_1 = \{v_3\}, q_2 = \{v_5\} $ in Fig.~\ref{AC-PP}(a). The public graph is represented by black edges, while the private graph edges are shown in blue for $q_1$ ($v_3$) and red for $q_2$ ($v_5$). Each vertex $v \in G'_{v_3}\cup G'_{v_5}$ is associated with a set of attributes, which are the union of public and private attributes. Note that, only vertices connected by private edges have corresponding private attributes.
%While in Fig.~\ref{AC-PP}(b),  $G'_{v_5}$ illustrates the public-private graph when only considering query node $v_5$. 
%$H$ is the subgraph found in $G'_{v_5}$ under a 3-core requirement, representing the community that shares the maximal common attributes \{CNN, GAN\} with query node $v_5$.
%\end{example}

\begin{example}
Consider the attributed public-private graph $G'_{v_5}$, $q = \{v_5\} $ in Figure~\ref{AC-PP}(b). $H$ is a 3-core subgraph found in $G'_{v_5}$, representing the community that shares the maximal common attributes \{CNN, GAN\} with query node $v_5$.  
\end{example}

\stitle{Application scenarios}. Our \ACQPP problem has several practical application scenarios for social recommendations and academic community search. Such public-private querying services can be supported in two setting. First, the service can be provided by the network platform, which owns the whole public network $G$ and allows the access to private graph $G_q$ only if the query is issued by the query user $q$. This follows the public-private setting proposed by Chierichetti et al. \cite{chierichetti2015efficient}.
%Previous studies, such as Chierichetti et al. \cite{chierichetti2015efficient} and Epasto et al. \cite{epasto2019device}, have also explored querying tasks in public-private social networks. %The former performs querying processing on the network platform with limited private data participation, while the latter focuses on fully on-device computation with strict privacy protection. 
In this case, the network data and public indexes 
could be done efficiently within the central computation by taking a small transforming cost. 
\blue{Second, following the fully on-device computation with strict privacy protection~\cite{epasto2019device}, our problem could be further extended to allowing public data $G$ and private data $G_q$ to be separately stored in the network platform and user $q$'s devices, respectively.  
Our developed techniques offer public and private indexing to support the extraction of a small candidate graph, which is transferred from the platform to the user devices. Finally, community candidate validation can be performed efficiently on a local device.}
%Last but not least, to fit with multiple users' requests, we further discuss the extension of our problem and algorithms to handle public-private community search with multiple query nodes in Section~VI.

\begin{algorithm}[t]
\small
%\caption{Basic Community Search ($G'_q$, $q$, $k$)}
\caption{Online-basic ($G'_q$, $q$, $k$)}
\label{algo.1}
\begin{flushleft}
\textbf{Input:} Public graph $G$, query $q$, private graph $G_q$,  integer $k$\\
\textbf{Output:} An attribute-maximized  $k$-core community $H$ containing $q$
% \begin{algorithmic}[1]
% \State Construct a pp-graph $ G'_q \leftarrow G \cup G_q$;
% \State Reduce $G'_q$ to a $k$-core subgraph;
% \State $d \leftarrow |attr'(q)|$;
% \While{$d\geq1$}
%      \State Find $S \subseteq attr'(q)$; 
%      \State $|S| \leftarrow d $;
%      \State Check $G'_q$ of whether there exists a community $H$ satisfying $k$-core containing $q$;
%      \If{such $H$ exists}
%         \State \textbf{break};
%      \EndIf
%      \State $d \leftarrow d-1$;
% \EndWhile
% \State \Return the community $H$
% \end{algorithmic}

% \begin{algorithmic}[1]
% \State Construct the pp-graph $G'_q \gets G \cup G_q$;
% \State Reduce $G'_q$ to its $k$-core subgraph;
% \For{each attribute $\alpha \in attr'(q)$}
%     \State Collect vertices $\mathcal{V}_\alpha \gets \{ v \in G'_q \mid \alpha \in attr(v) \}$;
% \EndFor
% \For{$d = |attr'(q)|$ down to $1$}
%         \For{each subset $S \subseteq attr'(q)$ with $|S| = d$}
%         \State Collect vertices $\mathcal{V}_S \gets \bigcap_{\alpha \in S} \mathcal{V}_\alpha$;
%         \State Construct candidate graph $G'_{q}[\mathcal{V}_S]$ induced by $\mathcal{V}_S$;
%         \State Check if $G'_q[\mathcal{V}_S]$ contains a $k$-core community $H$ with $q \in H$;
%        \If{such $H$ exists} 
%        \State break;
%        \EndIf
%        \EndFor
% \EndFor
% \State \Return the community $H$
% \end{algorithmic}

\begin{algorithmic}[1]
\State Construct the pp-graph $G'_q \gets G \cup G_q$;
\State Reduce $G'_q$ to its $k$-core subgraph;
\For{each attribute $\alpha \in attr'(q)$}
    \State Collect vertices $\mathcal{V}_\alpha \gets \{ v \in G'_q \mid \alpha \in attr(v) \}$;
\EndFor
\For{$d = 1$ to $|attr'(q)|$}
    \State $update \gets \textbf{False}$;
    \For{each subset $S \subseteq attr'(q)$ with $|S| = d$}
        \State $\mathcal{V}_S \gets \bigcap_{\alpha \in S} \mathcal{V}_\alpha$;
        \State Construct candidate graph $G'_q[\mathcal{V}_S]$;
        \State Check if $k$-core community $H^*$ in $G'_q[\mathcal{V}_S]$ with $q \in H^*$;
        \If{such $H^*$ exists}
            \State $H \gets H^*$; $update \gets \textbf{True}$;
        \EndIf
    \EndFor
    \If{$update = \textbf{False}$}
        \State \textbf{break};
    \EndIf
\EndFor
\State \Return the community $H$
\end{algorithmic}

\end{flushleft}
\label{online1}
\end{algorithm}

%% file: basic.tex
\section{basic Online search solutions}
\label{SectionIV}
% \begingroup
% \setlength{\parskip}{0.5em} % 在此设置段落间距
% \setlength{\parindent}{0pt} % 取消段落缩进

%For the \ACQPP problem proposed in this paper, we introduce 
%online search algorithms to address community search tasks for query nodes in public-private graphs.

In the following, we introduce 
online search algorithms for \ACQPP in public-private graphs.
%to address community search tasks for query nodes in public-private graphs.

\stitle{Straightforward online search.}
Algorithm 1 of Online-basic consists of four steps. 
Firstly, construct the public-private graph $G'_q$ by combining the public graph $G$ and the private graph $G_q$ of the query node $q$, then reduce it to its $k$-core subgraph to ensure structural cohesiveness (lines 1-2). 
Next, for each attribute $\alpha$ in the query node's attribute set $attr'(q)$, collect the set of vertices $\mathcal{V}_\alpha$ in $G'_q$ that contain this attribute (lines 3-4).
Then, start from the 1-attribute sets and iteratively increase their sizes from 1 to $|attr'(q)|$. 
To obtain candidate node sets, we consider each attribute subset $S\subseteq attr'(q)$ of size $d$, and compute $\mathcal{V}_S$ as the intersection of all $\mathcal{V}_\alpha$ where $\alpha \in S$. (lines 5-8).
Finally, induce the subgraph $G'_q[\mathcal{V}_S]$ from the nodes in $\mathcal{V}_S$, and perform $k$-core validation (lines 8-9). If such a community $H^*$ is found, update $H$ with $H^*$ (lines 11-12). If no valid community is found for any subset of size $d$, the algorithm terminates and returns $H$ (lines 13–15).

% for each attribute combination $S$, we identify the set of nodes in $G'_q$ that satisfy these attributes and perform $k$-core validation on them.
% %If the subgraph formed by these nodes and their connections can be decomposed into a $k$-core, 
% If the subgraph contains a $k$-core, 
% the community with the maximum shared attributes is successfully found. 
% %If no valid community is found at the current $d$,
% Otherwise, we decrement $d$ and repeat the process. 
%Detailed pseudocode is presented in Algorithm~\ref{online1}. 

% A straightforward method is to enumerate all subsets $S \subseteq attr(q)$ for the query node $q$, with the size of each subset $|S|$ ranging from $1$ to $|attr(q)|$. For each subset %$attr_i(q)$
% , we search within the pp-graph $G'_q$ to find the $k$-core community that satisfies the largest number of these subsets. 

\stitle{Binary fast search algorithm}. 
% However, we found that Online-basic is time-consuming. To accelerate the search process, we propose an improved binary search algorithm, Online-binary. The main difference between Online-basic and Online-binary lies in the search strategy. Online-basic performs a sequential search by decrementing $d$ from the maximum dimension, while Online-binary starts the search at the middle value $\frac{|attr'(q)|}{2}$. If a $k$-core community satisfying the current $d$-dimensional attribute combination is found, the search continues in the higher half of the range. Otherwise, it proceeds in the lower half. This process continues until a community with the most shared attributes is found, and no valid community with more shared attributes can be identified in the higher binary search range. Detailed pseudocode is presented in Algorithm~\ref{online2}. 
However, Online-basic is time-consuming due to its sequential search strategy. To improve efficiency, we propose Online-binary, which adopts a binary search approach. It begins at the middle value $|attr'(q)/2|$ and adjusts the search range based on whether a valid $k$-core community is found, shifting upward if successful, or downward otherwise. This continues until the community with the largest common attributes is identified.

The main issue with these algorithms is the high cost of enumerating attribute combinations. %To address this, we aim to develop an efficient solution. 
%By analyzing the \ACQPP problem, we identify two fundamental steps in its resolution: $k$-core decomposition and the selection of candidate attribute combinations. However, these two steps are inherently independent, and our problem is set in the context of public-private graphs, which further complicates the issue.  Therefore, our goal is to reduce the search space by identifying frequent attribute patterns and avoiding redundant $k$-core validations.
To address this, we propose a novel search framework as follows. 

\stitle{\blue{Our PP-FP framework overview.}}
\blue{Given a query vertex $q$ with its public-private attributed graph, our framework aims to efficiently identify a $k$-core community that maximizes the common attributes shared by all members. The framework is built upon two complementary index structures, a private PP-FP-tree index constructed for the query vertex $q$, and a public index built over the entire public graph.
The query process consists of three phases. In Phase-I, the PP-FP-tree is used to identify candidate nodes in the private neighborhood that share the maximal common attributes with $q$. In Phase-II, the common attributes selected from the PP-FP-tree are queried over the public index to expand candidate nodes. In Phase-III, candidates from Phase-I and Phase-II are merged and validated, finding a community that meets both attribute maximality and structure constraint.
}

%% file: index.tex
\section{Public-private index construction}
In this section,
%based on the previously outlined algorithm requirements, 
we introduce our public-private index construction, which consists of two main components.
\begin{itemize}
\renewcommand{\labelitemi}{\fontsize{14}{14}\selectfont\textbullet} 
\item \textbf{Component I: private PP-FP-tree index construction.} Construct a compact PP-FP-tree index for each query node in its private graph, to quickly identify the combination of the most frequently occurring attributes.

\item \textbf{Component II: public graph index construction.} Build a coreness tree index for the entire public graph to efficiently identify nodes that can potentially contribute to forming $k$-core and satisfying attribute combinations.

\end{itemize}

% As proven in Lemma~\ref{core}, when $core_{\mathcal{N}_{n+1}(q)} \leq core_{\mathcal{N}_n(q)}$, meaning that expanding to the $(n+1)$ hop-neighborhood no longer increases the coreness value, so we only need to retain nodes $v \in \mathcal{N}_n(q)$ and proceed with the node coreness list and attribute map construction.
\subsection{Preliminaries of PP-FP-Tree Index}

% \begin{table}[h]
%     \vspace{-0.2cm}
%     \hspace{-0.1\linewidth}
%     \centering
%     \begin{minipage}{0.4\linewidth}
%         \centering
%         \begin{tabular}{ccc}
%             \toprule
%             \multicolumn{2}{c}{Query: $q$} & \multicolumn{1}{c}{\{a, b, c, d, e\}} \\ \midrule
%             $v_1$ & 4 & \{b, c, d, e\} \\ 
%             $v_2$ & 3 & \{a, c, e\}  \\ 
%             $v_3$ & 3 & \{b, c, d\} \\ 
%             \bottomrule
%         \end{tabular}
%         \caption*{(a) Node attribute list}
%     \end{minipage}%
%     \quad
%     \hspace{0.03\linewidth}
%     \begin{minipage}{0.4\linewidth}
%         \centering
%         \begin{tabular}{cc}
%             \toprule
%             Attribute ($\alpha$) & Vertex sets ($\Theta_{\alpha}$) \\ \midrule
%             c & $v_1$, $v_2$, $v_3$\\
%             b & $v_1$, $v_3$\\
%             d & $v_1$, $v_3$ \\
%             a & $v_2$ \\ 
%             e & $v_1$, $v_2$ \\ 
%             \bottomrule
%         \end{tabular}
%         \caption*{(b) Node attribute map}
%     \end{minipage}
%     \vspace{-0.1cm}
%     \caption{Preliminaries of PP-FP-Tree Index.}
%     \label{list}
%     \vspace{-0.1cm}
% \end{table}

\begin{figure}[h]
\vspace{-0.4cm}
  \centering
  \includegraphics[width=\linewidth]{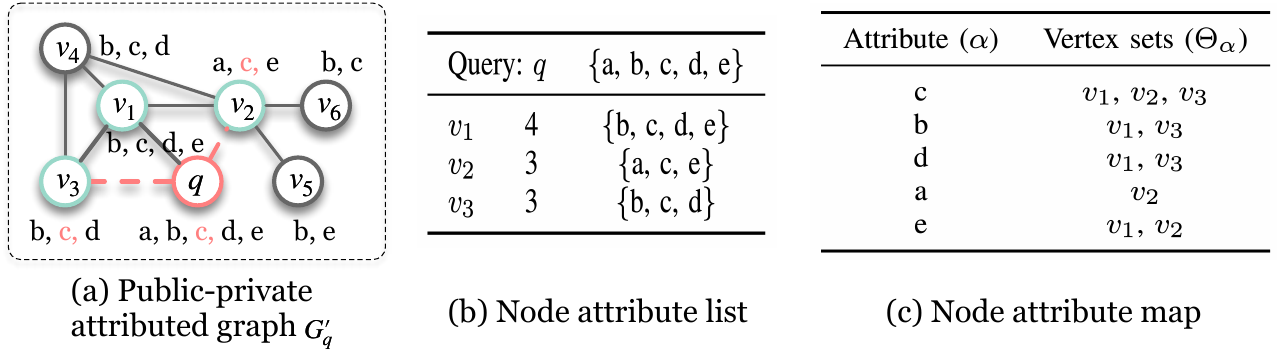}
  \caption{Preliminaries of PP-FP-Tree Index.}
  \label{preliminaries}
\vspace{-0.2cm}
\end{figure}

%Now, we introduce the algorithm for PP-FP-tree index construction on public-private graph in 1-hop neighborhood, 
Now, we introduce the preliminaries for PP-FP-tree index construction. 
%on public-private graph in 1-hop neighborhood,
%The 1-hop neighborhood, which is the subgraph of $G'_q$ induced by all public and private neighbors $N'(q)=N(q)\cup N_p(q)$.
In the public-private graph, we focus on the 1-hop neighborhood, which is the subgraph of $G'_q$ induced by the union of public-private neighbors of the query node $q$, i.e., $N'(q) = N(q) \cup N_p(q)$.
The process consists of two steps. First, we collect all attributes of $q$, i.e., $attr'(q)$. For each neighbor $u\in N'(q)$, these vertices are sorted in decreasing order of shared attributes with the query vertex $q$, i.e.,  $|attr'(u) \cap attr'(q)|$. This results in the node attribute list.
\begin{example}
    First, we obtain the 1-hop neighborhood of the query node $q$ from the public-private attributed graph in Figure~\ref{preliminaries}(a), denoted as $N'(q) = \{v_1, v_2, v_3\}$. Then, these nodes in $N'(q)$ are sorted in descending order based on the number of common attributes they share with $q$. As a result, we obtain the node attribute list in Figure~\ref{preliminaries}(b).
\end{example}

Second, for each attribute $\alpha\in attr'(q)$, we list the set of vertices containing the attribute $\alpha$, denoted as
$\Theta_{\alpha} = \{u \in N'(q) \mid \alpha \in attr'(u)\}$. Then, the vertices $u$ in $\Theta_{\alpha}$ are sorted in decreasing order of $|attr'(u) \cap attr'(q)|$. This forms the node attribute map.

%\xin{add an example shown in Figure II(b)}
\begin{example}
    For each attribute in Figure~\ref{preliminaries}(b), where $attr'(q) = \{a,b,c,d,e\}$, we list each set of vertices containing the attribute $\alpha \in attr'(q)$. When $\alpha = c$, the corresponding vertex set is $\Theta_{\alpha} = \{v_1, v_2, v_3\}$, sorted in descending order based on the number of common attributes with $q$. The final result is shown in Figure~\ref{preliminaries}(c).
\end{example}

The detailed construction of the node attribute list and node attribute map can be found in Algorithm~\ref{algo3}.

% \begin{figure}
%   \centering
% \includegraphics[width=0.3\linewidth]{figures/example-2-new.pdf}
%   \vspace{-0.2cm}
%   % \caption{An example of ACQ-PP Problem.}
%   \caption{An example of attributed public-private graph $G'_q$.}
%   \label{example-2}
% \vspace{-0.3cm}
% \end{figure}

\begin{figure*}
  \centering
  \includegraphics[width=\linewidth]{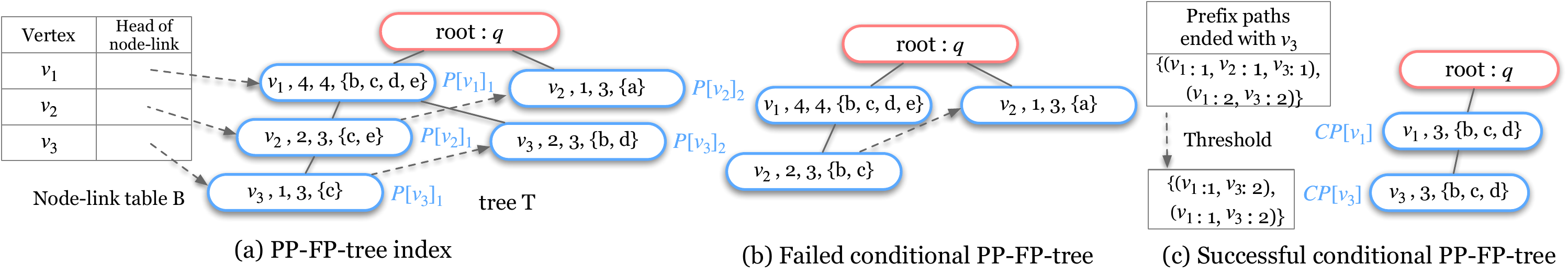}
  % \caption{An example of PP-FP-tree index for querying $q=v_1$ and $k=2$ to search 2-core community with at least 3 vertices.}
  \caption{An example of PP-FP-tree index for querying $q$ and $k=2$ to search 2-core community with at least 3 vertices.}
  \vspace{-0.2cm}
  \label{fptree}
\end{figure*}

\subsection{Index Construction of PP-FP-Tree} 
\label{inspired}
Now, we present a novel index of public-private frequent pattern tree called PP-FP-tree, \blue{which is inspired by the classic FP-tree~\cite{han2000mining} originally proposed for frequent itemset mining. Our PP-FP-tree extends the FP-tree by integrating multiple features of private edges and attributes, the structural coreness, and common attributes. 
%to the public-private network setting, and is used to efficiently identify candidate vertex sets with common attributes.
} For a query vertex $q$, its PP-FP-tree is built on the public-private graph $G'_q$. 

Our PP-FP-tree consists of two data structures: one tree $T$ and one node-link table $B$.
First, the tree $T$ rooted by $q$ consists of a few nodes $P[u]_x$, where the root is the center query node $q$ and the neighbor $u\in N(q)\cup N_p(q)$ for an integer $x\in \mathbb{Z}^+$. For any tree node $P[u]_x$, the prefix path of $P[u]_x$ is the path from $P[u]_x$ to root $q$, denoted as \kw{Prefix}-\kw{Path}$(P[u]_x)$. 
Second, the table $B$ consists of several structured chains, where each chain links to a public and private neighbor of $q$, i.e., $B[u]=\{$\text{PP-FP-tree node} $P[u]_x: u\in N(q)\cup N_p(q)\}$ for public-private neighbor $u$  in $G'_q$ and the integer indicator $1\leq x\leq |B[u]|$. Here, the number $x$ indicates the insertion order of $u$'s tree nodes into the tree $T$.
%represents the order number of inserting the tree node of $u$ into the tree $T$. 

\begin{algorithm}[t]
\small
\caption{Private indexing: nodes and attributes extraction for PP-FP-tree construction}
%Node attribute list and node attribute map 
%($q$, $G'_q$)}
\label{algo.3}
\begin{flushleft}
\textbf{Input:} A query node $q$, and the public-private graph $G'_q$ \\
\textbf{Output:} A node attribute list $nal_q$ and a node attribute map $nam_q$ of the query node $q$
\begin{algorithmic}[1]
\State Get public-private neighbors $N'(q)$ of the query node $q$ in $G'_q$;
\State Obtain $attr'(q) \leftarrow attr(q) \cup attr_p(q)$;
\For{each node $v \in N'(q)$}
    \State $attr \leftarrow attr'(v) \cap attr'(q)$; 
    %\Comment{Get the intersection with the attribute of the query node}
    %\State $count \leftarrow |attr|$;

    \If{$attr \neq \emptyset$}
        \State Build a list node $l[v] \leftarrow (v, |attr|, attr)$;
        \State Add $l[v]$ into $nal_q$; 
    \EndIf
\EndFor
\State
Sort $nal_q$ by $(-l[v].|attr|)$; 
%\Comment{Sort the list nodes by the shared attribute number in descending order.}

\For{each list node $l[v] \in nal_q$}
    \For{each attribute $\alpha \in l[v].attr$}
        \State Add $v$ into the vertex set $\Theta_{\alpha}$ in $nam_q$;
    \EndFor
\EndFor
\State \Return  the attribute list $nal_q$ and  attribute map $nam_q$;
%node attribute list $nal_q$ and node attribute map $nam_q$;
\end{algorithmic} 
\end{flushleft}
\label{algo3}
\end{algorithm}

Besides the tree structure of $T$ rooted by node $q$, each tree node $P[u]_x$ is also associated with three elements, representing the additional attribute information as follows:

\squishlisttight
\item $\boldsymbol{P[u]_x.\kw{Prefix}\textit{-}\kw{Attr}}$: is the set of the common attributes shared by all tree nodes along the prefix-path from $P[u]_x$ to the root $q$. 
% Then, $P[u]_x.\kw{Prefix}\textit{-}\kw{Attr}$ $= \bigcap_{P[w]_y\in \kw{Prefix}\textit{-}\kw{Path}(P[u]_x)} attr'(w)$.
Then, $P[u]_x.\kw{Prefix}\textit{-}\kw{Attr}$ $= \bigcap_{P[w]_y\in \kw{Prefix}\textit{-}\kw{Path}(P[u]_x)} $ \blue{$attr'(P[w]_y)$}.
%\item\textbf{{$P[u]_x.\kw{Prefix}\textit{-}\kw{AttrNum}$}}:
\item $\boldsymbol{P[u]_x.\kw{Prefix}\textit{-}\kw{AttrNum}}$:
is the number of common attributes in $P[u]_x.\kw{Prefix}\textit{-}\kw{Attr}$, i.e., $P[u]_x.\kw{Prefix}\textit{-}\kw{AttrNum} = |P[u]_x.\kw{Prefix}\textit{-}\kw{Attr}|$. 
\item
$\boldsymbol{P[u]_x.\kw{Overall}\textit{-}\kw{AttrNum}}$: is the number of common attributes between $u$ and root $q$, i.e., $P[u]_x.\kw{Overall}\textit{-}\kw{AttrNum}= |attr'(u) \cap attr'(q)|$. 
\squishend

%\xin{add a example}

%{root q}

\begin{example}
% In Figure~\ref{fptree}(a), for $ P[v_5]_1 $, its prefix path denoted by $ \kw{Prefix}\textit{-}\kw{Path}(P[v_5]_1) = \{ P[v_6]_1, P[v_3]_1, P[v_5]_1 \}$. $P[v_5]_1$.\kw{Prefix}-\kw{Attr} represents the attributes shared by all tree nodes $\{P[v_6]_1, P[v_3]_1, P[v_5]_1\}$ in $\kw{Prefix}\textit{-}\kw{Path}(P[v_5]_1)$, which is \{A, B\}. Thus, \( P[v_5]_1.\kw{Prefix}\textit{-}\kw{AttrNum}%= |P[v_5]_1.\kw{Prefix}\textit{-}\kw{Attr}|
% = 2 \) . Meanwhile, \( P[v_5]_1.\kw{Overall}\textit{-}\kw{AttrNum} \) represents the total number of attributes shared between node \( v_5 \) and the query root node $v_1$, which can be found in Table~\ref{list} (a) as 4.

In Figure~\ref{fptree}(a), the prefix path of $P[v_3]_1$ is $\kw{Prefix}$\textit{-}\kw{Path}$(P[v_3]_1)$ = $\{ P[v_1]_1, P[v_2]_1, P[v_3]_1 \}$. $P[v_3]_1$.\kw{Prefix}-\kw{Attr} represents the attributes shared by all tree nodes in $\kw{Prefix}$\textit{-}\kw{Path}$(P[v_3]_1)$, which is \{c\}. Thus, $ P[v_3]_1$.$\kw{Prefix}$\textit{-}\kw{AttrNum}=~1. Meanwhile, $P[v_3]_1$.$\kw{Overall}$\textit{-}\kw{AttrNum} represents the total number of attributes shared between node $v_3$ and the query root $q$, which can be found in Figure~\ref{preliminaries}(b) as 3. Therefore, in Figure~\ref{fptree}(a), the value of $P[v_3]_1$ is $(1, 3, \{c\})$.
\end{example}
%$\{P[v_1]_1, P[v_2]_1, P[v_3]_1\}$ 

% \begin{table}[t]
%     \centering
%     \hspace{-0.15\linewidth}
%     \begin{minipage}{0.45\linewidth}
%         \centering
%         \begin{tabular}{ccccc}
%             \toprule
%             \multicolumn{2}{c}{Query: $v_1$} & \multicolumn{1}{c}{\{A, B, C, D, E, F\}} \\ \midrule
%             $v_6$ & 5 & \{A, B, C, E, F\} \\ 
%             $v_3$ & 4 & \{A, B, C, D\}  \\ 
%             $v_5$ & 4 & \{A, B, E, F\} \\ 
%             $v_2$ & 3 & \{A, B, C\}  \\ 
%             $v_4$ & 3 & \{A, B, C\} \\ 
%             \bottomrule
%         \end{tabular}
%         \caption*{(a)}
%     \end{minipage}%
%     \quad
%     \hspace{0.03\linewidth}
%     \begin{minipage}{0.28\linewidth}
%         \centering
%         \begin{tabular}{cc}
%             \toprule
%             Attribute ($\alpha$) & Vertex sets ($\Theta_{\alpha}$) \\ \midrule
%             A & $v_6$, $v_3$, $v_5$, $v_2$, $v_4$ \\
%             B & $v_6$, $v_3$, $v_5$, $v_2$, $v_4$ \\
%             C & $v_6$, $v_3$, $v_2$, $v_4$ \\
%             D & $v_3$ \\ 
%             E & $v_6$, $v_5$ \\ 
%             F & $v_6$, $v_5$ \\ \bottomrule
%         \end{tabular}
%         \caption*{(b)}
%     \end{minipage}
%     \caption{(a) Node attribute list, (b) node attribute map}
%     \label{list}
% \end{table}

%We sort all attributes $\alpha \in attr'(v)$ in descreasing order of attribute count $|\Theta_{\alpha}|$. 

We construct the PP-FP-tree using the vertex sets $\Theta_{\alpha}$ for each attribute $\alpha \in attr'(q)$. For each attribute $\alpha$, we sequentially insert every vertex $u \in \Theta_{\alpha}$ into the tree $T$ as $P[u]_x$ and update the three key information mentioned above.

%Second, we construct the PP-FP-tree using the vertex sets of $\Theta_{\alpha}$ for each attribute $\alpha \in attr'(q)$.
%For each attribute $\alpha$, we sequentially insert each vertex $u \in \Theta_{\alpha}$ into the tree $T$ as $P[u]_x$ and update the three key information items mentioned above.

%\stitle{Algorithm}. 
%Algorithm~\ref{algo4} outlines the details of PP-FP-tree construction method in Fig.~\ref{fptree}. For the node attribute map \( nam_q \), each vertex \( v \) in the vertex set \( \Theta_{\alpha} \) is processed sequentially (lines 2-3). For each vertex $v$, the structure of the PP-FP-tree is updated based on whether it shares prefix paths with previously added tree nodes. If no shared prefix paths exist, a new branch is created for the vertex $v$, adding it as a new tree node $P[v]_x$ with its $\kw{prefix}$-$\kw{AttrNum}$ initialized to 1 (lines 4-5). If shared prefix paths exist, the corresponding tree node $P[v]_x.\kw{prefix}$-$\kw{AttrNum}$ is incremented by 1 (lines 6-7). In either case, the shared attribute count, $\kw{Overall}$-$\kw{AttrNum}$, for the corresponding item $P[v]_x$ is incremented by 1 (line 8). 

Algorithm~\ref{algo4} outlines the details of PP-FP-tree construction in Figure~\ref{fptree}. \blue{We firstly initialize the PP-FP-tree $T$ rooted by the node $q$ (line 1).} \blue{For each attribute $\alpha$ in the node attribute map in Figure~\ref{preliminaries}(c), $r$ is initialized to the root (lines 2-3).} 
% and is updated sequentially as vertices $v\in\Theta_{\alpha}$ are processed.
\blue{When inserting a vertex $v$, we check whether $v$ is a child node of the current node $r$. If not, create a new tree node $P[v]_x.( \kw{Prefix}\textit{-}\kw{AttrNum}, \kw{Overall}\textit{-}\kw{AttrNum}, \kw{Prefix}\textit{-}\kw{Attr})$
initialized to $(1,|attr'(v)\cap attr'(q)|,\alpha)$ (lines 4-6); }
Otherwise, the corresponding $P[v]_x$ is updated with its $\kw{prefix}$-$\kw{AttrNum}$ incremented by 1 \blue{(lines 7-8).}
%, and its $\kw{prefix}$-$\kw{Attr}$ takes the union of $\alpha$ (lines 6-8). 
\blue{$r$ is then updated to $P[v]_x$ to continue the construction along the prefix path (line 9).} %In either case,  $P[v]_x$.$\kw{Overall}$-$\kw{AttrNum}$ is incremented by 1 (line 9). 
Finally, return the PP-FP-tree $T$ (line 10).

\begin{algorithm}[t]
\small
\caption{
Private indexing: PP-FP-tree index construction }
%Extraction Nodes and Attributs for PP-FP-tree 
% ($ncl_q, nam_q$)}
%($nam_q$)}
\label{algo.4}
\begin{flushleft}
% \textbf{Input:} A node coreness list $ncl_q$ and a node attribute map $nam_q$ of the query node $q$ \\
\textbf{Input:} A node attribute map $nam_q$ of the query node $q$ \\
\textbf{Output:} A PP-FP-tree $T$ for the query node $q$
\begin{algorithmic}[1]
\State \blue{Initialize PP-FP-tree $T$ rooted by the node $q$;}
\For{each vertex set $\Theta_{\alpha} \in nam_q$}
        \State \blue{Initialize $r
        \leftarrow root$};
    \For{each vertex $v \in \Theta_{\alpha}$ sequentially}
       % \If{$v$ does not follow $\kw{Prefix}$-$\kw{Path}(\blue{v})$ in $T$}%the existing prefix paths in $T$}
        \If{\blue{$r$ does not contain child node $v$}}
           \State Create a new tree node \blue{$P[v]_x = (1, $ $|attr'(v) \cap attr'(q)|, \alpha)$};% in $T$; 
           %\blue{$P[v]_x = (1, 1, \alpha)$};
        \Else 
           \State Increment $P[v]_x.\kw{Prefix}\textit{-}\kw{AttrNum}$ by 1;
        \EndIf
        \State \blue{Update $r \leftarrow P[v]_x$;}
    \EndFor
\EndFor
\State \Return the PP-FP-tree {$T$}

\end{algorithmic}
\end{flushleft}
\label{algo4}
\end{algorithm}

\begin{figure}[t]
  \centering
  \includegraphics[width=\linewidth]{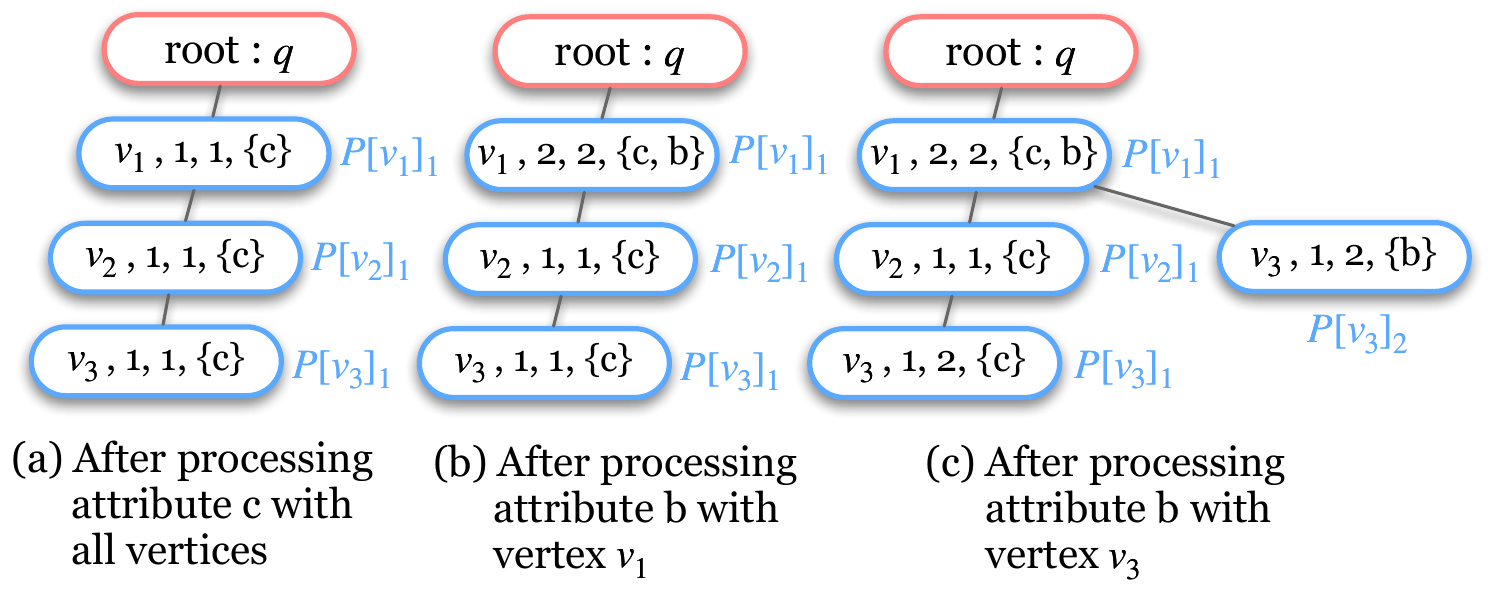}
  \caption{\blue{An example on PP-FP index construction.}}
  \vspace{-0.3cm}
  \label{index_construction}
\vspace{-0.3cm}
\end{figure}

\begin{example}
\blue{Figure~\ref{index_construction} shows the process of PP-FP-tree construction. It illustrates two cases in Algorithm~\ref{algo4}, creating a new tree node (lines 5-6) and updating an existing tree node (lines 7-8). Consider the node attribute map in Figure~\ref{preliminaries}(c). First, we insert vertices $\{v_1, v_2, v_3\}$ with attribute $c$, which results in Figure~\ref{index_construction}(a). Second, consider the vertices $\{v_1, v_3\}$ with attribute $b$, and $r$ is initialized to $root$. Since $r$ has the child node $v_1$, thus $\{b\}$ is added to $P[v_1]_1$ in Figure~\ref{index_construction}(b). $r$ is then updated to $P[v_1]_1$. For the next vertex $v_3$, since $P[v_1]_1$ has no child node for $v_3$, a new tree node $P[v_3]_2$ is created as the child node of $P[v_1]_1$ in Figure~\ref{index_construction}(c).}  
\end{example} 

In the following, we make two useful observations.
%as follows. 

\begin{observation} If $P[v]_x.\kw{Prefix}\textit{-}\kw{AttrNum} = k$, all tree nodes appear in the \kw{Prefix}-\kw{Path}$(P[v]_x)$ contain at least $k$ common attributes with the query node $q$. 
\label{lemma2} 
\end{observation}

%\begin{observation} If the length of the prefix path denoted by $|\kw{Prefix}\textit{-}$ $\kw{Path}(P[v]_x)| = n$, which means all of the $n$ nodes in this prefix path share the same attributes of $P[v]_x.\kw{Prefix}\textit{-}\kw{Attr}$. 
%\label{height} 
%\end{observation}

\begin{observation} If the length of the prefix path $|\kw{Prefix}\textit{-}\kw{Path}(P[v]_x)| = n$, then all $n$ nodes in this prefix  path share the attributes in 
$P[v]_x.\kw{Prefix}\textit{-}\kw{Attr}$.
\label{height}
\end{observation}

In Figure~\ref{fptree}(a), since $P[v_3]_1.\kw{Prefix}\textit{-}\kw{AttrNum} = 1$, by Observation~\ref{height}, the tree nodes in \kw{Prefix}-\kw{Path}$(P[v_3]_1)=\{P[v_1]_1$, $P[v_2]_1$, $P[v_3]_1\}$ all share one common attribute with the root node $q$. 
According to Observation~\ref{lemma2}, \blue{$|\kw{Prefix}\textit{-}\kw{Path}(P[v_3]_1)| = 3$ indicates that three tree nodes shares the attribute $c$.}
%$|\kw{Prefix}-\kw{Path}(P[v_3]_1)|$ contains three tree nodes, indicating that three tree nodes share the attributes in $P[v_3]_1.\kw{Prefix}\textit{-}\kw{Attr}$, which is \{c\}.

% \end{example}

\subsection{Conditional PP-FP-Tree}
% We note that the PP-FP-tree cannot support all queries. For example, if we want to find the 3-core community $H_2$ with 3 attributes in Figure~\ref{example-2}(c), which requires at least three nodes sharing 3 common attributes with $v_1$, we cannot obtain this sorely based on Observation~\ref{height}. In example~\ref{P5}, we mentioned that in \kw{Prefix}-\kw{Path}$(P[v_4]_1)$, we can identify five nodes that share 2 common attributes with $v_1$, but we cannot find three nodes that share 3 common attributes with $v_1$.
%We note that the PP-FP-tree cannot support all queries. For example, if we want to find the 2-core community $H$ with 3 attributes in Fig.~\ref{Framework}, which requires at least two nodes sharing 3 common attributes with $q$, we cannot conclude this solely based on Observations \ref{lemma2} and \ref{height}. 
We note that the PP-FP-tree cannot support all queries.
For instance, when searching for the 2-core community $H$ with maximal attributes in Figure~\ref{Framework}, the result cannot be obtained solely based on Observations~\ref{lemma2} and~\ref{height}.
%we cannot obtain this sorely based on Observations \ref{lemma2} and \ref{height}. 
% In example~\ref{P5}, we mentioned that in \kw{Prefix}-\kw{Path}$(P[v_4]_1)$, we can identify five nodes that share 2 common attributes with $v_1$, but we cannot find three nodes that share 3 common attributes with $v_1$.
%This limitation arises because the attribute set of a node may be distributed across different branches of the PP-FP-tree. Therefore, we need to introduce the conditional PP-FP-tree to more effectively and comprehensively mine frequent nodes (i.e., nodes with high attribute overlap). The following Observation~\ref{union} forms the basis of the conditional PP-FP-tree.
Since the attributes of a node may be distributed across different branches of the PP-FP-tree, it's hard to reconstruct the complete attribute overlap for each node. To address this limitation, we construct a conditional PP-FP-tree to better identify nodes with strong attribute overlap with the query node. 
Observation~\ref{union} forms the basis for constructing this conditional tree.

\begin{observation}
    Given the PP-FP-tree of the query node $q$ represented by $T$, the attribute set of a node $v$ can be represented as:
    \[
    attr'(v) \cap attr'(q) = \bigcup_{P[v]_x\in T} {P[v]_x.\kw{Prefix}\textit{-}\kw{Attr}}
    \]
    \label{union}
\end{observation}

%\begin{example}
    In Figure~\ref{fptree}(a), according to Observation~\ref{union}, the attribute set of $v_3$ can be represented by the $attr'(v_3) \cap attr'(q)= P[v_3]_1.\kw{Prefix}\textit{-}\kw{Attr} \cup P[v_3]_2.\kw{Prefix}\textit{-}\kw{Attr}$. %Similarly, the attribute set of $v_5$ can be expressed as $attr(v_5) = \kw{P_3[v_5]_1.attr}\cup \kw{P_2[v_5]_2.attr}$.
%\end{example}

%\begin{definition}[Requirements of conditional PP-FP-tree construction] Given a query node $q$, after constructing its PP-FP-tree $T$, if a tree node $P[v]_x \in T$ satisfies all of the following conditions:

%(I) $P[v]_x.\kw{Overall}\textit{-}\kw{Attr} \geq$ required common attribute value, 

%(II) $P[v]_x.\kw{Prefix}\textit{-}\kw{Attr} <$ required common attribute value,

%(III) $|\kw{Prefix}\textit{-}\kw{Path}(P[v]_x)|\geq k$, which means at least $k$ nodes share $P[v]_x.\kw{Prefix}\textit{-}\kw{Attr}$ common attributes with the root node. We consider $v$ to be a potential candidate node and will proceed to construct $v$'s conditional PP-FP-tree.
%\label{requirements}
%\end{definition}

To extract all possible combinations of common attributes for a subgraph containing vertex $v$, we motivate to construct a conditional PP-FP-tree by collecting all paths from the leaf node $v$ to the root $q$. However, we do not need to enumerate all possible conditional PP-FP-trees, we can use the following requirements to reduce such constructions.   % $P[v]_x$

% \begin{definition}[Requirements for Conditional PP-FP-tree Construction]
% Given a query node $q$, after constructing its PP-FP-tree $T$, a tree node $P[v]_x \in T$ is considered for conditional PP-FP-tree construction if all of the following conditions hold:

% (I) $P[v]_x.\kw{Overall}\textit{-}\kw{Attr} \ge$ required common attribute value;

% (II) $P[v]_x.\kw{Prefix}\textit{-}\kw{Attr} <$ required common attribute value;

% (III) $|\kw{Prefix}\textit{-}\kw{Path}(P[v]_x)| \ge k$, 
% meaning at least $k$ nodes share the attributes in $P[v]_x.\kw{Prefix}\textit{-}\kw{Attr}$ with the root node $q$.

% Such a node $v$ will then be used to construct its conditional PP-FP-tree.
% \label{requirements}
% \end{definition}

\begin{definition}[Requirements for Conditional PP-FP-tree Construction]
Given a query node $q$ and its PP-FP-tree $T$, a tree node $P[v]_x \in T$ is considered for conditional PP-FP-tree construction if all the following conditions hold:

\squishlisttight
\item \blue{$P[v]_x.\kw{Overall}\textit{-}\kw{Attr} \ge \ell$;}

\item \blue{$P[v]_x.\kw{Prefix}\textit{-}\kw{Attr} < \ell$;}

\item $|\kw{Prefix}\textit{-}\kw{Path}(P[v]_x)| \ge k$, which means at least $k$ nodes share the prefix attributes of $P[v]_x$ with query node $q$.
\squishend

\blue{Here, $\ell$ denotes the required number of common attributes.} Such a node $v$ is then used to construct its conditional PP-FP-tree.
\label{requirements}
\end{definition}

We first extract all prefix paths in tree $T$ that end with node $v$, denoted as $\kw{Prefix}$\textit{-}$\kw{Path}(v)$. Based on these paths $\kw{Prefix}$\textit{-}$\kw{Path}(P[v]_x) \in \kw{Prefix}$\textit{-}$\kw{Path}(v)$, we construct a conditional PP-FP-tree that preserves only high-frequency nodes sharing significant attribute overlap with $v$. This facilitates efficient mining of frequent attribute patterns and accurate identification of nodes that share the most attributes with the query node.

%First, we need to extract all prefix paths ending with $v$ from the tree $T$ denoted by $\kw{Prefix}\textit{-}\kw{Path}(v)$. Next, we use all prefix paths $\kw{Prefix}\textit{-}\kw{Path}(P[v]_x) \in \kw{Prefix}\textit{-}\kw{Path}(v)$ to construct a new conditional PP-FP-tree, which contains only the frequent node sets with high attribute overlap with $v$. 
%This allows efficient identification of frequent attribute patterns and accurately locate nodes in the public-private graph that share the most attributes with the query node. %associated with specific nodes. 
%By introducing conditional trees into the PP-FP-tree structure, we can comprehensively identify nodes in the public-private graph that share the most attributes with the query node, enhancing the precision of community search.

% Then, we apply two filtering steps, one is the coreness, and each term in the prefix paths must satisfy $\kw{P(v).core}\geq k$. Any terms that do not meet this requirement will not be added to the conditional PP-FP tree. The other is the threshold, which is specified to be the $k$-th highest frequency of the node that appears simultaneously with the node $v$. Terms that do not satisfy will also not be added to the conditional PP-FP tree. Finally, we get the conditional PP-FP tree. Now we give the Definition~\ref{Condfp} of conditional PP-FP tree and the Example~\ref{conditional} to illustrate this better.

\begin{algorithm}[t]
\caption {Public indexing: global graph index construction %($G$)
}
\label{algo.5}
\small
\begin{flushleft}
\textbf{Input:} An attributed public graph $G = (V, E, A)$\\
\textbf{Output:} The coreness tree index $coreT$ %and graph attribute map $gam$ of the public graph
\begin{algorithmic}[1]
\State Initialize an empty coreness tree index $coreT$; %, an empty graph attribute map $gam$;
\State Perform $k$-core decomposition to compute $\text{coreness}(v)$ for each node $v \in G$;
\For{each connected component $U \subseteq G$}
    \State Initialize an empty branch for connected component $U$;
    \For{each coreness value $t$ in $U$}
        \State Build a coreness tree node $u_t \leftarrow$ $ u_t \cup \{ v \in U \mid $ $ \text{coreness}$ $(v)$ $= t\}$;
        \State Assign an attribute map $m_t$ to the tree node $u_t$;
        \For{each attribute $\alpha \in attr(u_t)$}
        \State Add $\alpha$ to $m_t$;
        \For{each node $v \in u_t$}
        \If{$\alpha \subseteq attr(v)$}
        \State Add $v$ to vertex set $\theta_{\alpha}$ for attribute $\alpha$ in $m_t$;
        \EndIf
        \EndFor
        \EndFor
        \State Insert $(u_t, m_t)$ into branch $U$ in ascending order by $t$;
    \EndFor
    \State Insert the branch $U$ into $coreT$;
\EndFor

\State \Return $coreT$;
\end{algorithmic}
\end{flushleft}
\label{publicindexal}
\end{algorithm}

\begin{definition}[Conditional PP-FP-tree] 
 Given a PP-FP-tree $T$ rooted by $q$ and a vertex $v \in N'(q)$, the conditional PP-FP-tree of $v$, denoted as $T_v$, is defined as follows. First, extract all prefix-paths $\kw{Prefix}\textit{-}\kw{Path}(P[v]_x)$ from $T$ to form a subtree rooted at $q$. Then, obtain the common attribute set $C_v= attr'(q)\cap attr'(v)$ and prune all tree nodes without attribute in $C_v$ from the extracted subtree, and merge the remaining paths to obtain the conditional PP-FP-tree $T_v$. 
\label{Condfp}
\end{definition}

%Note that all tree nodes $P[u]_x$ can be easily retrieved by a look-up of table $B[u]$. Based on the conditional tree $T_u$, we can extract all tree nodes' corresponding vertices to form a candidate set for $k$-core community validation.

% \xin{add an example. Extract $v_5$ and $v_6$.}

\begin{example} 
% In Figure~\ref{fptree}(c), we build a conditional PP-FP-tree for $v_5$. We first obtain all its prefix paths denoted by $\kw{Prefix}\textit{-}\kw{Path}(v_5)$, which include $\{v_6: 2, v_3: 2, v_5: 2\}$ and $\{v_6: 2, v_5: 2\}$. Then, we obtain the common attribute $C_{v_5}= attr'(v_5)\cap attr'(q)$. Obviously, $C_{v_5} \not\subseteq attr'(v_3)$, so we delete node $v_3$.
% %The threshold needs to be set to the $k$-th highest $CP[v_c].freq$. If the $k$-th highest frequency $>0$ (meaning there are at least $k$ nodes), then we can successfully build $v_5$'s conditional PP-FP tree.
% In $v_5$'s conditional PP-FP-tree, we record the frequency of simultaneous appearances of $v_c$ and $v_5$ in all prefix paths $\kw{Prefix}\textit{-}\kw{Path}(v_5)$ , which also represents the overlapping attributes between them. Specifically, in Figure~\ref{fptree}(c), we obtain $CP[v_6]=(v_6, 4), CP[v_5]=(v_5, 4)$, indicating that $v_6$, $v_5$ and $v_1$ share a total of 4 common attributes: $\{A, B, E, F\}$.

In Figure~\ref{fptree}(c), we build a conditional PP-FP-tree $T_{v_3}$ for $v_3$. We first obtain $\kw{Prefix}\textit{-}\kw{Path}(v_3)$ from the tree $T$, including $\{v_1: 1, v_2: 1, v_3: 1\}$ and $\{v_1: 2, v_3: 2\}$. 
Then, we obtain the common attribute $C_{v_3}= attr'(v_3)\cap attr'(q)$. Since $C_{v_3} \not\subseteq attr'(v_2)$, node $v_2$ is pruned. 
Finally, the remaining paths are merged to form $T_{v_3}$, where node frequencies indicate co-occurrence with $v_3$. 
As shown in Figure~\ref{fptree}(c), we obtain $CP[v_1]=(v_1, 3), CP[v_3]=(v_3, 3)$, indicating that $v_1$, $v_3$ and $q$ share 3 common attributes $\{b, c, d\}$.

%In $v_3$'s conditional PP-FP-tree, we record the frequency of simultaneous appearances of nodes and $v_3$ in all prefix paths $\kw{Prefix}\textit{-}\kw{Path}(v_3)$, which also represents the overlapping attributes between them. Specifically, in Figure~\ref{fptree}(c), we obtain $CP[v_1]=(v_1, 3), CP[v_3]=(v_3, 3)$, indicating that $v_1$, $v_3$ and $q$ share a total of 3 common attributes: $\{b, c, d\}$.

\label{conditional}
\end{example}
%Given a node $v$, its conditional PP-FP tree denoted as $\kw{CP(v)}$ is built by extracting all prefix paths that include $\kw{P(v)}$ in the PP-FP tree. A threshold is specified to identify frequent paths, and any nodes $v_t$ with $\kw{P(v_t)}.core<k$ are excluded. These results in the final conditional PP-FP tree. 

% ///In the conditional PP-FP tree of $v_4$, we find that $\{v_4, v_2, v_3, v_1\}$ all share three attributes, indicating that they have the potential to a $k$-core community with three common attributes. Therefore, we then consider them as candidate nodes for the subsequent $k$-core verification.

\subsection{Public Graph Index Construction}

The construction of the coreness tree index involves several steps. First, perform $k$-core decomposition to determine the coreness of each node in the public graph. Next, identify the connected component each node belongs to, grouping nodes within the same connected component each node belongs to, grouping nodes within the same connected component $U$ into the same branch. Finally, within each branch, nodes are assigned to different tree nodes $u_t$ based on their coreness $t$. Besides this, we equip each tree node $u_t$ with an attribute map $m_t$. This map records each attribute that appears in the nodes of the tree node $u_t$, along with the corresponding nodes where each attribute occurs.

% The attribute map is constructed by organizing each attribute $\alpha$ present in the graph along with its corresponding vertex set $\theta_{\alpha}$.
This allows efficient retrieval of nodes satisfying coreness and sharing specific attributes simultaneously, facilitating attribute-based queries within the public graph. Detailed pseudocode is presented in Algorithm~\ref{publicindexal}.

\begin{figure}[!t]
    \centering
    \vspace{-0.2cm}
    % 第一张图
    \begin{subfigure}{\linewidth}
        \centering
        \includegraphics[width=\linewidth]{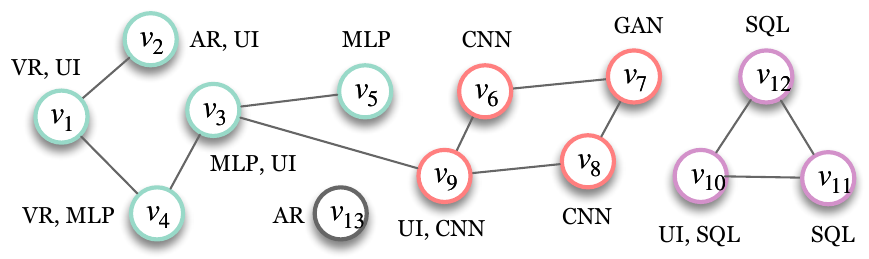}
        \caption*{(a) Attributed public graph $G$}
    \end{subfigure}
    \vspace{-0.2cm}  % 调整图之间距离
    % 第二张图
    \begin{subfigure}{\linewidth}
        \centering
        \includegraphics[width=\linewidth]{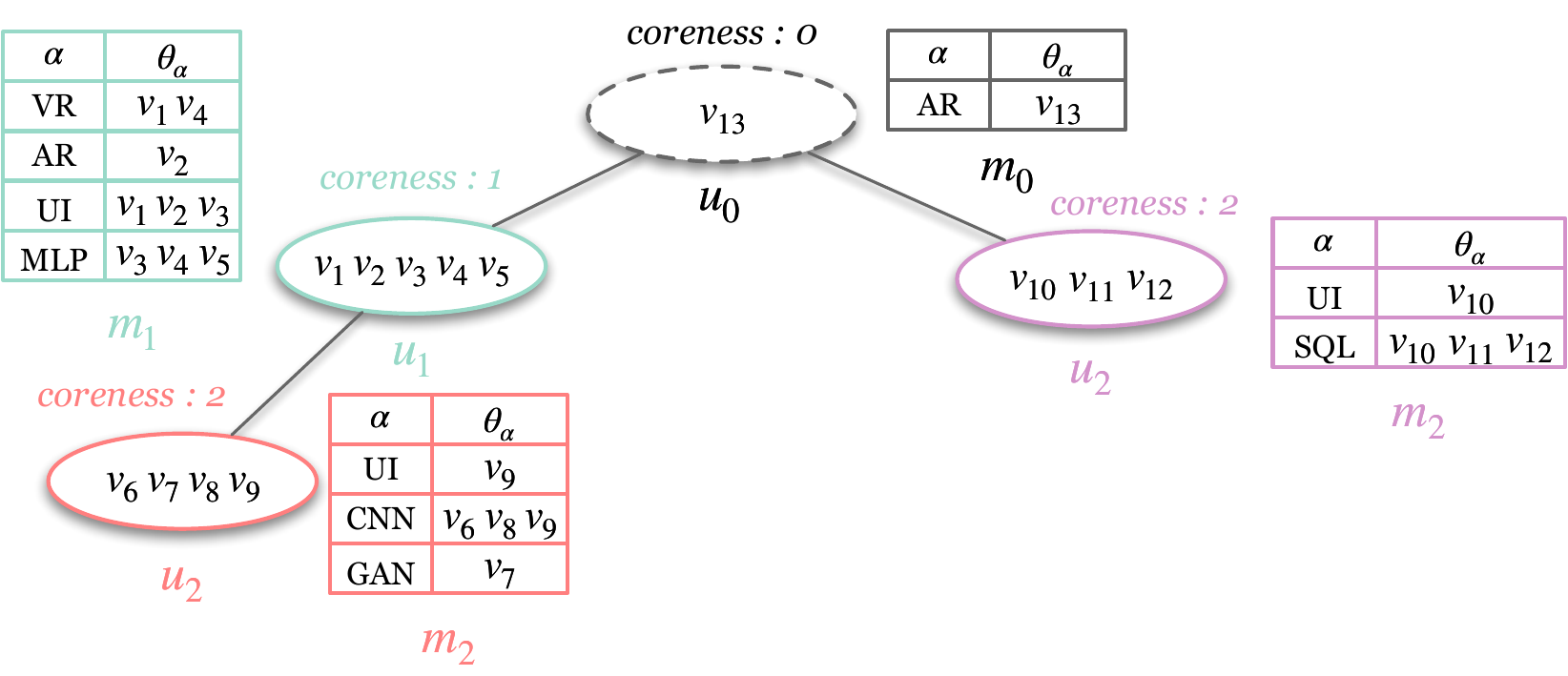}
        \caption*{(b) Coreness tree index $CoreT$}
    \end{subfigure}
    \vspace{-0.3cm}
    \caption{Public indexing construction for public graph $G$.}
    \label{public_graph_index}
    \vspace{-0.4cm}
\end{figure}

\begin{example}
To construct the coreness tree index $coreT$ in Figure~\ref{public_graph_index}(b) for the public graph $G$ in Figure~\ref{public_graph_index}(a), we first perform a $k$-core decomposition on $G$ to obtain the coreness of each node. 
Next, nodes belonging to the same connected component $U$ are grouped into the same branch of $coreT$, with isolated nodes (i.e., nodes with coreness 0) serving as the root node. As a result, one branch contains tree nodes $u_0$: $\{v_{13}\}$, $u_1$: $\{v_1, v_2, v_3, v_4, v_5\}$, $u_2$: $\{v_6, v_7, v_8, v_9\}$, while another branch contains the tree node $u_2$: $\{v_{10}, v_{11}, v_{12}\}$, as they belong to a different connected component in the graph.
Each tree node $u_t$ maintains an attribute map $m_t$, which records, for each attribute appearing in the nodes of $u_t$, the set of nodes in which the attribute occurs.

\end{example}

%\begin{example}
%Each tree node $u_t$ maintains an attribute map $m_t$, which records all attributes appearing in $u_t$'s nodes and their corresponding occurrences.
%\end{example}

\section{PP-FP-tree querying algorithm}
% This section introduces our PP-FP tree query algorithm, which identifies communities in public and private graphs through our proposed PP-FP tree structure. The algorithm consists of four key phases, as follows. 

% After introducing our designed public-private graph index, this section further presents the PP-FP-tree query algorithm developed based on it, enabling efficient community identification in public-private graphs. The algorithm consists of the following three key phases.

In this section, we introduce a public-private community querying algorithm based on the public-private graph index. The key idea of our PP-FP-tree based querying algorithm consists of three phases: \emph{attribute selection}, \emph{candidate subgraph expansion}, and \emph{community validation}.

%\begin{itemize}
%\renewcommand{\labelitemi}{\fontsize{14}{14}\selectfont\textbullet} 

\begin{figure*}
\vspace{-0.2cm}
  \centering
  \includegraphics[width=\linewidth]{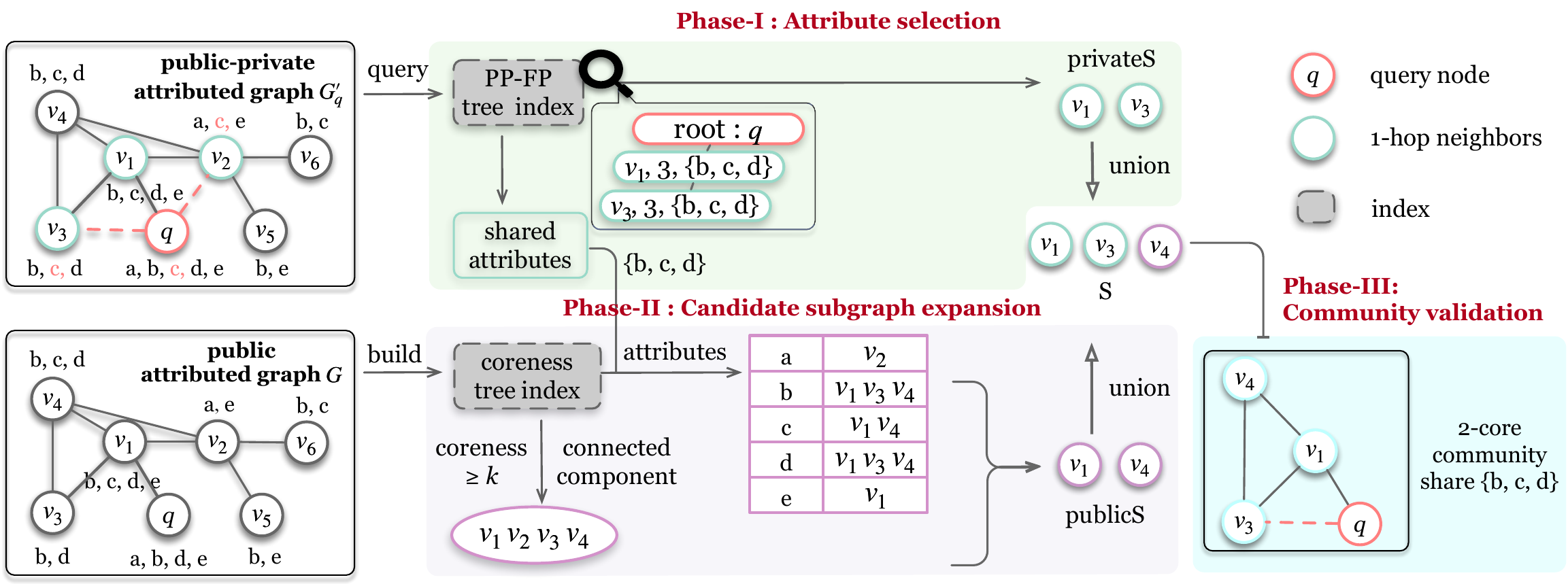}
  \caption{PP-FP-tree query algorithm framework with $k = 2$.}
  \label{Framework}
\vspace{-0.3cm}
\end{figure*}

\squishlisttight
% \item \textbf{Phase-I: Attribute set selection.} 
\item \textbf{Phase-I: Attribute selection via PP-FP-trees.} 
We first extract the largest set of common attributes from our constructed PP-FP-trees, where a possible candidate subgraph of $k$ nodes associated with such attributes exists. 
%Identify the largest shared attribute set $attr(privateS)$ from the PP-FP-tree index of the query node that appears at least $k$ times, with its corresponding node set $privateS$.
% \item \textbf{Phase-II: Attribute map intersection.} Use the graph attribute map to find the intersection of nodes that match the selected attribute set to form $attrSet$.
% \item \textbf{Phase-III: Coreness set collection.} First, identify the branch in the coreness tree index where the query node is located (i.e., the connected component it belongs to). Then, within this branch, take the union of all nodes with the coreness of at least $k$ to form $coreSet$.
\item \textbf{Phase-II: 
Candidate subgraph expansion in public graph.} 
Taking the attributes from Phase-I, we expand the candidate subgraph locally in public graph by involving those new nodes that have the coreness of at least $k$.
%First, identify the branch in the coreness tree index where the query node is located (i.e., the connected component it belongs to). Then, within this branch, take the union of nodes with the coreness of at least $k$ and sharing the attribute set $attr(privateS)$ to form the set $publicS$.

\item \textbf{Phase-III: Public-private community validation.} 
% First, take the intersection of $attrSet$ and $coreSet$ to obtain the set $publicS$. Second, take the union of $publicS$ and $privateS$ to obtain the set $S$, and perform $k$-core decomposition on the subgraph formed by the nodes in $S$ and their connections in the public-private graph to identify the $k$-core community with maximal common attributes.
%Take the union of sets $publicS$ and $privateS$ to obtain the set $S$, and perform $k$-core decomposition on the subgraph formed by the nodes in $S$ and their connections in the public-private graph to identify the $k$-core community with maximal common attributes.
%Take the union of the node sets $publicS$ and $privateS$ to form $S$, and perform $k$-core decomposition on the subgraph of $G'_q$ induced by $S$ to identify the $k$-core community with maximal common attributes.
Finally, we validate the candidate subgraph peeled by the core decomposition algorithm to meet the requirements of $k$-core and common attributes. 
%\end{itemize}
\squishend

%\subsection{Querying Phases I and II}
\subsection{Phases-I, II: Attribute Selection and Candidate Expansion}
%: Attribute set selection} 
%\subsection{Phase-I: Attribute set selection} 

% \stitle{Phase-I: Attribute set selection}. First, we search within the PP-FP-tree index of the query node's public and private neighbors to find the largest attribute set shared by at least $k$ nodes. 
% We require that at least $k$ nodes share the attribute set since for the query node $q$ to form a $k$-core community.
\stitle{Phase-I: Attribute selection via PP-FP-trees. }%Selection in private graph.} 
First, we search the PP-FP-tree index in the query node's public-private neighborhood to find the largest attribute set $attr'(privateS)$ shared by at least $k$ nodes, which forms the set $privateS$. 

Algorithm~\ref{search} outlines the details of Phase-I. We initialize $privateS$ as an empty set. %First, we take the total number of attributes shared between the first node in the first branch with $q$ as $N$ (lines 1-2).
\blue{First, we set $N$ as the maximum number of attributes shared with $q$ among all tree nodes in $T$ (lines 1-2).}
Starting from $N$, we iteratively search the PP-FP-tree to find a node set $privateS$ sharing $N$ attributes (lines 3-4). If $|privateS|\geq k$, we continue with the validation step in Algo.~\ref{algo.7}. If a feasible community $H$ is found, we return $H$ (lines 5-7). If $|privateS|< k$, we check whether the conditions for constructing a conditional PP-FP-tree are satisfied. If so,
we construct the conditional PP-FP-Tree $T_v$ by Def.~\ref{Condfp} to obtain a candidate set $privateS$;
%we invoke Algo.~?? to build. 
If the new $|privateS|\geq k$, we continue with the validation step in Algo.~\ref{algo.7}. If a feasible community $H$ is found, we return $H$ (lines 8-12). If no valid community is found, we decrement $N$ by 1 and repeat the process (line 13).

\begin{example}
%In Fig.~\ref{Framework}, the PP-FP-tree index of the public-private attributed graph $G'_q$ is illustrated in Fig.~\ref{fptree}(a). 
%We apply Algo.~\ref{search} for attribute selection. Starting from $N=4$ (the overall attribute number of $P[v_1]_1$), we search the PP-FP-tree to find a node set $privateS$ sharing $N$ attributes. However, no other nodes have the same overall attribute number, resulting in $|privateS| < k$. Therefore, we reduce $N$ to 3. Since $P[v_2]_1$ meets the requirements for conditional PP-FP construction, we attempt to build its conditional PP-FP tree but fail, as shown in Fig.\ref{fptree}~(b). We then proceed to $P[v_3]_1$, which also satisfies the conditions, and successfully construct its conditional PP-FP tree, as illustrated in Fig.\ref{fptree}~(c).
%Through conditional PP-FP-tree mining in Fig.~\ref{fptree}(c), where nodes $v_1$ and $v_3$ share the attribute set $\{b,c,d\}$. These nodes form the set $privateS$, representing the selection in private graph.

We illustrate the process of attribute selection based on PP-FP-tree in Phase-I in Figure~\ref{Framework}. We first start from $N=4$ %by checking the index $P[v_1]_1$ and find the candidate set of four attributes $\{b, c, d, e\}$.
\blue{since it is the maximal common attribute number in the tree in Figure~\ref{fptree}(a).} But, there exist no other nodes sharing the same attributes, resulting in $|privateS| < k$. Thus, we reduce $N$ by one to check $N=3$. 
We check the tree node $P[v_2]_1$ but cannot build its conditional PP-FP-tree shown in Figure~\ref{fptree}(b). Then, we proceed to the tree node $P[v_3]_1$ and find three common attributes in Figure~\ref{fptree}(c). This indicates a feasible community $H$ with the node set of $privateS=\{v_1,  v_3\}$, which satisfies the 2-core requirement.
\end{example}

\begin{algorithm}[t]
\caption{
PP-FP-tree based Attributed Community Search 
%Search $k$-core community with maximum attributes ($k$, $q$, $T$)
}
\small
\label{algo.6}
\begin{flushleft}
\textbf{Input:} PP-FP-tree $T$, query $q$, integer $k$\\
\textbf{Output:} A public-private attributed community $H$
%satisfied $k$-core queried by the node $q$
\begin{algorithmic}[1]
\State 
% Initialize potential max attribute $pmaxa$, max attributes $maxa$, temp max attributes $tmaxa \leftarrow 0$, %coreness map $cmap$, 
% branches $b$, selected nodes $snodes$, temp selected nodes $tsnodes$, the set $built \leftarrow \emptyset$,
Initialize $privateS \leftarrow \emptyset$;
%found\leftarrow \text{false}, 
% \For{each $line$ in $ncl_q$}
% \State $cmap[n.id] \leftarrow ncl_q\{step, coreness\}$;
% \EndFor
% \For{each $line$ in $T_q$}
% \For{each $line$ in $ncl_q$}
% \State Create node $n$ with \{$id$, $count$, $totalcount$, $hop$, $coreness$\};
% \EndFor
% \EndFor
%\State $N \leftarrow %P[v]_x. \kw{Overall}\textit{-}\kw{AttrNum}$ of the first node in the first branch of $T$;
\State \blue{$N \leftarrow \max_{P[u]_x \in T} {P[u]_x.\kw{Overall}\textit{-}\kw{AttrNum}}$};

\While {$N>1$}
% \For{each $branch$ in $b$}
% \For{each tree node $P[v]_x$ in $branch$}
% \If{$P[v]_x.\kw{Prefix}\textit{-}\kw{AttrNum} = N$}
% \State $S \leftarrow S \cup v$;
\State Extract a set $privateS$ of candidate vertices with at least $N$ common attributes, i.e., $privateS$ = $\{w \mid w \in \kw{Prefix}\textit{-}\kw{Path}(v), P[w]_x.\kw{Prefix}\textit{-}\kw{AttrNum} \geq N\}$;
% \State Extract a set $privateS$ of candidate vertices with at least $N$ common attributes, i.e.,
% \[
% privateS = \left\{w \mid 
% \begin{aligned}
%     & w \in \kw{Prefix}\textit{-}\kw{Path}(v), \\
%     & P[w]_x.\kw{Prefix}\textit{-}\kw{AttrNum} \geq N
% \end{aligned}
% \right\}
% \]
\If{$|privateS|\geq k$}
\State Validate community $H$ based on $privateS$ in Algo.~\ref{algo.7};
%\State Validate the feasibility of $k$-core community $H$ expanding from the seed $S$ using Algorithm 8;
\If{$\exists$ a feasible community $H$}
\Return $H$;
\EndIf
\EndIf

%Requirements of conditional PP-FP-tree construction 
\If{$P[v]_x$ satisfies the conditions of building conditional PP-FP-tree by Def.~\ref{requirements}}
\State %Invoke Algo.~\ref{algo.5} to
Search candidate set by constructing conditional PP-FP-Tree $T_v$ by Def.~\ref{Condfp} and obtain a candidate set $privateS$;
\EndIf
\If{$|privateS|\geq k$}
\State Validate community $H$ based on $privateS$ in Algo.~\ref{algo.7};
\If{$\exists$ a feasiable community $H$}
\Return $H$;
\EndIf
\EndIf
%\ElsIf{$P[v]_x$ satisfies the conditions of building conditional PP-FP tree}
% \State Invoking Algorithm~\ref{algo.5};
% \State $
% \begin{aligned}
% \end{aligned}$
% \If{$|S|\geq k$}
% \State Apply Algorithm 8;
% \EndIf
% \EndFor
% \If{$found$}
% \Return the community $H$
% \Else \State $S \leftarrow S-1$;
% \EndIf
%\State $N \leftarrow N-1$;
\State Decrement $N$ by 1;
\EndWhile
\State \Return {$\emptyset$};
\end{algorithmic}
\end{flushleft}
\label{search}
\end{algorithm}

\stitle{Phase-II: Candidate subgraph expansion in public graph. }%Expansion in public graph
In this phase, we need to identify the nodes in the public graph $G$ that could expand the $k$-core community with maximal attributes with $q$. First, we locate the branch in the coreness tree index $coreT$ where the query node $q$ is positioned, which represents the connected component $U$ that $q$ belongs to in the public graph. Second, using the attribute set $attr'(privateS)$ identified in the Phase-I, we search within the connected component $U$ for nodes that satisfy the conditions of coreness $\geq k$ and sharing the attribute set $attr'(privateS)$. These nodes form the set $publicS$, representing the nodes in the public graph that could potentially expand the $k$-core community.

\begin{algorithm}[t]
%\caption{Validate candidate set $(snodes, commattr, G'_q,q,k)$}
\small
\caption{Community validation using candidate $privateS$}
%set $(snodes, commattr, G'_q,q,k)$}
\label{algo.7} 

\begin{flushleft}
%\textbf{Input:} The candidate node set $S$, the common attribute of the set $attr(S)$, the query node $q$, the graph attribute map $gam$, the coreness tree index $coreT$, the public-private graph $G'_q$, and integer $k$ 
\textbf{Input:} 
% candidate $privateS$,  query $q$, public index,  private PP-FP-tree index
candidate $privateS$, query $q$, coreness tree index $coreT$
\\
\textbf{Output:} A boolean value indicating the feasibility of finding community $H$ by $privateS$ 

\begin{algorithmic}[1]
\State Initialize 
%$attrSet \leftarrow \emptyset, coreSet \leftarrow \emptyset, 
$publicS \leftarrow \emptyset$, $S \leftarrow \emptyset$;   
% \For {each attribute $\alpha \in attr(privateS)$}
% \State Get vertex sets $\theta_{\alpha}$ from $gam[\alpha]$;
% \If{$attrSet = \emptyset$}
% \State $attrSet \leftarrow \theta_{\alpha}$;
% \Else
% \State $attrSet \leftarrow attrSet \cap \theta_{\alpha}$;
% \EndIf
% \EndFor
%\State $attrS \leftarrow S \cup attrS$;
\State $publicS \leftarrow $ get the union of nodes in the tree node $u_t$ with coreness $t\geq k$ and sharing $attr(privateS)$ in the connected component $\{U\mid q\in U\}$ from coreness tree index $coreT$;
% \State $publicS \leftarrow attrSet \cap coreSet$;
\State $S \leftarrow privateS \cup publicS$;
\If{$|S| \geq k + 1$}
    \State \textbf{if}  a non-empty $k$-core community $H$ exists in the induced subgraph $G'_q[S]$ containing $q$ \textbf{then return} True;
    % \State Initialize $G[S]$ as a new graph;
    % \For{each $nodex \in S$}
    % \For{each $nodey \in S$}
    %     \If{$G'_q$ has an edge $(nodex, nodey)$}
    %     \State Add an edge $(nodex, nodey)$ to $G[S]$;
    %     \EndIf
    % \EndFor
    % \EndFor
% \State Perform $k$-core decomposition on $G[S]$;
% \If{$G[S]$ contains node $q$}
% \State \Return True;
%\EndIf
\EndIf
\State \Return False;
\end{algorithmic}
\end{flushleft}
\end{algorithm}

\begin{example}
Consider the public attributed graph in Figure~\ref{Framework}. To expand the $k$-core community with $q$, we first locate the connected component containing $q$ where all nodes have coreness $\geq k$. We then check whether these nodes share the attributes $\{b, c, d\}$ from Phase-I. Only $\{v_1, v_4\}$ satisfy this condition, forming the set $publicS$.
\end{example}

% \subsection{Phase-III: Candidate community validation}
\subsection{Phase-III: Public-Private Community Validation}
\label{validation}

%Algorithm~\ref{algo.7} outlines the details of phase II and phase III. Two sets are initialized: $publicS$, the candidate nodes from the public graph, and $S$, the final candidate set (line 1). $publicS$ is determined by collecting nodes with coreness $\geq k$ that share the attribute set $attr(privateS)$ within the connected component of $q$ in the coreness tree index $coreT$ (line 2). The final set $S$ is then obtained as the union of $privateS$ and $publicS$ (line 3). If $|S| \ge k+1$, a non-empty subgraph $H$ in the induced subgraph $G'_q[S]$ is extracted to verify the feasibility of the $k$-core community containing $q$, and $H$ is returned as the result.

%\stitle{Phase-III: Candidate community validation. } 
\stitle{Phase-III: Candidate community validation. } 
%Algorithm~\ref{algo.7} presents the details of phase-II and phase-III. 
Algorithm~\ref{algo.7} presents the details of Phase-III. 
Two sets are initialized: $publicS$, representing the set of candidate nodes in the public graph; $S$, representing the final candidate set (line 1). 
$publicS$ is determined as the union of node sets with coreness $\geq k$ and sharing attribute set $attr'(privateS)$ in the connected component where $q$ is located from coreness tree index $coreT$ (line 2). 
The final set $S$ is then obtained as the union of $privateS$ and $publicS$ (line 3). 
%representing the node set from the public-private graph 
If the size of $S$ is at least $k+1$, and a non-empty $k$-core community $H$ containing $q$ exists in the induced subgraph $G'_q[S]$, return True (lines 4-5). Otherwise, return False (line 6).

\begin{example}
We obtain $privateS$ from Phase-I and $publicS$ from Phase-II. By combining them, we have $S = privateS \cup publicS = \{v_1, v_3, v_4\}$. Since $|S| > 2$, we perform community validation and obtain a 2-core community sharing $\{b, c, d\}$.
\end{example}

\subsection{\blue{Complexity Analysis}}
\label{seccomplexity}
\blue{In the following, we analyze the complexity of PP-FP-tree index construction and search algorithms.
We denote the size of vertices, edges, and attributes in public graph $G$ as $n=|V|$, $m=|E|$, and $t=|\sum_{v\in V} attr(v)|$, respectively. Assume that a connected graph $G$ with $n\in O(m)$. The number of private edges is $m_p= \sum_{v\in V} |N_p(v)|$. Thus, the total number of public and private edges is denoted as $\hat{m}=m+m_p$. The maximum number of public attributes and private attributes are denoted as $\tau_{max} =\max_{v\in V} |attr(v)|$ and $\rho_{max} =\max_{q\in V, v\in N_p(q)} |attr_p(v)|$, respectively. %The public-private neighborhood and attribute of the query node $q\in V$ are denoted as $N'(q)=N_p(q)\cup N(q)$ and $A'(q)=attr(q)\cup attr_p(q)$, respectively.
%The public-private neighborhood and attribute of query $q\in V$ are denoted as $N'(q)=N_p(q)\cup N(q)$ and $A'(q)=attr(q)\cup attr_p(q)$, respectively.
The public-private  attribute of $q$ is $A'(q)=attr(q)\cup attr_p(q)$.}

\stitle{\blue{Complexity of index construction.}} \blue{
We analyze the index construction in terms of public indexing and private indexing.\\
First, we consider the public indexing in Algorithm~\ref{algo.5}. The core decomposition takes $O(m)$ time and $O(m+t)$ space for storing the entire graph $G$. To construct connection components in $k$-cores, it makes use of union-find structure taking $O(m\alpha(n))$ time and $O(n)$ space, where $\alpha(n)$ is an inverse Ackermann function usually with $\alpha(n)\leq 5$ in practice. Moreover,  Algorithm~\ref{algo.5} takes $O(t)$ time to reorganize the nodes' attribute %list
map. Overall, Algorithm~\ref{algo.5} takes $O(m\alpha(n))$ time  and $O(n+m+t)\subseteq O(m+t)$ space. \\
Second, we consider the private indexing in Algorithm~\ref{algo.3}. For each vertex $q\in V$, it constructs the PP-FP-tree by the one-time scanning of all attributes of nodes in $q$'s public and private 
%graphs.
neighborhood.
It takes $O(\sum_{v\in N(q)\cup N_p(q)} |attr(v)\cup attr_p(v)|)$ time. Thus, the total time of constructing the PP-FP-tree of all nodes $q\in V$ is $O(\sum_{q\in V} \sum_{v\in N(q)\cup N_p(q)} |attr(v)\cup attr_p(v)|)$
$=O((\rho_{max}+\tau_{max}) \cdot  \sum_{q\in V} \sum_{v\in N(q)\cup N_p(q)} 1) $
$=O((\rho_{max}+\tau_{max}) \cdot (m+m_p) )$ $=O((\rho_{max}+\tau_{max}) \hat{m} )$ time. 
As Algorithm~\ref{algo.6} can release the memory of PP-FP-tree index after the construction of each vertex, it takes $O(m+t+m_p+n\rho_{max})$ $=O(\hat{m}+t+n\rho_{max})$  space. 
%The total index size of private PP-FP-tree takes $O((\rho_{max}+\tau_{max}) \hat{m} )$ space.
In summary, our index construction takes $O(m\alpha(n)+(\rho_{max}+\tau_{max}) \hat{m}))$ time and $O(\hat{m}+t+n\rho_{max})$ space, which produces the index size of $O((\rho_{max}+\tau_{max}) \hat{m})$ space. 
%The sorting issue can be done in linear time using bin sort. 
}

%PP-FP-tree algorithm in terms of the public graph index construction and the private graph index construction. 
% mi The public index organizes nodes by their coreness and connected components in the public graph. Each tree node is equipped with an attribute map. The map records the attributes appears in this tree node along with the corresponding vertex sets. Therefore, the space complexity of the public graph is $O(n+t)$. }
%\blue{index size% $O(nA_{max}) $
%space complexity}%$O(m+nA_{max})$}
%\\mi\blue{Let $attr'(q)$ represent the public-private attributes of $q$. The space complexity of the PP-FP-tree index is $O(\sum_{q\in {V_q}}|N'(q)|\cdot|attr'(q)|)$. As a result, the total index size is $O(|V|+\sum_{v\in V}|attr(v)|+ \sum_{q\in {V_q}}|N'(q)|\cdot|attr'(q)|)$ $\subseteq O(\sum_{q\in{V_q}}|N'(q)|\cdot|attr'(q)|)$. Space complexity of PP-FP querying algorithm is $O(|V|+|E|+|A|+\sum_{v\in {V}}|N'(q)|\cdot|attr'(q)|)$}

\stitle{\blue{Complexity of query processing.}} 
\blue{We analyze the complexity of public-private attributed community search in Algorithm~\ref{algo.6}. The whole framework consists of three phases. The Phase-I selects candidate attributes $A_q^{\ast}$ in the PP-FP-tree of query vertex $q$. Due to the problem NP-hardness, the number of $A_q^{\ast}$ is bounded by the possible combination of $A'(q)$, i.e., $|A_q^{\ast}| \in O(2^{|A'(q)|})$, which shares the same theoretical computation with Inc-S~\cite{fang2016effective}
and Dec~\cite{fang2017effective}.
Our PP-FP-tree techniques significantly reduce the practical enumeration of $A'(q)$. Based on $A_q^{\ast}$, Algorithm~\ref{algo.7} expands $A_q^{\ast}$ to a local subgraph $\hat{G}$ in Phase-II and adopts the community validation by a linear peeling of $k$-core decomposition in Phase-III. 
These two phases take $O(|E(\hat{G})|+|A(\hat{G})|)$ time, where $|E(\hat{G})|$ and $|A(\hat{G})|$ are the size of edges and attributes in $\hat{G}$, respectively. 
%These two phases take $O(m_{\hat{G}}+A_{\hat{G}})$ time, where $m_{\hat{G}}=|E(\hat{G})|$ and $A_{\hat{G}}=|A(\hat{G})|$ are the size of edges and attributes in $\hat{G}$, respectively. 
These two phases take $O(m_{\hat{G}}+n_{\hat{G}}(\rho_{max}+\tau_{max}))$ time, where $m_{\hat{G}}$ and $n_{\hat{G}}$ are the size of edges and nodes in $\hat{G}$, respectively.
Here, $m_{\hat{G}}\ll \hat{m}$ and $n_{\hat{G}} \ll n$ for a small graph $\hat{G}$ in practice.
%Here, the candidate $\hat{G}$ is a quite small graph w.r.t. $G$ in practice.
%(\rho_{max}+\tau_{max}) \hat{m}
%Here, $m_{\hat{G}}\ll \hat{m}$ 
As a result, Algorithm~\ref{algo.6} takes $O(|A_q^{\ast}| (m_{\hat{G}}+n_{\hat{G}}(\rho_{max}+\tau_{max})))$ time and $O(\hat{m}+t+n\rho_{max})$ space. 
}

%% file: experiment.tex
\section{experiments} 
\label{sectionexp}
%In this section, we present the experiment results. The first part introduces the setup of our experiments, the second part presents the experimental results, and the third part explores a case study.

%\subsection{Setup}

\stitle{Datasets.}
We use nine real-world datasets of public-private networks in Table~\ref{dataset}. In the five DBLP datasets~\cite{huang2018pp}, each vertex represents an author where the attributes are the authors' frequent research keywords. \blue{The number of attributes per author ranges from 1 to 20.} The edges represent the co-authorship relationships between authors \footnote{\url{https://github.com/samjjx/pp-data}}. 
%For each author, we extract up to 20 frequent keywords from the titles of their publications to serve as their attributes. 
%Additionally, we consider published articles as public information, representing the public graph, while ongoing collaborations are treated as private information, known only to the authors, representing the private graph. 
We treat the co-authorship in published works as public graphs and the ongoing collaborations as private graphs.  
%The Facebook dataset includes 10 ego-networks \cite{leskovec2012learning}, each represented by its ego-user $E$, where $E\in\{0, 107, 348,$ $414,686,698,1684,1912,3437,3890\}$. The vertex attributes of these ego networks are derived from real user-profiles and anonymized. Each ego network gives multiple real communities called friendship circles. Since there is no private graph in the original dataset, we convert the original dataset into the public-private graph by converting 50\% of the edges of the query nodes into private edges and assigning one to three private common attributes to the nodes connected by these edges in the ego-networks. The last two datasets are YouTube and LiveJournal social networks. Since the datasets do not contain node attributes, we assign two attributes to the nodes in each of the top 5000 ground-truth communities. Additionally, we randomly select 10\% of the nodes to add noise attributes.
The Facebook dataset includes 10 ego-networks \cite{leskovec2012learning}, each represented by its ego-user $E$, where $E\in\{0,107,348,414,686,698,1684,1912,3437,3980\}$. Each ego-network contains multiple real communities as the social circles. \blue{The number of attributes per node ranges from 1 to 30.}
\blue{The last three datasets YouTube, LiveJournal, and Orkut, each contain 5000 ground-truth communities. The number of attributes per node ranges from 2 to 50.} We randomly select 10\% of the nodes to add noise attributes.
To generate private information in these four datasets, we convert 50\% of the edges of the query nodes into private edges and change 1-3 attributes as the private attributes of nodes.

\noindent \textbf{Compared methods.} 
%To evaluate the effectiveness and efficiency of community search and index construction, we compare five methods as follows:
We compare 5 methods as follows.
% \begin{itemize}
\squishlisttight
    \renewcommand{\labelitemi}{\large\textbullet} % 或者使用 \Huge 或 \LARGE
   
    % \item Basic-g \cite{fang2017effective}: is an online keyword community search algorithm that identifies communities by progressively expanding candidate sets within the $k$-core subgraph.
    
    % \item Basic-w \cite{fang2017effective}: 
    % Similar to Basic-g, Basic-w searches on  entire graph instead of the $k$-core subgraph only.
    
    % \item Inc-T \cite{fang2016effective}: 
    % extends Inc-S by dynamically adjusting  candidates for dynamic updates and query optimization.
    
    \item Online-basic: 
    is our proposed Algorithm 1 of a straightforward online search.
    
    \item Online-binary: is our proposed algorithm of an online search with binary search.

    \item Inc-S \cite{fang2016effective}: is an index-based static keyword community search algorithm that improves efficiency by incremental candidate expansion and pruning.

    \item Dec \cite{fang2017effective}: is an index-based community search algorithm that verifies candidate keyword sets decrementally.
    %in a decremental manner to enable early pruning and reduce cost.

    % \item ATC \cite{huang2016attribute}: 

    \item PP-FP: is our proposed  PP-FP-tree based querying algorithm, which is publicly available on GitHub \footnote{\url{https://github.com/csyqchen/PP-FP}}.
    %is our proposed efficient three-phase query algorithm based on the PP-FP-tree index, which is publicly available on GitHub \footnote{\url{https://github.com/csyqchen/PP-FP}}.
% \end{itemize}
\squishend

\begin{table*}[t!]
 % \vspace{-0.4cm}
    \centering
    \begin{tabular}{ccccccccc}
    \toprule
    Datasets & $|V|$ & $|E|$ & $|V_{private}|$ & $|E_{private}|$ & \blue{Graph Size} & \blue{Index Size} & \blue{Construction Time} & \blue{Memory Usage}\\ 
    \midrule
    DBLP2013 & 2,221,139 & 5,432,667 & 1,265,175 & 6,007,245 & \blue{1.5GB} & \blue{3.6GB} & \blue{1455.5s} & \blue{2.0GB} \\
    DBLP2014 & 2,221,139 & 6,186,831 & 1,150,642 & 5,322,474 & \blue{1.4GB} & \blue{5.5GB} & \blue{2026.9s} & \blue{2.1GB}\\
    DBLP2015 & 2,221,139 & 7,012,003 & 1,018,652 & 4,518,645 & \blue{1.2GB} & \blue{5.0GB} & \blue{1921.1s} & \blue{2.3GB}\\
    DBLP2016 & 2,221,139 & 7,864,133 & 870,054 & 3,628,517 & \blue{1.1GB} & \blue{4.3GB} & \blue{1440.5s} & \blue{2.5GB}\\
    DBLP2017 & 2,221,139 & 8,794,753 & 690,588 & 2,658,750 & \blue{0.9GB} & \blue{3.5GB} & \blue{1989.8s} & \blue{2.7GB}\\
    Facebook & 4,039 & 88,234 & 172 & 1,200 & \blue{1.7MB} & \blue{0.6MB} & \blue{5.8s}& \blue{4.3MB}\\
    YouTube & 1,134,890 & 2,987,624 & 702,958 & 2,183,530 & \blue{0.3GB} & \blue{3.5GB} & \blue{1006.3s} & \blue{0.6GB}\\
    LiveJournal & 2,798,962 & 34,681,189 & 1,798,200 & 8,404,127 & \blue{1.5GB} & \blue{11.7GB} & \blue{4175.7s} & \blue{3.4GB}\\
    \blue{Orkut} & \blue{3,072,441} & \blue{117,185,083} & \blue{1,698,008} & \blue{6,469,106} & \blue{2.6GB} & \blue{12.0GB} & \blue{4658.1s} & \blue{11.3GB}\\
    \bottomrule
    \end{tabular}
    \caption{Datasets used in our experiments.}
    \label{dataset}
    \vspace{-0.4cm}
\end{table*}

Note that Inc-S and Dec are two community search methods originally designed for public attributed graphs. For comparison, we treat the public-private attributed graph as a dynamic graph. Specifically, we first dynamically update the keywords and edges in the indexes of Inc-S and Dec to match the public-private graph, and then perform community search based on the updated indexes.

% \begin{figure*}[t] % 使用 figure* 使其跨越双栏
% \centering
% \hspace{0.05\textwidth}    
%     \includegraphics[width=\textwidth]{exp_figures/dblp_iteration.pdf} 
%     \label{fig:fig1}
%     \vspace{-0.5cm}
%     \caption{Validation iteration results on DBLP datasets.}
% \label{fig:validation}
% \vspace{-0.3cm}
% \end{figure*}
\vspace{-0.1cm}
\begin{figure*}[!t]
    \centering
    \hspace{-0.02\textwidth}

    % ---------- First row ----------
    \begin{subfigure}[t]{0.242\textwidth}
        \centering
        \includegraphics[width=\textwidth, height=3.2cm]{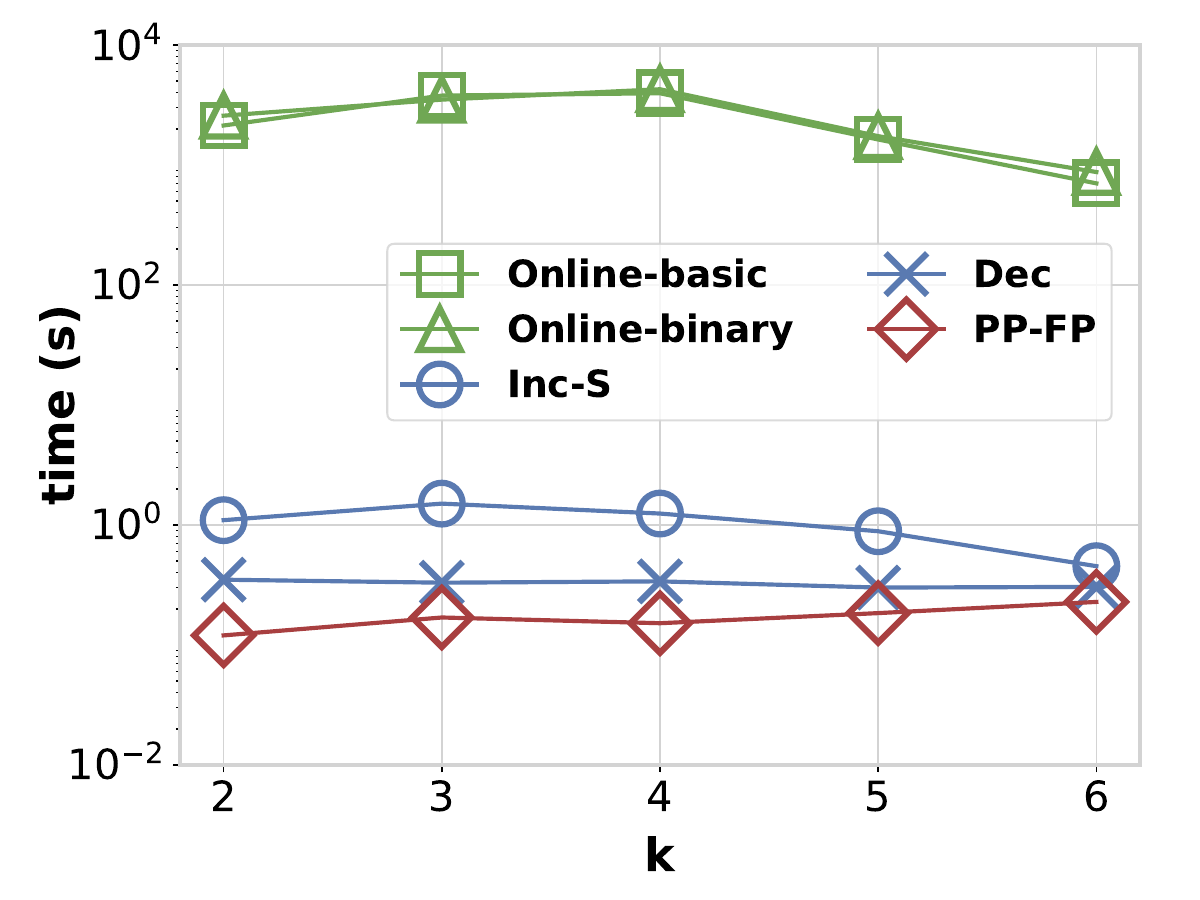}
        \caption{DBLP2013}
        \label{fig:eff-2013}
    \end{subfigure}
    \begin{subfigure}[t]{0.242\textwidth}
        \centering
        \includegraphics[width=\textwidth, height=3.2cm]{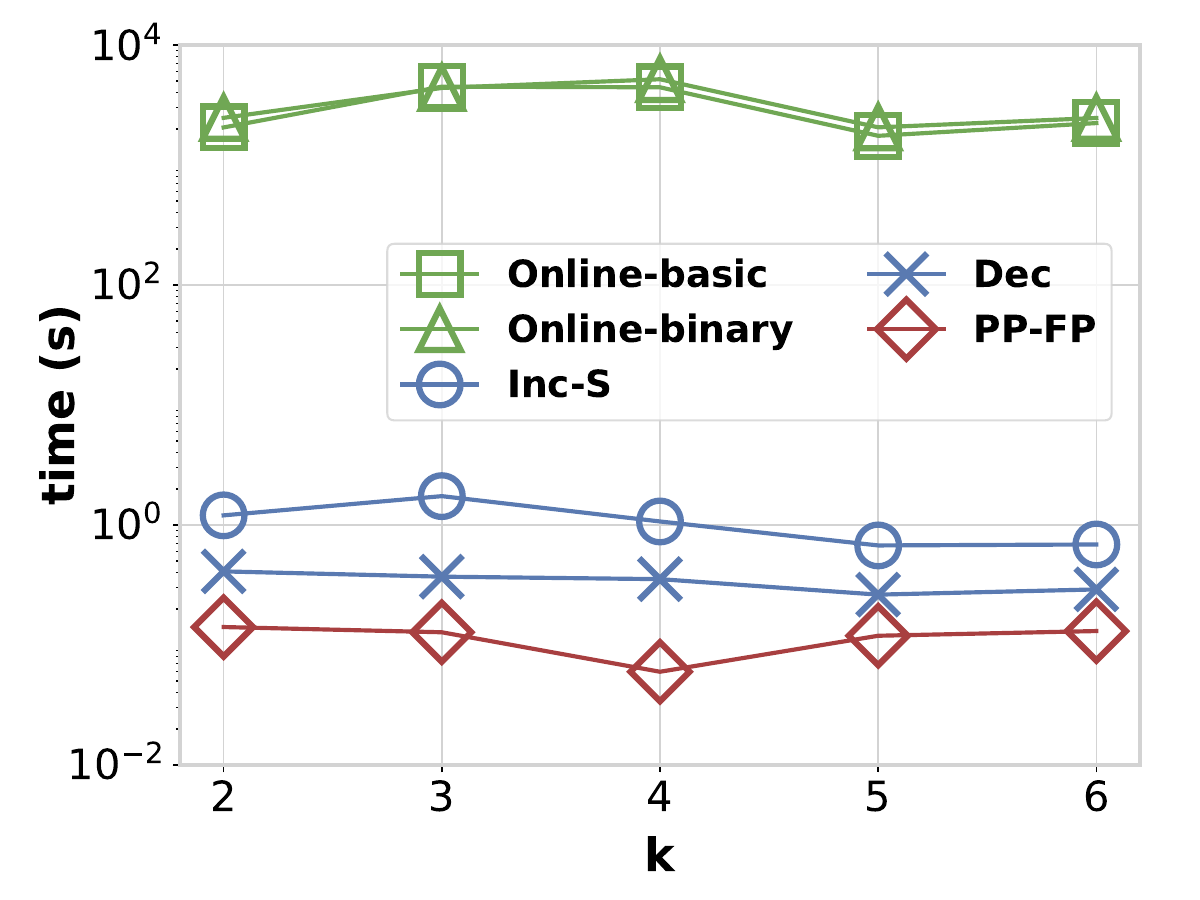}
        \caption{DBLP2014}
        \label{fig:eff-2014}
    \end{subfigure}
    \begin{subfigure}[t]{0.242\textwidth}
        \centering
        \includegraphics[width=\textwidth, height=3.2cm]{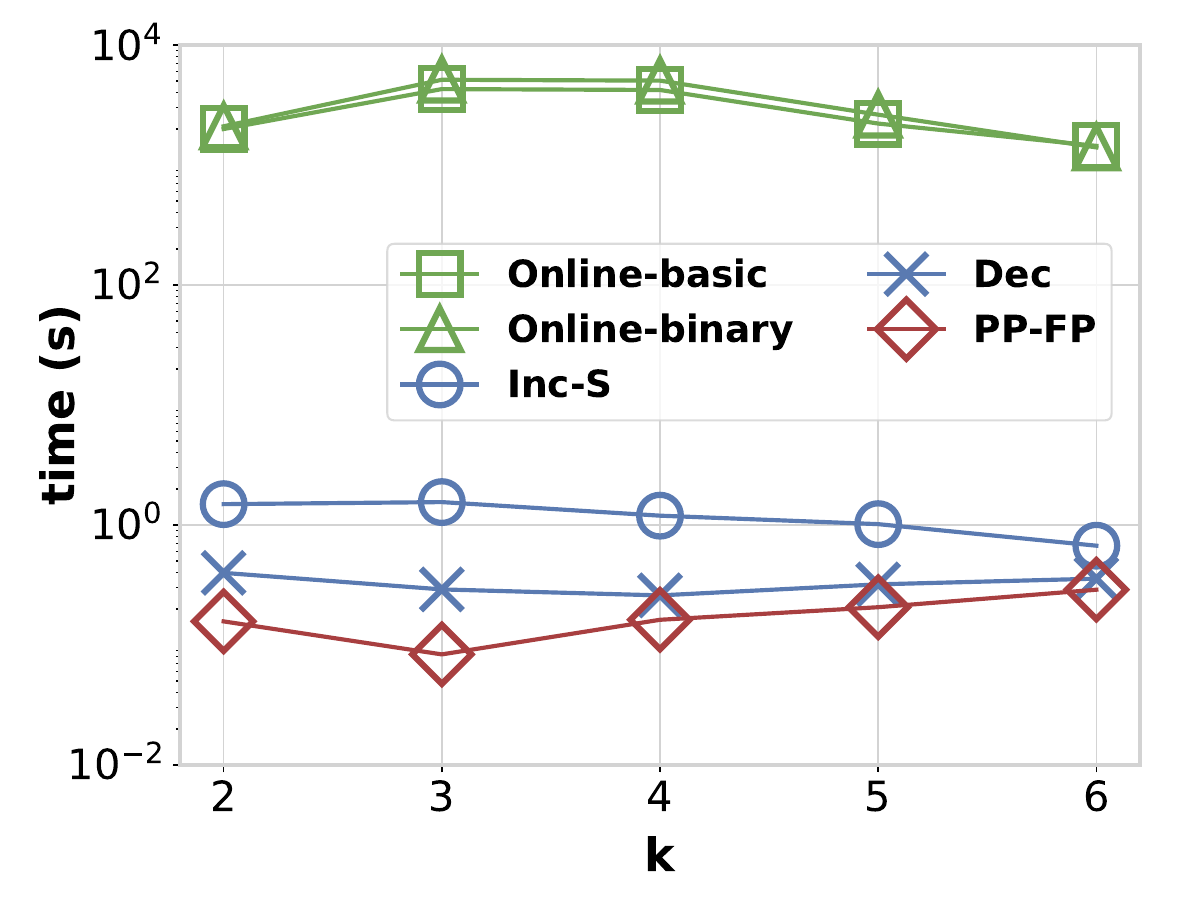}
        \caption{DBLP2015}
        \label{fig:eff-2015}
    \end{subfigure}
    \begin{subfigure}[t]{0.242\textwidth}
        \centering
        \includegraphics[width=\textwidth, height=3.2cm]{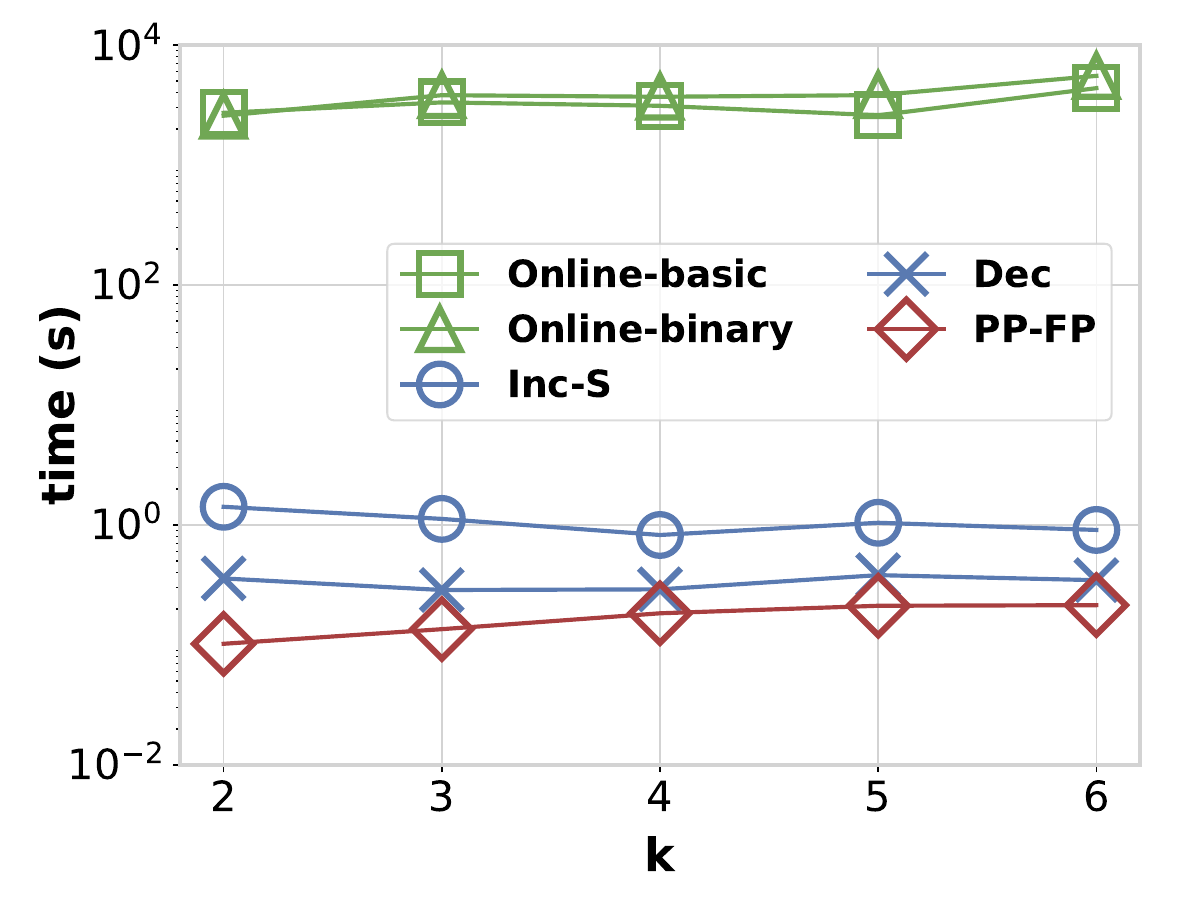}
        \caption{DBLP2016}
        \label{fig:eff-2016}
    \end{subfigure}

    %\vspace{-0.25cm}

    % ---------- Second row ----------
    \begin{subfigure}[t]{0.242\textwidth}
        \centering
        \includegraphics[width=\textwidth, height=3.2cm]{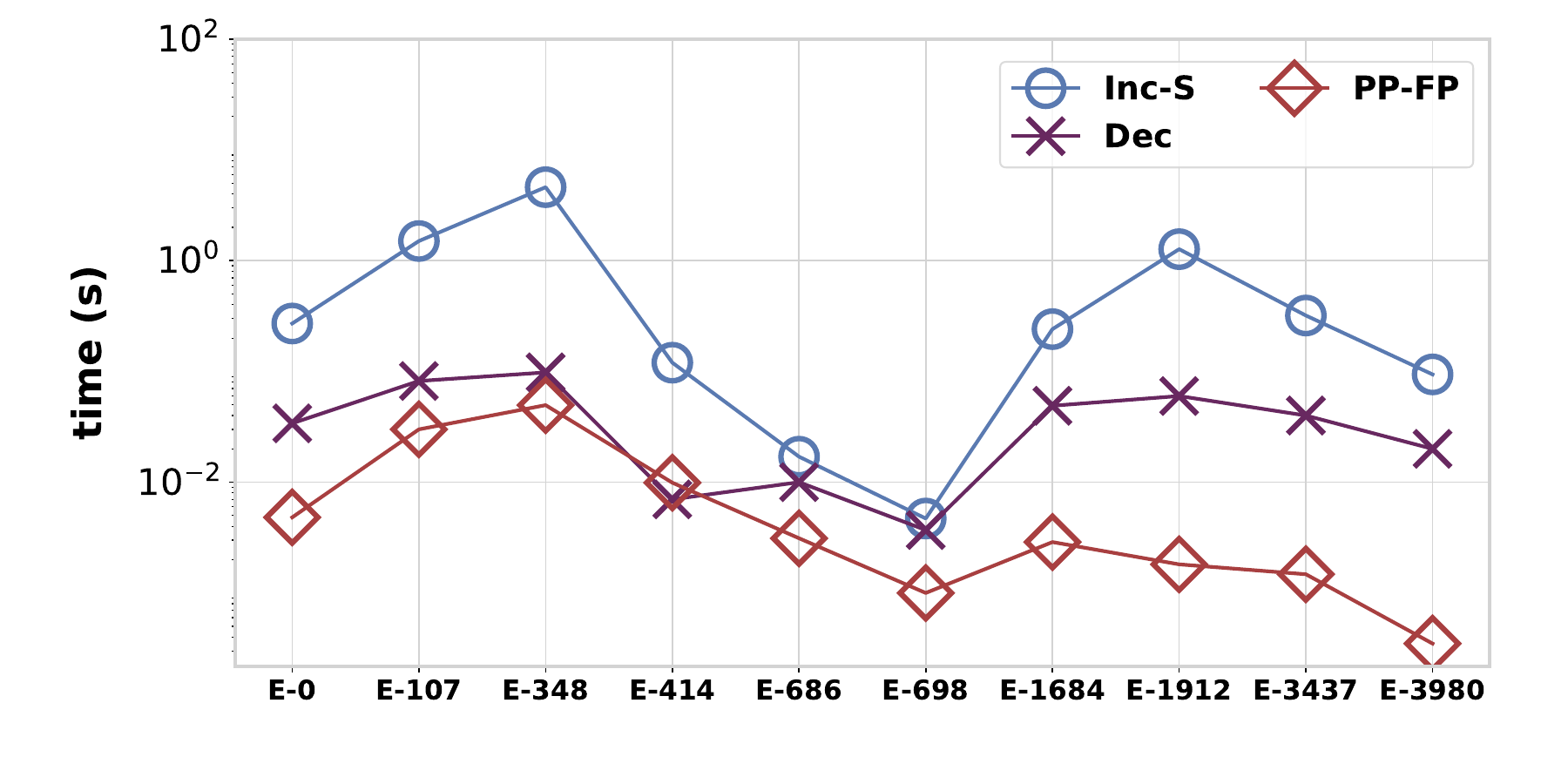}
        \caption{Facebook}
        \label{fig:eff-facebook}
    \end{subfigure}
    \begin{subfigure}[t]{0.242\textwidth}
        \centering
        \includegraphics[width=\textwidth, height=3.2cm]{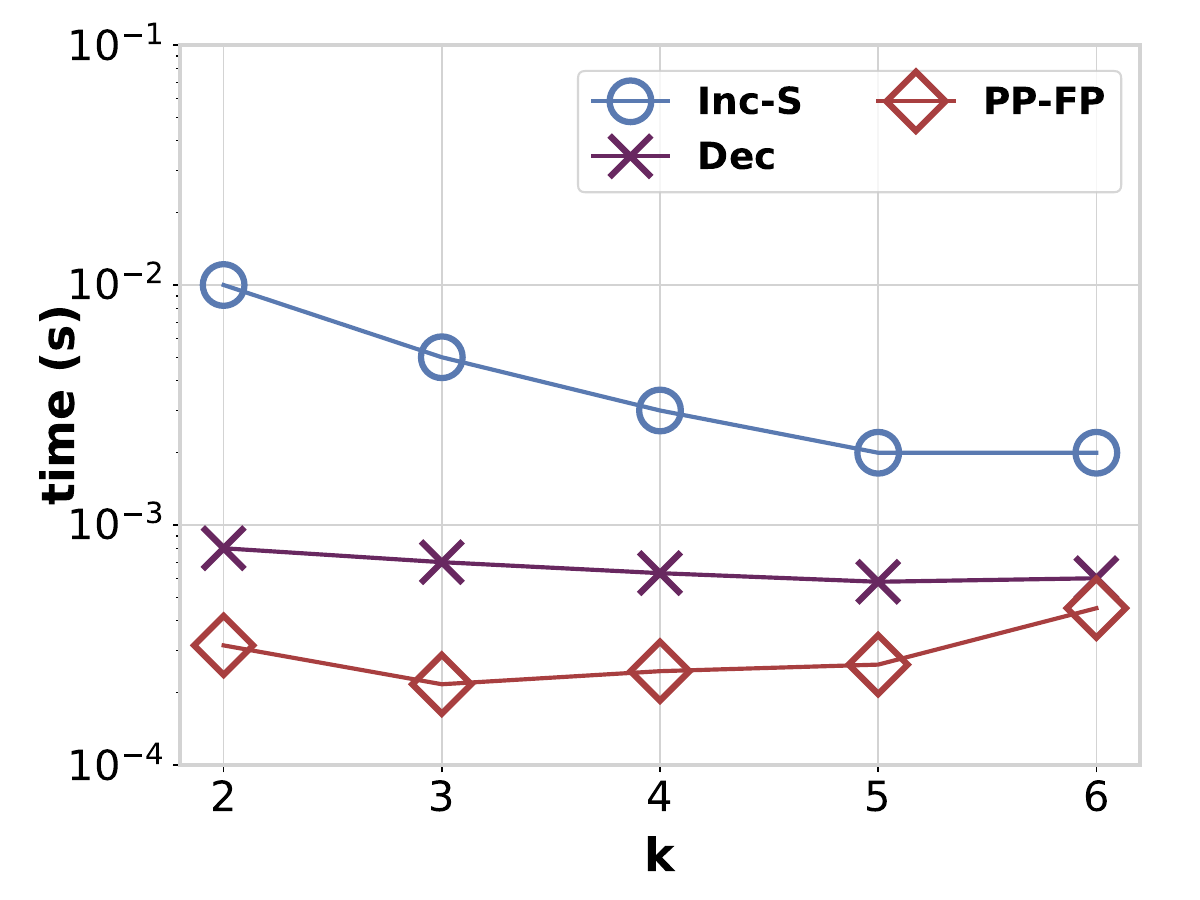}
        \caption{LiveJournal}
        \label{fig:eff-livejournal}
    \end{subfigure}
    \begin{subfigure}[t]{0.242\textwidth}
        \centering
        \includegraphics[width=\textwidth, height=3.2cm]{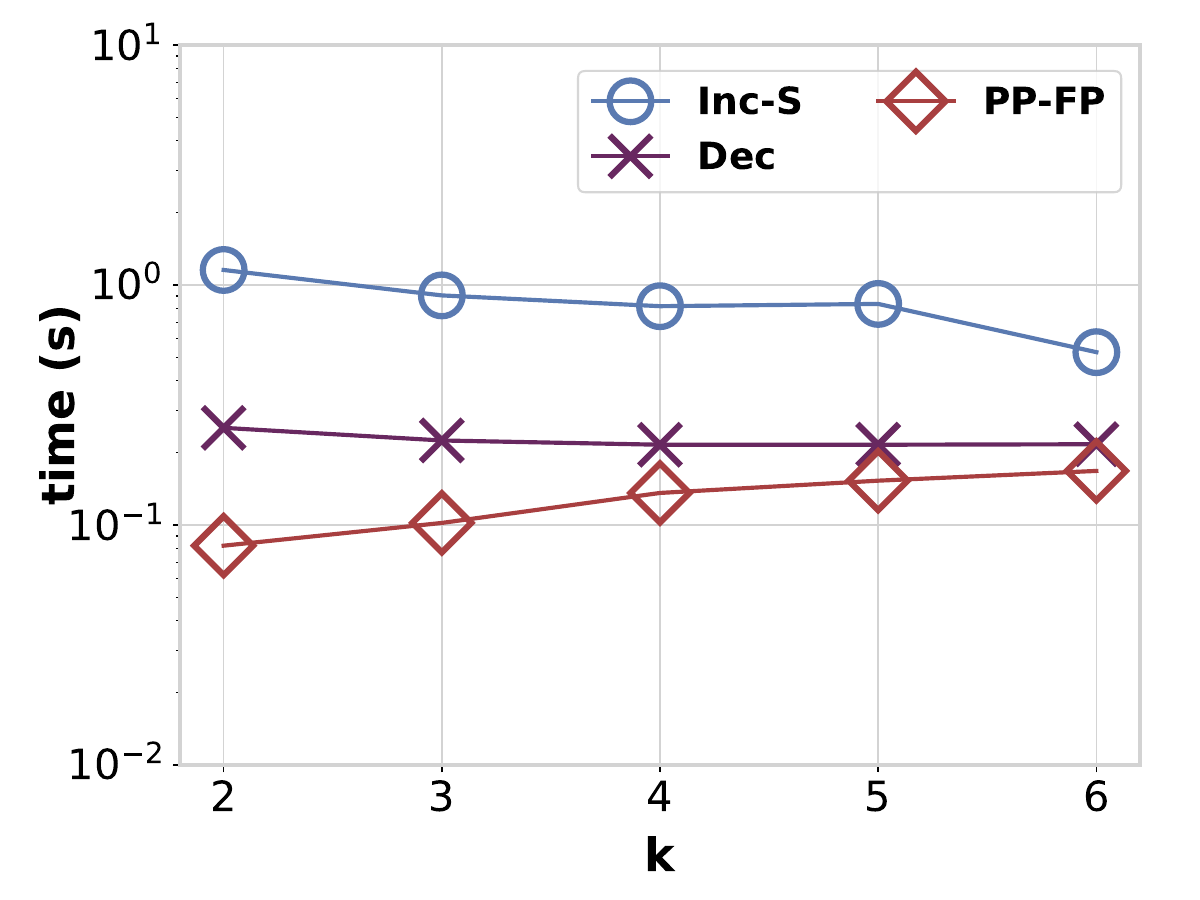}
        \caption{YouTube}
        \label{fig:eff-youtube}
    \end{subfigure}
    \begin{subfigure}[t]{0.242\textwidth}
        \centering
        \includegraphics[width=\textwidth, height=3.2cm]{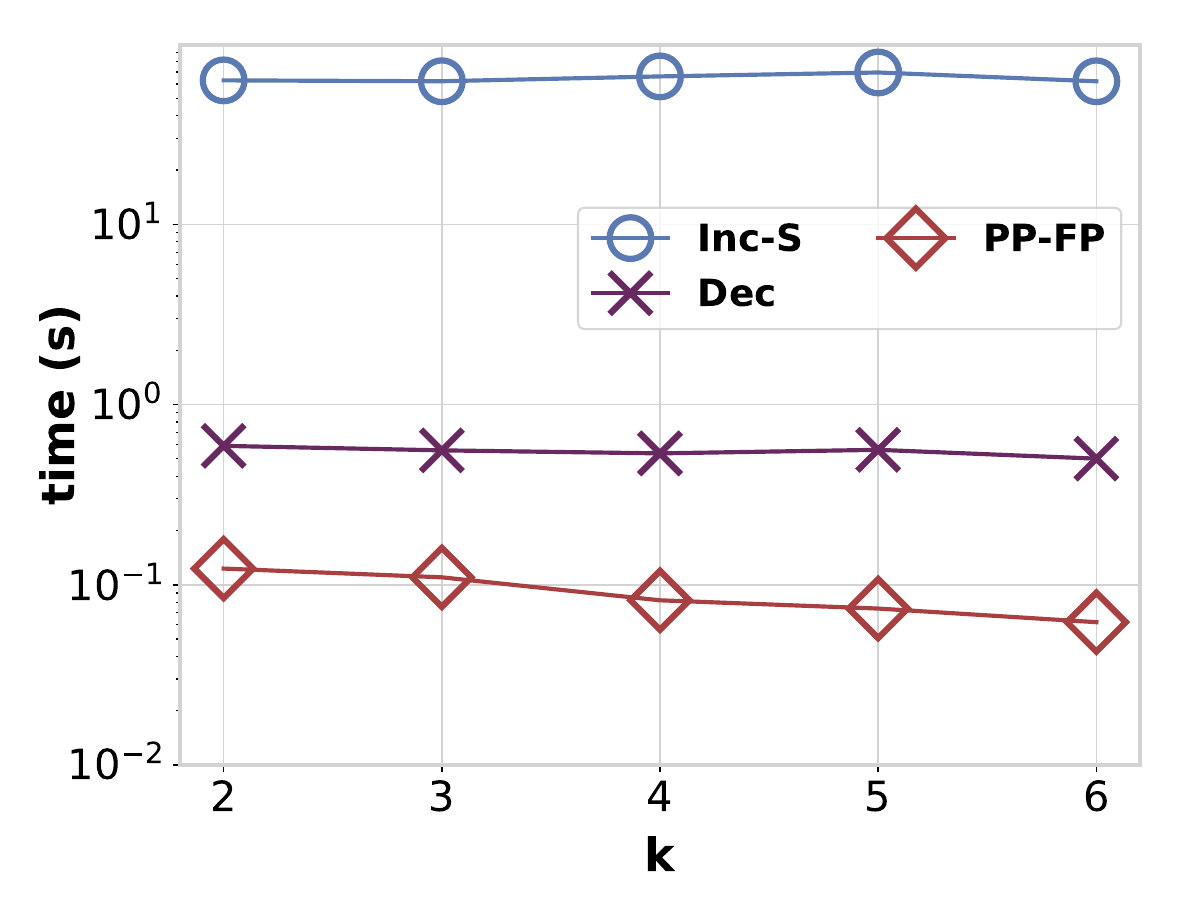}
        \caption{\blue{Orkut}}
        \label{fig:eff-orkut}
    \end{subfigure}

    \caption{Community search efficiency results on all datasets.}
    \label{fig:efficiency}
    \vspace{-0.3cm}
\end{figure*}

\noindent \textbf{Parameters and evaluation metrics.} 
% To evaluate \ACQPP, we firstly evaluate the efficiency of all community search methods on all datasets by comparing the running time. 
% Secondly, we assess the validation iteration results of four methods on five DBLP datasets.
% Thirdly, we assess the quality of the discovered communities. Since the DBLP datasets do not provide ground-truth communities, we use the Online-basic algorithm, an exhaustive graph search algorithm, as the benchmark. We randomly select query nodes from each dataset to evaluate \ACQPP queries.
% The quality is evaluated by comparing the number of shared attributes in the discovered community $H$, denoted as $|attr(H)|$. 
% For the Facebook, LiveJournal, and Youtube datasets, which provide ground-truth communities, we randomly select query nodes from the ground-truth community $\widehat{H}$ and use different algorithms to discover the community $H$. The quality of the discovered community is then compared using the F1 score, defined as $F1(H, \widehat{H}) = \frac{2 \cdot \text{prec}(H, \widehat{H}) \cdot \text{recall}(H, \widehat{H})}{\text{prec}(H, \widehat{H}) + \text{recall}(H, \widehat{H})}$ \cite{huang2016attribute}. 
% Additionally, we generate synthetic datasets to analyze the scalability of different algorithms. In the scalability test, we set the default value of $k$ to 3. 
We evaluate \ACQPP in terms of quality and efficiency. 
%For quality evaluation, we test the number of common attributes and F1-score on the networks with ground-truth communities. 
For quality evaluation, we use two metrics.
For datasets without ground-truth communities (DBLP), we measure the number of common attributes in the discovered community $|\text{attr}'(H)|$; for datasets with ground-truth communities (Facebook, LiveJournal, and YouTube, \blue{Orkut}), we adopt the F1-score between the discovered community $H$ and the ground-truth community $\widehat{H}$, defined as $F1(H, \widehat{H}) = \frac{2 \cdot \text{prec}(H, \widehat{H}) \cdot \text{recall}(H, \widehat{H})}{\text{prec}(H, \widehat{H}) + \text{recall}(H, \widehat{H})}$~\cite{huang2016attribute}. 
%In this case, we randomly select query nodes $q$ from each ground-truth community $\widehat{H}$ and apply different algorithms to discover the corresponding community $H$. 
For efficiency, we evaluate the runtime %both and iteration cost. Specifically, we compare the runtime 
of all methods across datasets.
%and report the number of validation iterations.
%For quality, we use two strategies. On DBLP datasets without ground-truth communities, we use the Online-basic method as the benchmark and measure the number of shared attributes in the discovered community $H$, denoted as $|attr(H)|$. On Facebook, LiveJournal, and Youtube datasets with ground-truth communities, we randomly select query nodes from the ground-truth community $\widehat{H}$ and use different algorithms to discover the community $H$. The quality of the discovered community is then compared using the F1 score, defined as $F1(H, \widehat{H}) = \frac{2 \cdot \text{prec}(H, \widehat{H}) \cdot \text{recall}(H, \widehat{H})}{\text{prec}(H, \widehat{H}) + \text{recall}(H, \widehat{H})}$ \cite{huang2016attribute}. %synthetic datasets and 
\blue{Furthermore, we evaluate scalability on 
efficiency across private attribute ranges and private edge ranges on real-world datasets.} The default parameter is $k = 3$.
%Furthermore, we evaluate scalability on synthetic datasets and efficiency across private edge ranges. We set the default parameter $k =3$.
%\subsection{Performance Evaluation} 
%\subsubsection{Efficiency Analysis}

\stitle{Exp-1: Community search efficiency evaluation.}
We compare the efficiency of our algorithm PP-FP with other existing community search methods from two categories: non-indexed search methods (i.e., Online-basic, Online-binary) and indexed methods (i.e., Inc-S~\cite{fang2016effective}, Dec~\cite{fang2017effective}) as shown in Figs.~\ref{fig:efficiency}(a)–(h). %Figs.~6(a)–(h) compare the efficiency of the baselines with our PP-FP algorithm in discovering attributed communities. 
The running time of Online-basic and Online-binary is generally longer due to the exhaustive enumeration of candidates%attribute sets
. In contrast, two index-based methods Inc-S and Dec, maintain relatively high efficiency. Nevertheless, PP-FP consistently achieves the fastest time, with only a slight increase as $k$ grows, demonstrating superior scalability and high efficiency.

\stitle{\blue{Exp-2: Index construction cost and memory usage.}} \blue{We evaluate the overhead of our method PP-FP in terms of graph size, index size, index construction time, and memory usage. The results shown in Table~\ref{dataset} demonstrate that the total index size and memory consumption grow steadily with the graph size, indicating PP-FP is practical for large attributed graphs.}

\begin{figure*}[t]
    \centering

    \begin{subfigure}[t]{0.242\textwidth}
        \centering
        \includegraphics[width=\textwidth, height=3.2cm]{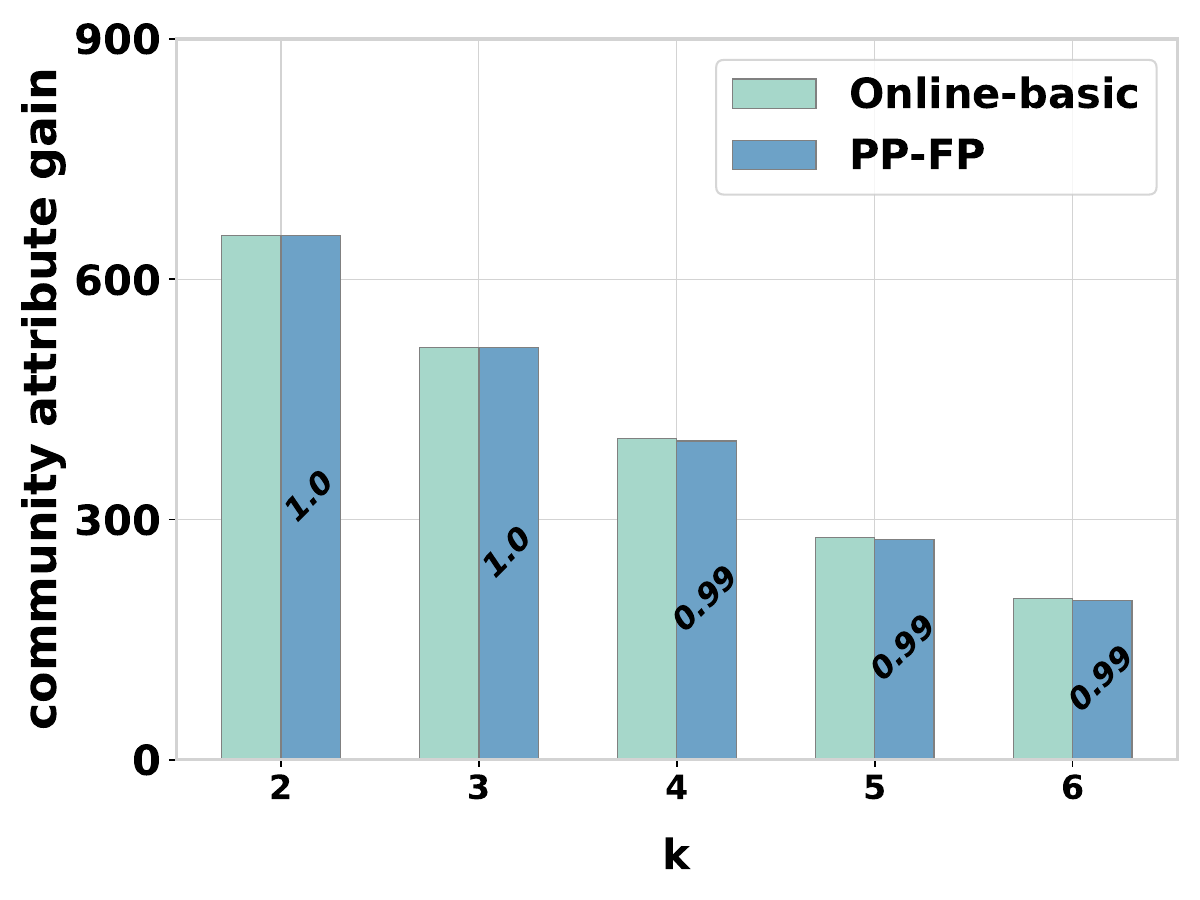}
        \caption{DBLP2013}
        \label{fig:qual-2013}
    \end{subfigure}
    \begin{subfigure}[t]{0.242\textwidth}
        \centering
        \includegraphics[width=\textwidth, height=3.2cm]{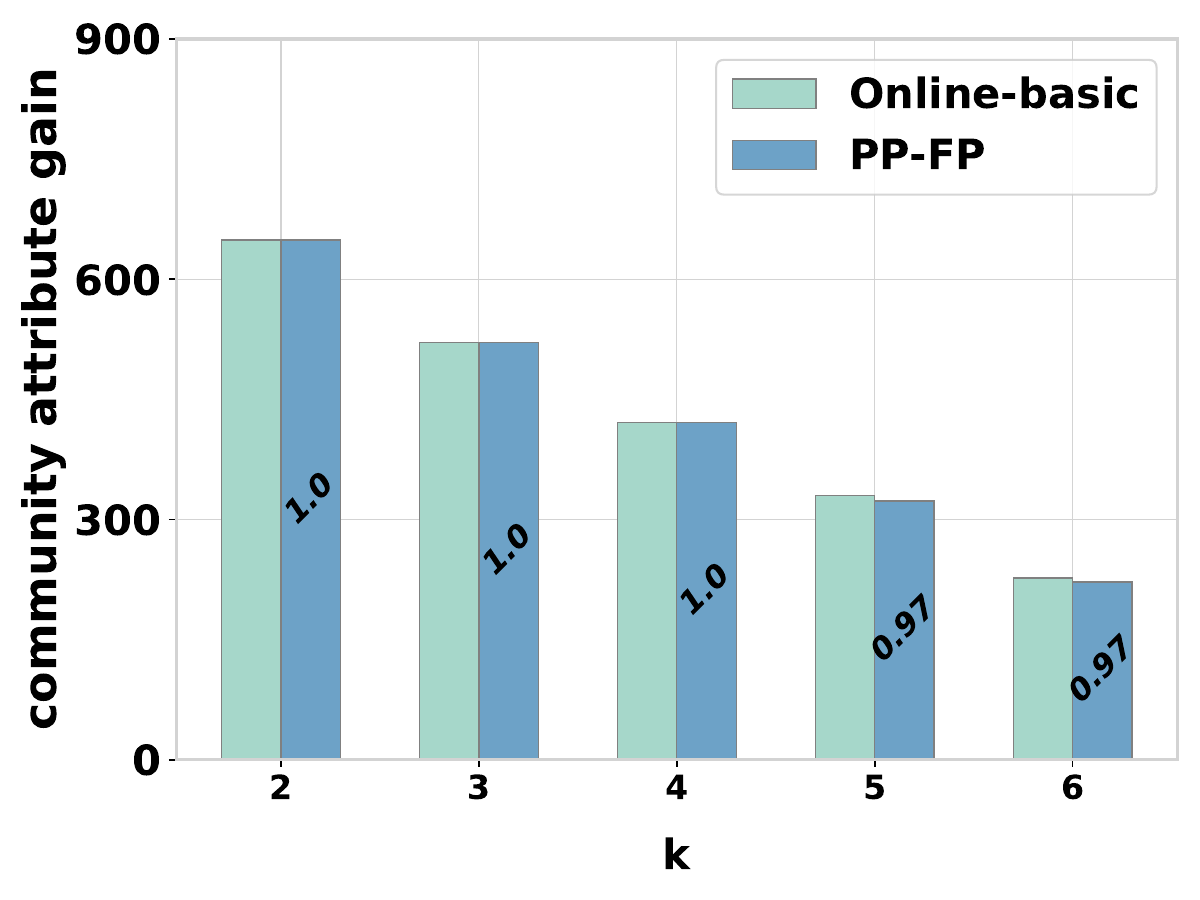}
        \caption{DBLP2014}
        \label{fig:qual-2014}
    \end{subfigure}
    \begin{subfigure}[t]{0.242\textwidth}
        \centering
        \includegraphics[width=\textwidth, height=3.2cm]{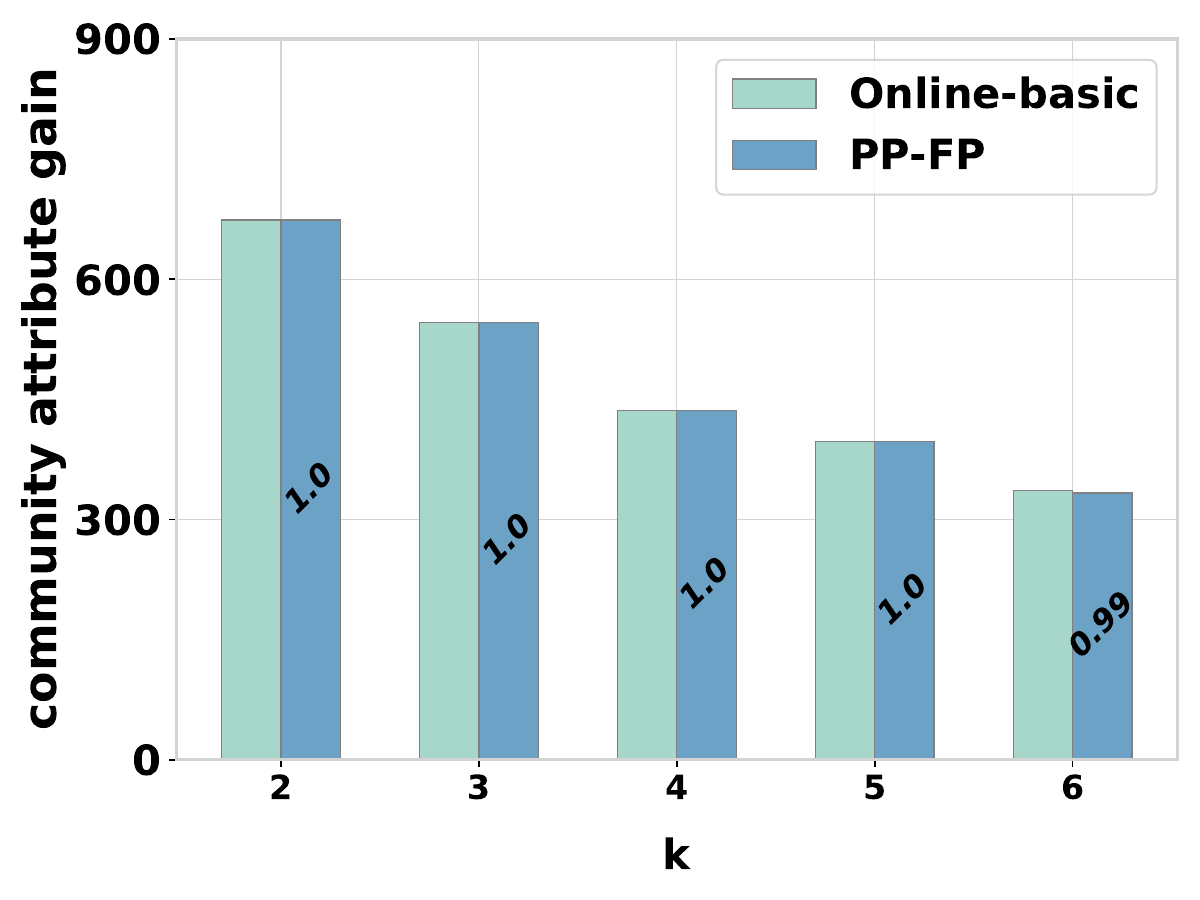}
        \caption{DBLP2015}
        \label{fig:qual-2015}
    \end{subfigure}
    \begin{subfigure}[t]{0.242\textwidth}
        \centering
        \includegraphics[width=\textwidth, height=3.2cm]{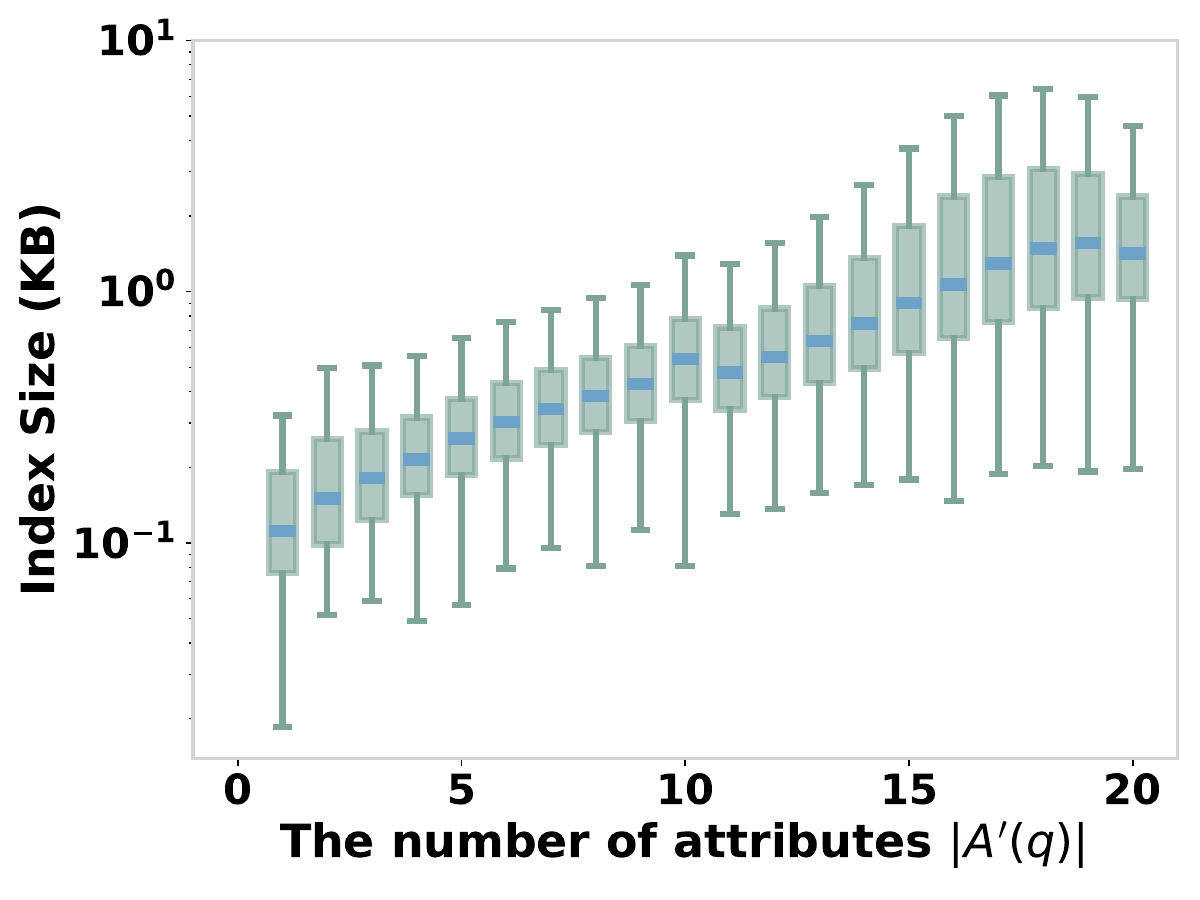}
        \caption{\blue{Sensitivity test}}
        \label{fig:sensitivity}
    \end{subfigure}

    \caption{\blue{Quality score evaluation on DBLP datasets (a)-(c) and sensitivity test of index size (d).}}
    \label{fig:quality_sensitivity}
    \vspace{-0.5cm}
\end{figure*}

\begin{figure*}[t]
\centering

\begin{minipage}[t]{0.242\textwidth}
  \centering
  \includegraphics[width=\textwidth,height=3.2cm]{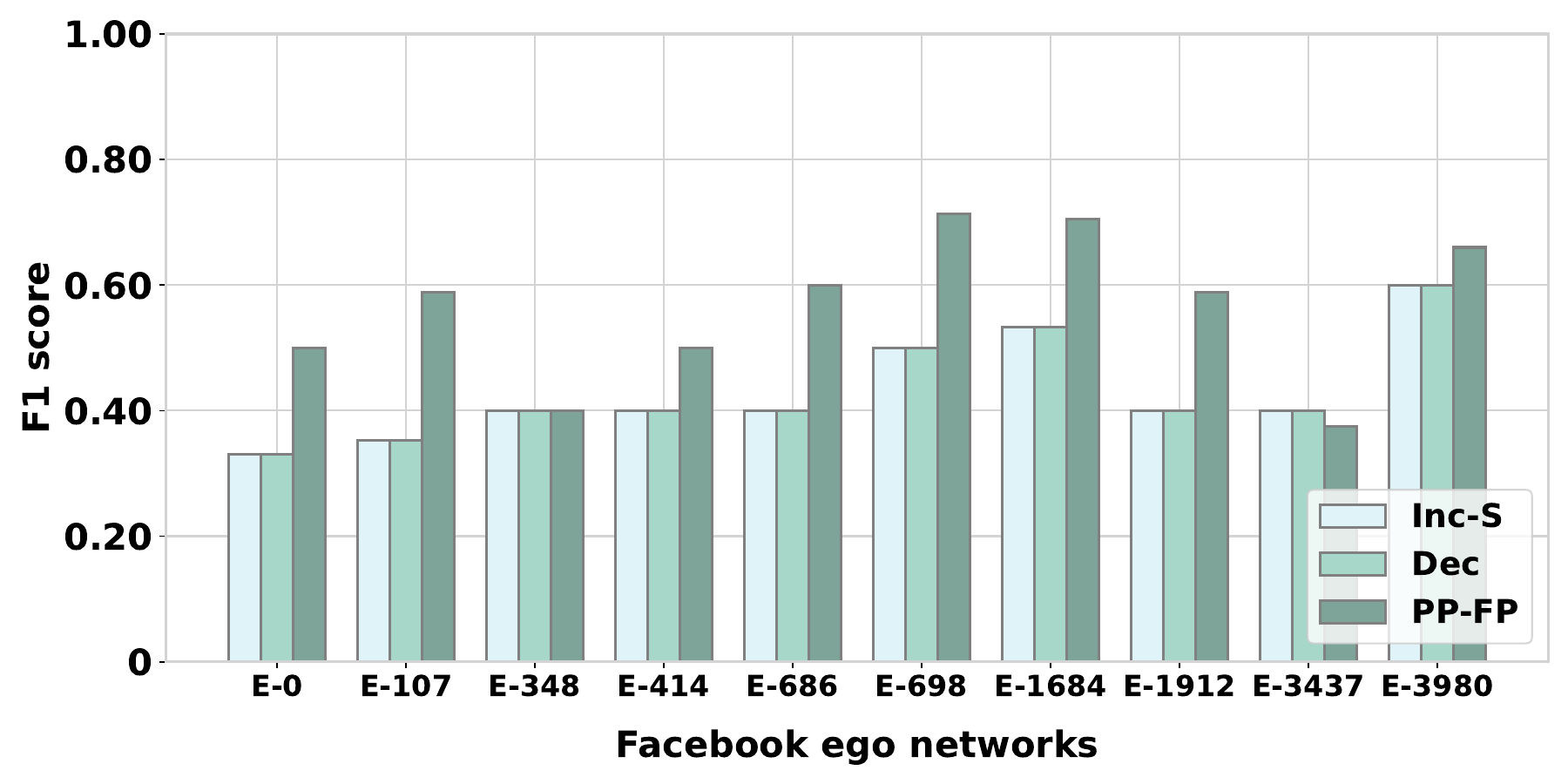}
  \vspace{-0.6cm}
  \captionof{figure}{F1-scores on Facebook.}
  \label{f1:fb}
\end{minipage}
\hfill
\begin{minipage}[t]{0.242\textwidth}
  \centering
  \includegraphics[width=\textwidth,height=3.2cm]{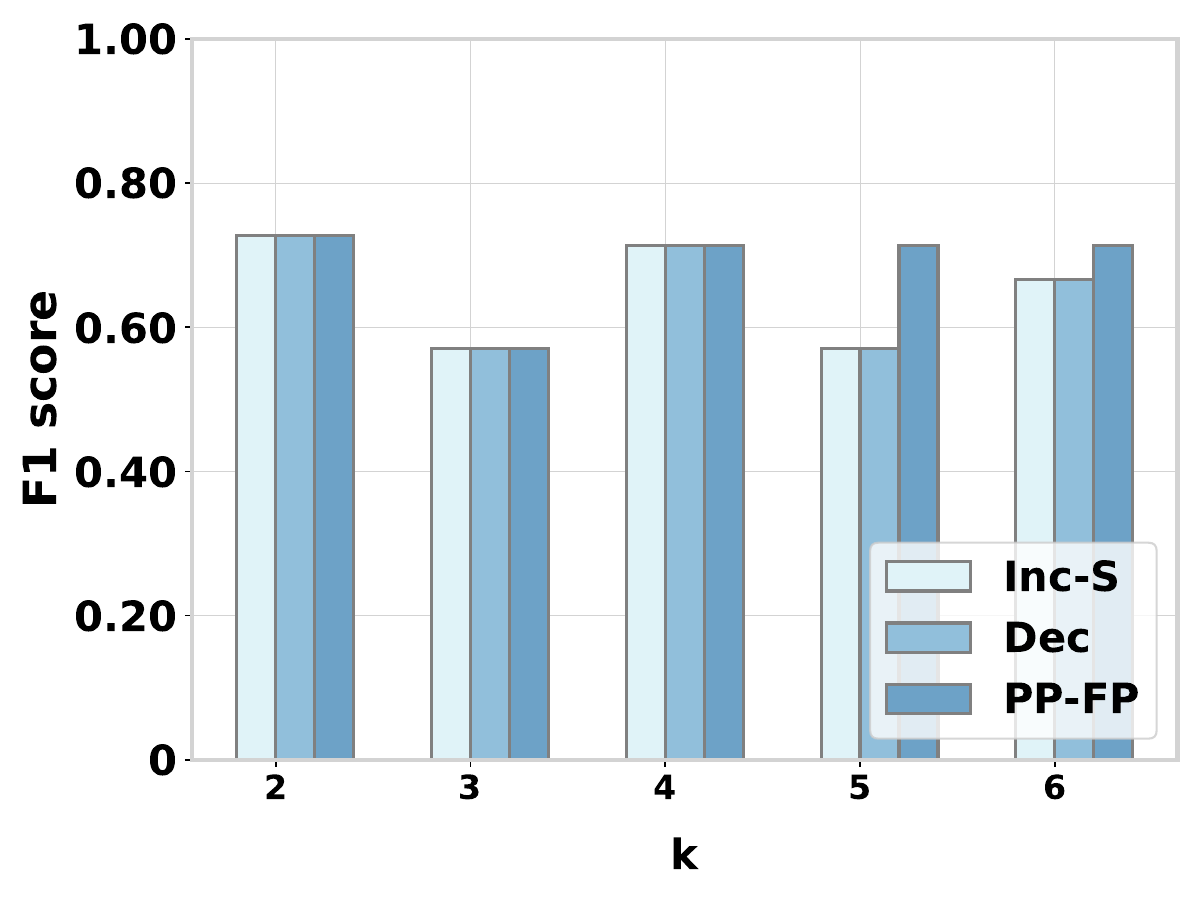}
  \vspace{-0.6cm}
  \captionof{figure}{F1 on LiveJournal.}
  \label{f1:lj}
\end{minipage}
\hfill
\begin{minipage}[t]{0.242\textwidth}
  \centering
  \includegraphics[width=\textwidth,height=3.2cm]{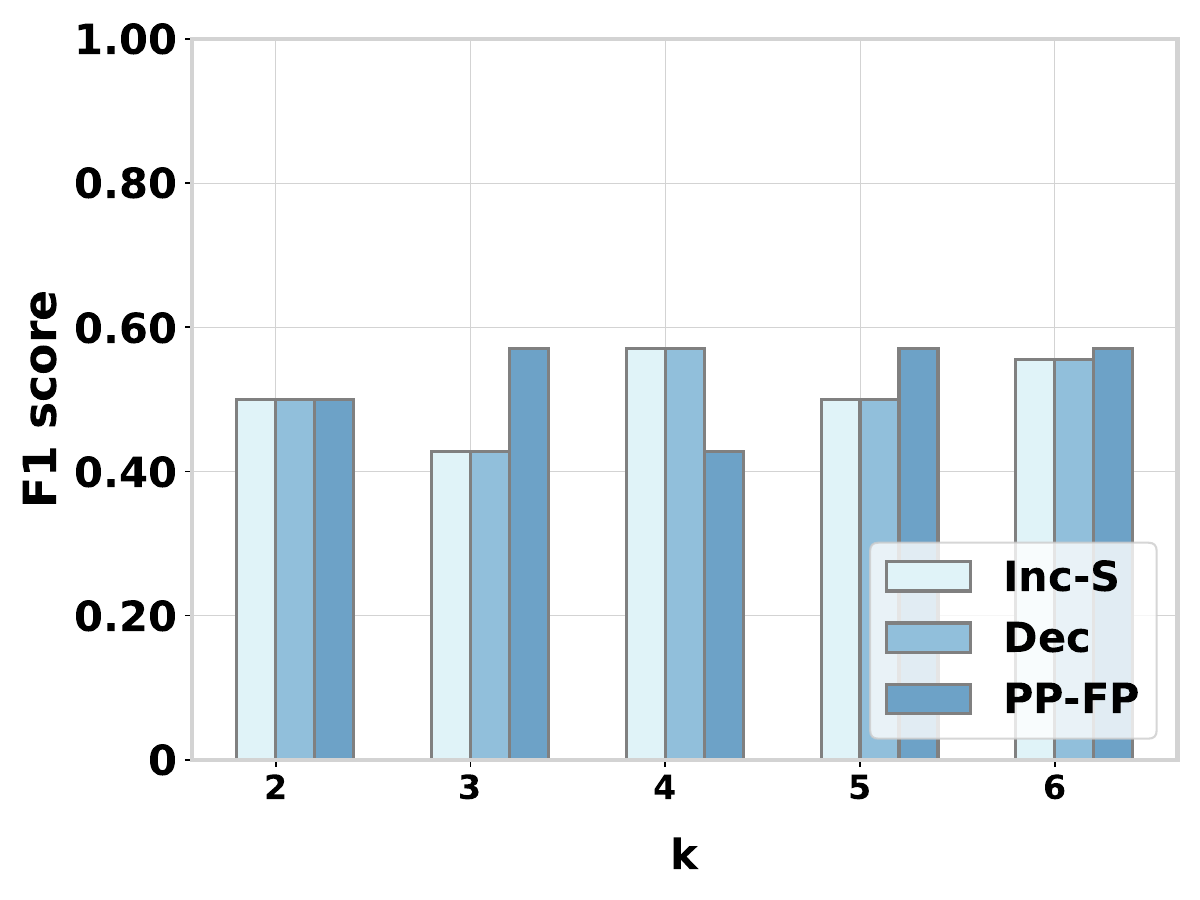}
  \vspace{-0.6cm}
  \captionof{figure}{F1 on YouTube.}
  \label{f1:yt}
\end{minipage}
\hfill
\begin{minipage}[t]{0.242\textwidth}
  \centering
  \includegraphics[width=\textwidth,height=3.2cm]{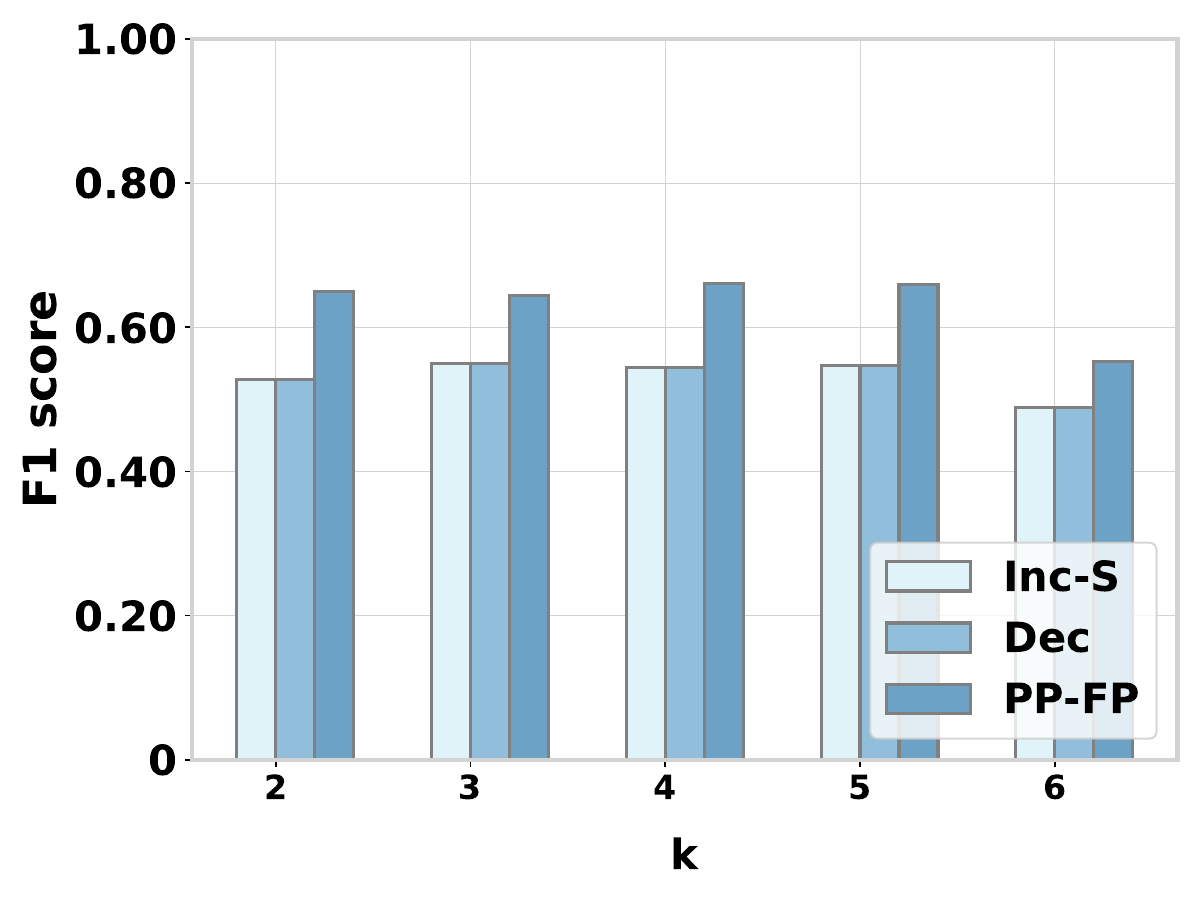}
  \vspace{-0.6cm}
  \captionof{figure}{\textcolor{black}{F1-scores on Orkut.}}
  \label{f1:orkut}
\end{minipage}

\vspace{-0.25cm}
\end{figure*}

\noindent \textbf{Exp-3: Quality evaluation.} 
%For the DBLP datasets, which lacks ground-truth communities, we measure the quality score by the common attributes in the discovered community, represented by $|attr(H)|$. To ensure a comprehensive evaluation, we randomly select 100 query nodes from the DBLP datasets, each with its private edges in the range of $[0, 500]$. 
%As shown 
We evaluate the quality of the discovered communities in Figures~\ref{fig:quality_sensitivity}-\ref{f1:orkut}. Figures~\ref{fig:quality_sensitivity}(a)–(c) show that PP-FP achieves almost the same community attribute gains as Online-basic while running five orders of magnitude faster. 
Although PP-FP performs slightly worse on a few query nodes because some attributes extend beyond their public-private neighborhoods, its average community attribute gain remains above 0.97, demonstrating a good balance between accuracy and efficiency.
%As shown in Fig.~9, for a small portion of query nodes, our PP-FP method performs slightly worse than Online-basic, which performs exhaustive enumeration. This is because some query nodes’ attributes may extend beyond their direct public and private neighbors, and PP-FP focuses on attributes within this neighborhood. So in rare cases, it may miss certain relevant attributes from distant neighbors, leading to a lower score compared to Online-basic. However, our algorithm is five orders of magnitude faster than Online-basic while maintaining a quality score above 0.97, highlighting its significant advantage in fast and high-quality community search.
%For the Facebook, LiveJournal, and YouTube datasets with ground-truth communities, we randomly select query nodes from the ground-truth communities provided by these three datasets. As shown in Figs.~10-12, we select the Inc-S and Dec algorithms for community quality comparison with our PP-FP algorithm. We observe that F1 score of the communities identified by our PP-FP consistently outperforms those identified by Inc-S and Dec. This demonstrates the superior effectiveness of our method in capturing high-quality communities.
% \textbf{For datasets with ground-truth communities (Facebook, LiveJournal, and YouTube), 
% we evaluate quality using the F1-score between the discovered and the ground-truth communities. As shown in Figs.~10-12, PP-FP consistently achieves higher F1-scores than Inc-S and Dec, demonstrating the superior effectiveness of our method in identifying high-quality communities.}
We report the F1-scores between the discovered and the ground-truth communities in Figures~\ref{f1:fb}-\ref{f1:orkut}. The results show that PP-FP consistently achieves higher F1-scores than Inc-S and Dec, demonstrating the superior effectiveness of our method in identifying high-quality communities.

\begin{figure}[t]
\vspace{-0.3cm}
\centering

\begin{subfigure}[t]{0.241\textwidth}
    \centering
    \includegraphics[width=\textwidth, height=3cm]{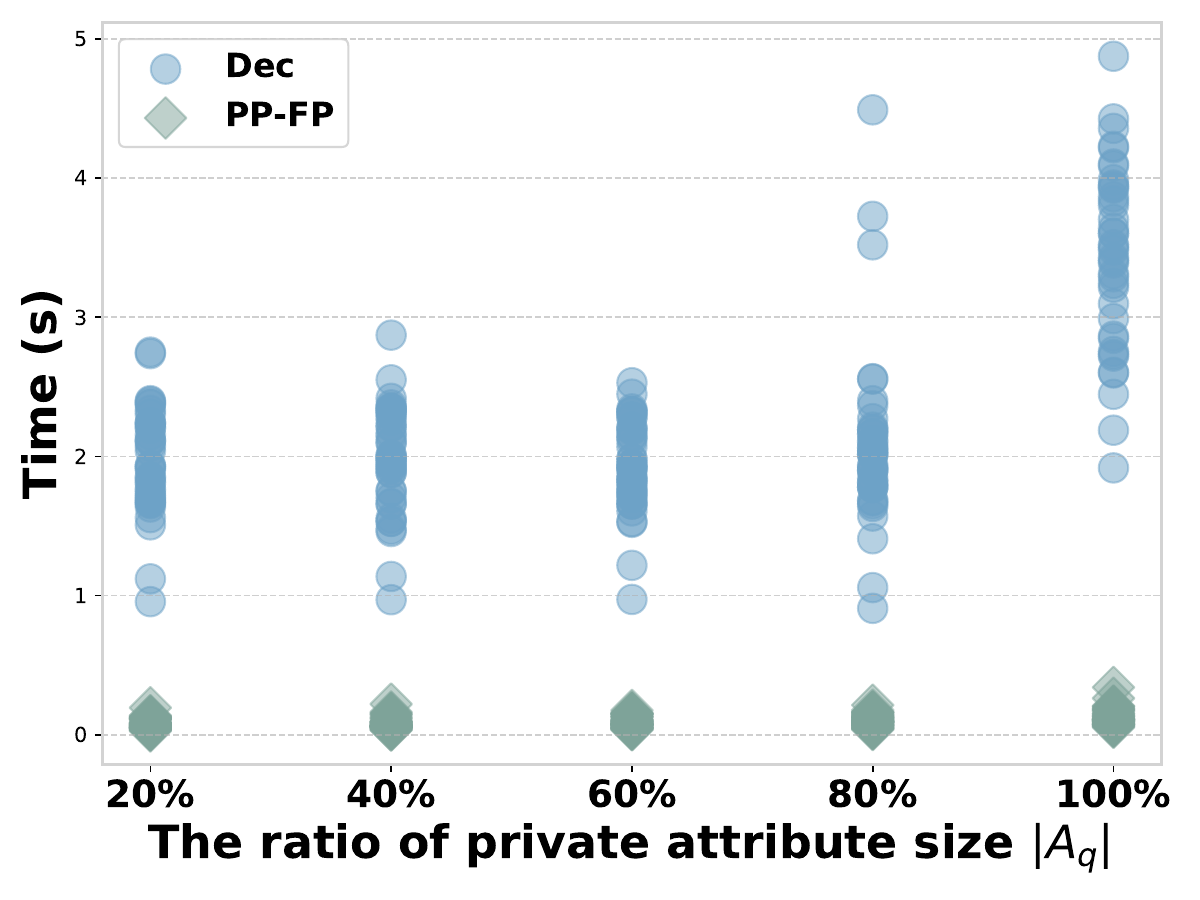}
    \vspace{-0.5cm}
    \caption{\textcolor{black}{On private attribute size $|A_q|$.}}
    \label{fig:scalability_attr}
\end{subfigure}
\hfill
\begin{subfigure}[t]{0.241\textwidth}
    \centering
    \includegraphics[width=\textwidth, height=3cm]{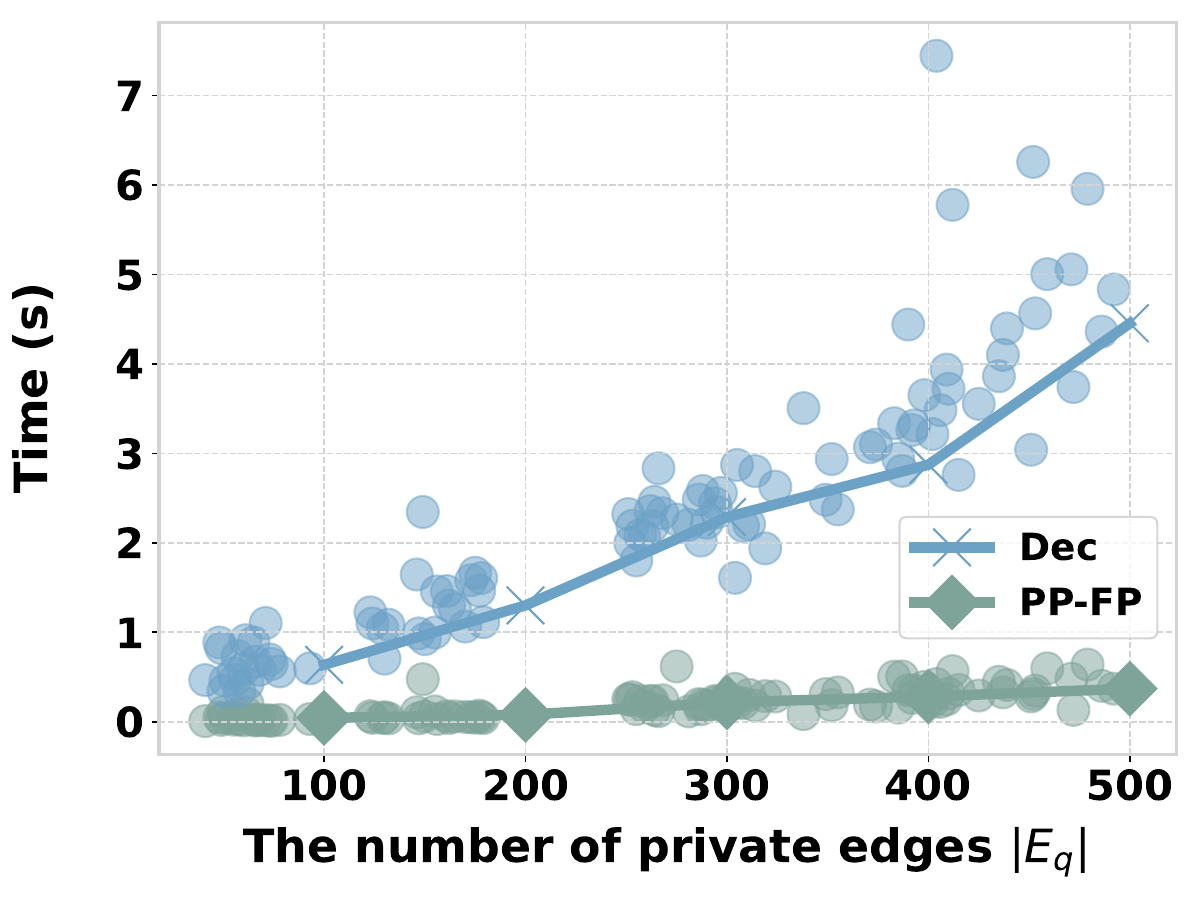}
    \vspace{-0.5cm}
    \caption{\textcolor{black}{On private edge size $|E_q|$.}}
    \label{fig:scalability_edge}
\end{subfigure}
\caption{\textcolor{black}{Scalability test.}}
\label{fig:scalability}
\vspace{-0.2cm}
\end{figure}

\begin{figure}[t]
\vspace{-0.2cm}
\centering

\begin{subfigure}[t]{0.241\textwidth}
    \centering
    \includegraphics[width=\textwidth, height=3cm]{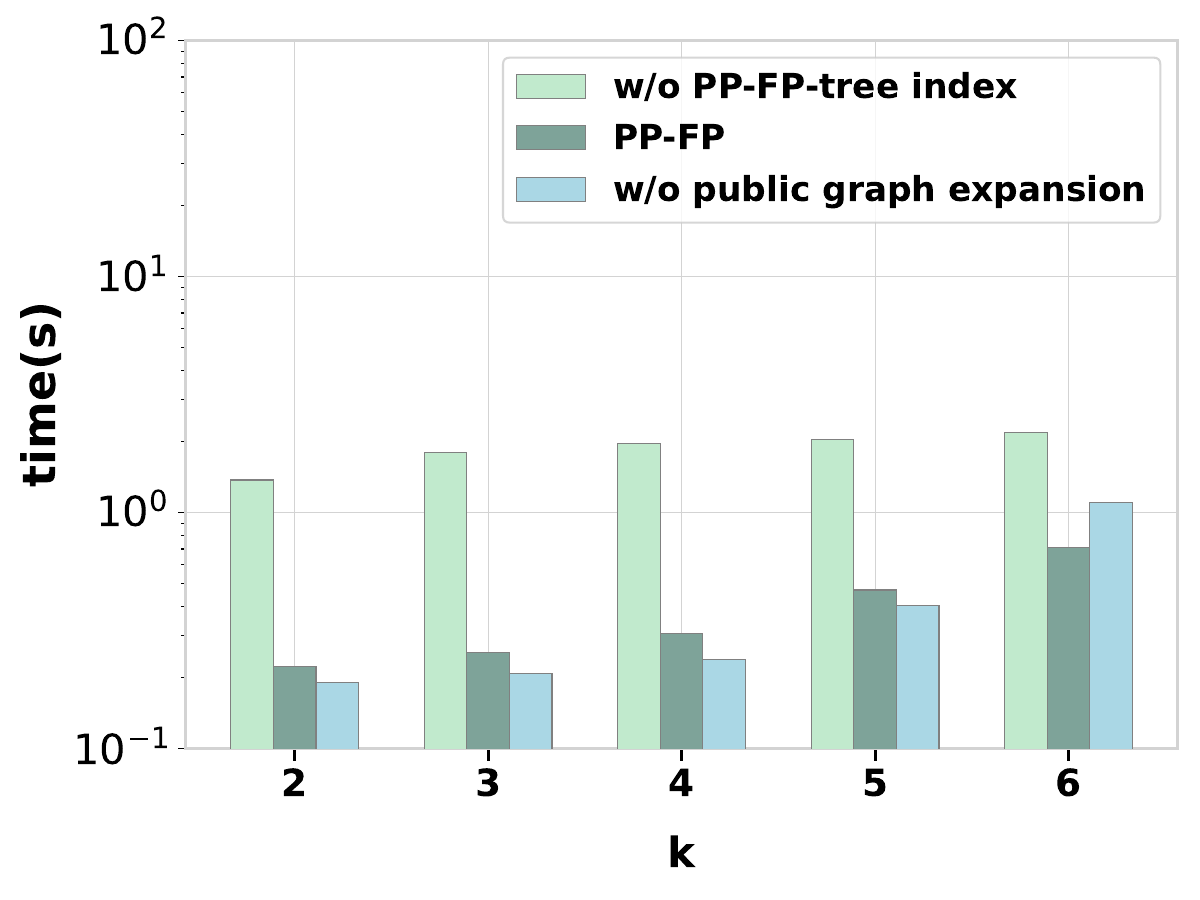}
    \vspace{-0.55cm}
    \caption{\textcolor{black}{Efficiency.}}
    \label{fig:ablation_eff}
\end{subfigure}
\hfill
\begin{subfigure}[t]{0.241\textwidth}
    \centering
    \includegraphics[width=\textwidth, height=3cm]{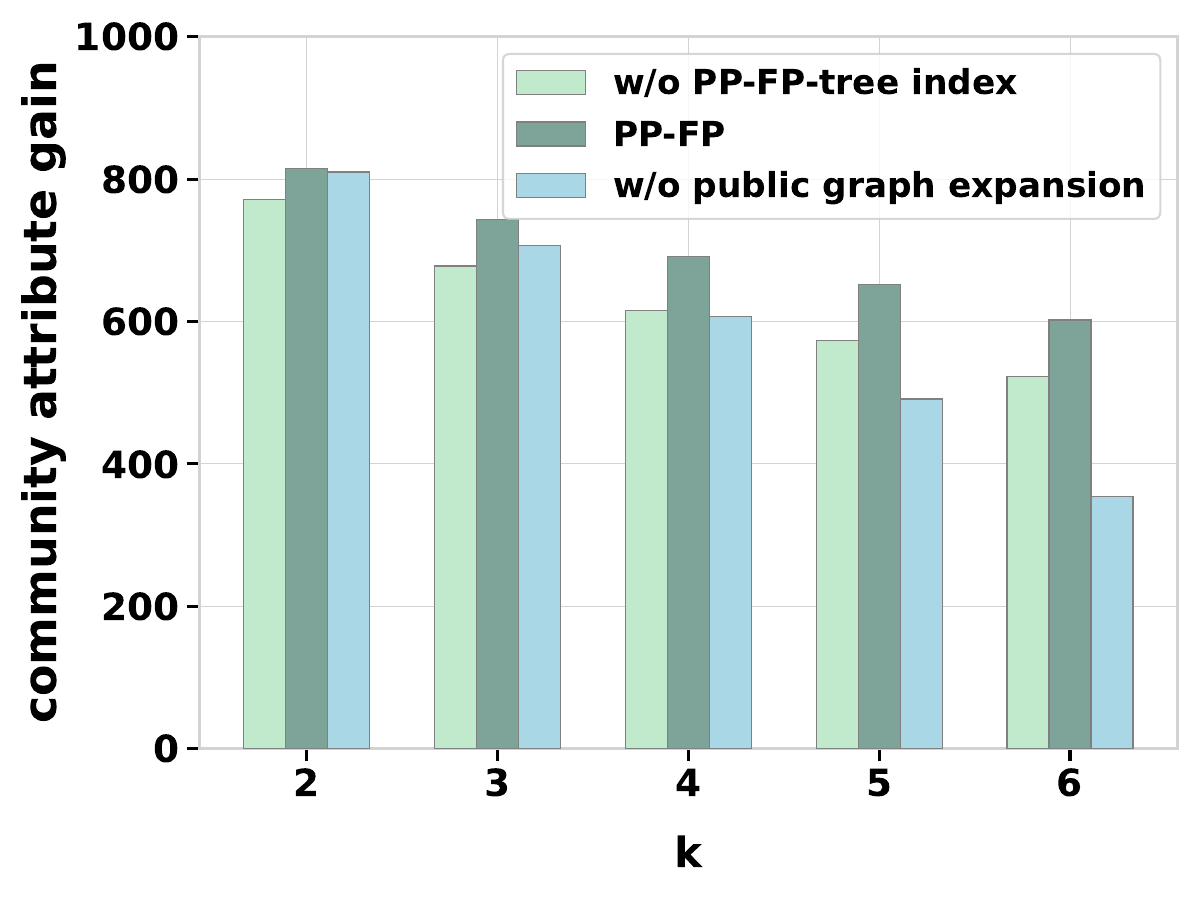}
    \vspace{-0.55cm}
    \caption{\textcolor{black}{Quality.}}
    \label{fig:ablation_qual}
\end{subfigure}
\caption{\textcolor{black}{Ablation study.}}
\label{fig:ablation}
\vspace{-0.6cm}
\end{figure}

% \begin{figure}[t]
% \vspace{-0.2cm}
%     \centering
%     \includegraphics[width=0.5\textwidth, height=3cm]{exp_figures/sensitivity.pdf}
%     \caption*{\blue{Fig.~15: Sensitivity test.}}
%     \label{fig:sensitivity}
%     \vspace{-0.5cm}
% \end{figure}

\noindent \textbf{Exp-4: Scalability test.}
We conducted scalability tests on DBLP2017 dataset to evaluate our method against the competitor by 
\blue{varying private attribute size and} 
private edge size.

\noindent\blue{\textbf{(I) Scalability test w.r.t. private attribute size $|A_q|$.}}
%\noindent\textbf{(I) Scalability test w.r.t. the whole graph size $|V|$.}
%\noindent \textbf{Exp-4: Scalability test.} 
%\blue{We first use the DBLP2017 dataset, with private attribute sizes of the query nodes ranging from 20\% to 100\%, and compare the runtime of our PP-FP with the most efficient baseline Dec.}
\blue{We first collect query nodes with private neighbors range from 200 to 250. For each query node, we vary the size of its private attribute set from 20\% to 100\%, and compare the runtime of our PP-FP with the most efficient baseline Dec in Figure~\ref{fig:scalability}(a). Since Dec needs to update its index after acquiring private information, its runs slower than our method PP-FP with sufficient private information.}

\noindent\textbf{(II) Scalability test w.r.t. private edge size $|E_q|$.}
%\stitle{Exp-5: Efficiency evaluation on private edges $|E_q|$.} 
We further examine how query performance is affected by the amount of private information.
%To evaluate how the performance of search methods scales with increasing private information
Specifically, we collect query nodes whose private edge counts range from 0 to 500 and compare the runtime of our PP-FP with Dec. 
As shown in Figure~\ref{fig:scalability}(b), the solid lines represent the average runtimes for both methods, while the circular markers indicate the runtime of each query node. PP-FP shown in green maintains a consistently low and stable runtime across all ranges, exhibiting little sensitivity to the growth in private edges, whereas Dec shown in blue slows down markedly as $|E_q|$ increases.
%These results demonstrate that PP-FP provides better scalability for public-private graph scenarios where query nodes involve substantial private information.

%Overall, PP-FP exhibits outstanding scalability in both the public and private graph dimensions, providing superior performance for public–private graph scenarios where query nodes involve substantial private information.
%In summary, our proposed PP-FP approach demonstrates a good scalability by \blue{private attribute size $|A_q|$} and private graph size $|E_q|$, which shows stable increment tend w.r.t. Dec.  
In summary, our proposed PP-FP approach demonstrates good scalability with respect to the private attribute size $|A_q|$ and the private graph size $|E_q|$, exhibiting a more stable increasing trend compared to Dec.
%As shown in Fig.~13, blue and green markers correspond to Dec and PP-FP, respectively, while the solid lines indicate average runtimes. It is evident that PP-FP maintains a consistently low and stable runtime across the entire range, showing little sensitivity to the amount of private information. In contrast, Dec’s runtime increases substantially with more private edges. 

\stitle{\blue{Exp-5: Sensitivity test.}} \blue{We evaluate the sensitivity of the PP-FP-tree index size to the number of attributes $|A'(q)|$ using all query nodes in the DBLP2017 dataset. For each query node $q\in V$, we record the size of its PP-FP-tree index and the number of attributes. Figure~\ref{fig:quality_sensitivity}(d) shows an approximately linear growth trend, indicating that the index size grows linearly with the number of attributes.}

\stitle{\blue{Exp-6: Ablation Study.~}}\blue{We conduct two ablation studies on DBLP2017 with 100 query nodes to evaluate the impact of different components of our method in efficiency and community quality.}

\noindent\textbf{\blue{Ablation 1: PP-FP-tree index.
}}\blue{We remove the PP-FP-tree index and enumerate candidate nodes with attributes from the public-private neighborhood, followed by Phase-II and Phase-III. This variant evaluates the effect of the PP-FP-tree.}

\noindent\textbf{\blue{Ablation 2: public graph expansion.
}}\blue{We remove the candidate subgraph expansion in public graph, using Phase-I directly followed by Phase-III. This variant evaluates the effect of subgraph expansion.}

\blue{Figure~14 presents the ablation results on efficiency and quality. Across all values of $k$, PP-FP achieves higher community attribute gain than the two ablation variants. PP-FP is consistently more efficient than the variant without PP-FP-tree index, as the tree index enables effective pruning by organizing nodes with shared attributes. When $k$ is 2 to 5, the variant without public graph expansion is more efficient than PP-FP, since it does not expand in the public graph. Interestingly, when $k$ is 6, PP-FP is faster than the variant without public graph expansion, since it is hard to identify a valid 6-core community within the public-private neighborhood, leading to both reduced efficiency and quality.}

\stitle{Exp-7: Case study}. We conduct a case study on the DBLP2017 dataset, focusing on a prominent researcher in computer vision, Ming-Hsuan Yang. We select him as the query node $q$, and search for 4-core communities with maximal common attributes.
%in the public graph $G$ and the public-private graph $G'_q$. 
Figure~15 presents our findings. 
In the public graph $G$, the identified 4-core community consists of eight nodes but is not highly cohesive. %, as it is not fully connected.
The query node is directly linked to only four of the remaining seven nodes, and the community shares only four attributes. %\{detection, image, learning, using\}. 
In contrast, the community identified in the public-private graph is a 4-core community %of five nodes 
forming a clique, representing %an extremely dense subgraph where 
all community members are tightly interconnected. 
Moreover, this community shares seven common attributes. %\{visual, object, detection, tracking, image, learning, using\}
Among these, the keywords highlighted in red \{visual, object, tracking\} are present only in the public-private graph. This clearly demonstrates that the public-private graph captures more comprehensive and up-to-date information. 
Additionally, in the public graph, neither Weiming Hu nor Philip Torr has a direct collaboration with Ming-Hsuan Yang. However, these connections are successfully captured in the public-private graph, reflecting their real-world collaboration at ECCV 2018. 
%This case highlights the advantages of using public-private graphs for community search. 
Compared to public graphs, public-private graphs facilitate the early discovery of new connections and enable the detection of more semantically meaningful, tightly connected, and timely communities.

\begin{figure}[t]
  \centering
  \vspace{-0.3cm}
 \includegraphics[width=\linewidth]{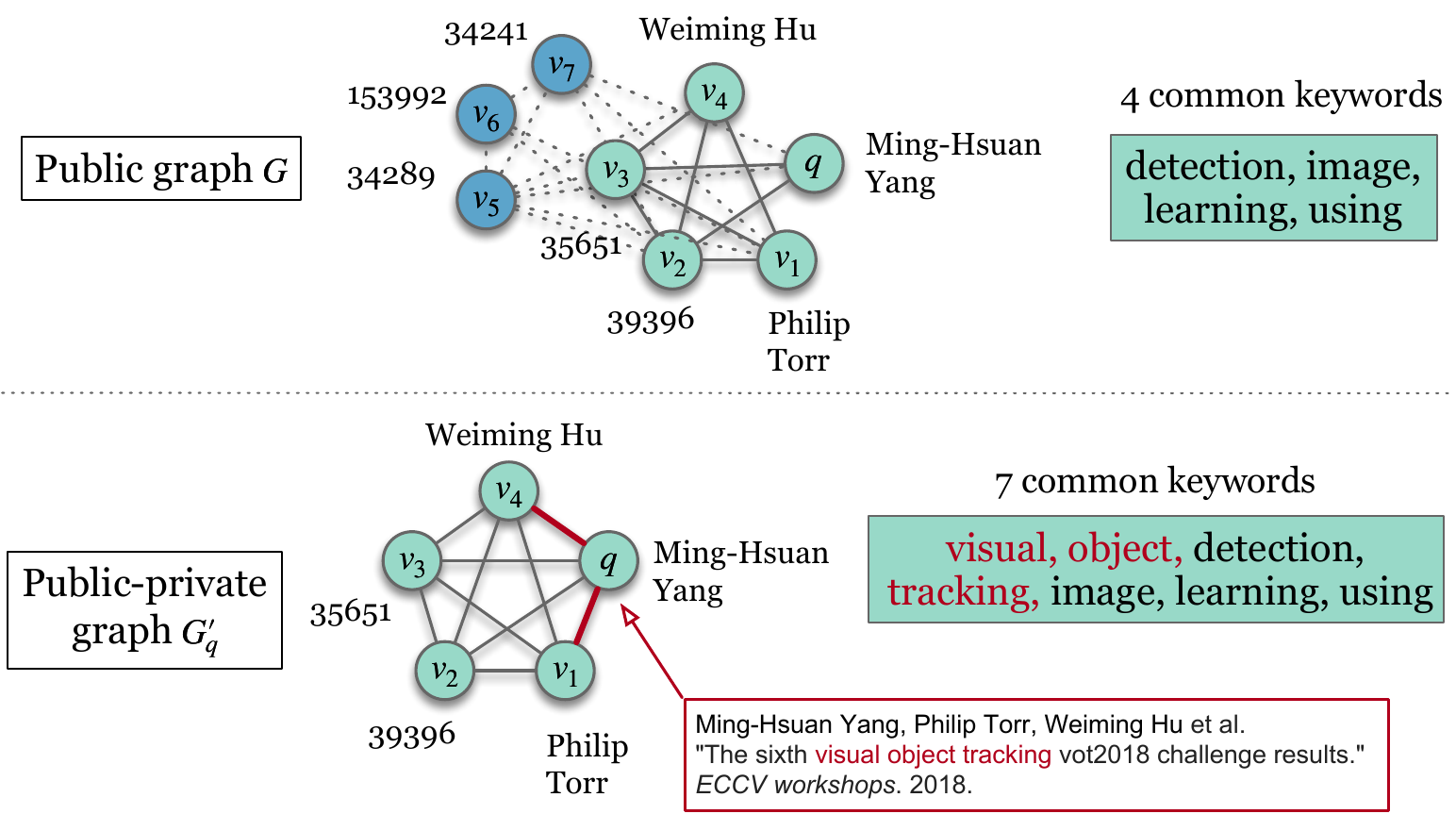}
  \caption{The 4-core attributed communities with maximum common attributes in the public and public-private graphs.}
  %DBLP graph in 2017. }
  \label{case}
\vspace{-0.5cm}
\end{figure}